\documentclass[aps,prd,reprint,
amsmath,amssymb,
]{revtex4-2}
\usepackage{acro} 
\usepackage{graphicx}
\usepackage{dcolumn}
\usepackage{bm}
\usepackage{hyperref}
\usepackage{float}
\usepackage{enumitem}

\newcommand{\PW}{1.00}         

\newcommand{\syidx}[1]{%
	\acs{#1}%
}

\DeclareAcronym{SYMed}{
	short=\rho,
	long=Energy density,
	tag=symbol
}
\DeclareAcronym{SYMedbg}{
	short=\bar{\rho},
	long=Energy density of background Universe,
	tag=symbol
}
\DeclareAcronym{SYMedavg}{
	short=\langle\bar{\rho}\rangle,
	long=Energy density averaged over oscillation period,
	tag=symbol
}
\DeclareAcronym{SYMpr}{
	short=p,
	long=Pressure,
	tag=symbol
}
\DeclareAcronym{SYMprbg}{
	short=\bar{p},
	long=Pressure of background Universe,
	tag=symbol
}
\DeclareAcronym{SYMpravg}{
	short=\langle\bar{p}\rangle,
	long=Pressure averaged over oscillation period,
	tag=symbol
}
\DeclareAcronym{SYMdenspar}{
	short=\Omega,
	long=Density parameter for component $i$,
	tag=symbol
}
\DeclareAcronym{SYMDE}{
	short=\Lambda,
	long=Dark Energy,
	tag=symbol
}
\DeclareAcronym{SYMH}{
	short=H,
	long=Hubble parameter in cosmic time,
	tag=symbol
}
\DeclareAcronym{SYMHc}{
	short=\mathcal{H},
	long=Hubble parameter in conformal time,
	tag=symbol
}
\DeclareAcronym{SYMG}{
	short=G,
	long=Gravitational constant,
	tag=symbol
}
\DeclareAcronym{SYMc}{
	short=c,
	long=Speed of light,
	tag=symbol
}
\DeclareAcronym{SYMa}{
	short=a,
	long=Scalefactor a(t),
	tag=symbol
}
\DeclareAcronym{SYMg}{
	short=g,
	long=FLRW metric,
	tag=symbol
}
\DeclareAcronym{SYMgbg}{
	short=\bar{g},
	long=FLRW metric of background,
	tag=symbol
}
\DeclareAcronym{SYMLdens}{
	short=\mathcal{L},
	long=Lagrangian density,
	tag=symbol
}
\DeclareAcronym{SYMsfbec}{
	short=\psi,
	long=wave function of BEC,
	tag=symbol
}
\DeclareAcronym{SYMsf}{
	short=\varphi,
	long=Scalar Field,
	tag=symbol
}
\DeclareAcronym{SYMsfbg}{
	short=\bar{\varphi},
	long=Scalar Field,
	tag=symbol
}
\DeclareAcronym{SYMsfperturb}{
	short=\phi,
	long=Perturbations of Scalar Field,
	tag=symbol
}
\DeclareAcronym{SYMvpot}{
	short=V,
	long=Potential of the Scalar Field Dark Matter,
	tag=symbol
}
\DeclareAcronym{SYMlambdasi}{
	short=\lambda,
	long=Strength of self-interaction,
	tag=symbol
}
\DeclareAcronym{SYMlambdasit}{
	short=\tilde{\lambda},
	long=Strength of self-interaction,
	tag=symbol
}
\DeclareAcronym{SYMdenscontrast}{
	short=\delta,
	long=The density contrast describes the relative deviation of density from the average density of the Universe,
	tag=symbol
}
\DeclareAcronym{SYMcurvpar}{
	short=\kappa,
	long={Curverture parameter: -1..open Universe, +1..closed Universe, 0..flat Universe},
	tag=symbol
}
\DeclareAcronym{SYMeos}{
	short=w,
	long={The equation of state (EOS) relates pressure to energy density},
	tag=symbol
}
\DeclareAcronym{SYMderivD}{
	short=D,
	long={The covariant derivative with respect to $\mu$},
	tag=symbol
}
\DeclareAcronym{SYMkron}{
	short=\delta_{ij},
	long={Kronecker delta},
	tag=symbol
}
\DeclareAcronym{SYMmpsynh}{
	short=h,
	long={Metric perturbation in synchronous gauge},
	tag=symbol
}
\DeclareAcronym{SYMmpsyneta}{
	short=\eta,
	long={Metric perturbation in synchronous gauge},
	tag=symbol
}
\DeclareAcronym{SYMmpnewpot}{
	short=\Phi,
	long={Metric perturbation in newtonian gauge -- potential},
	tag=symbol
}
\DeclareAcronym{SYMmpnewlapse}{
	short=\Psi,
	long={Metric perturbation in newtonian gauge -- laps function},
	tag=symbol
}
\DeclareAcronym{SYMct}{
	short=\tau,
	long={Conformal time},
	tag=symbol
}
\DeclareAcronym{SYMt}{
	short=t,
	long={Cosmological time},
	tag=symbol
}
\DeclareAcronym{SYMemt}{
	short=T,
	long={Energy momentum tensor},
	tag=symbol
}
\DeclareAcronym{SYMemtbg}{
	short=\bar{T},
	long={Energy momentum tensor for background},
	tag=symbol
}
\DeclareAcronym{SYMemtvelocity}{
	short=\theta,
	long={Velocity divergence of the energy-momentum tensor},
	tag=symbol
}
\DeclareAcronym{SYMemtstress}{
	short=\sigma,
	long={Anisotropic stress of the energy-momentum tensor},
	tag=symbol
}
\usepackage{empheq}
\newcommand{\EQidx}[3][]%
{%
	\ifx&#1&%
	\begin{equation} \label{eq:#2}%
	\csname #2\endcsname#3%
	\end{equation}
	\else%
	\begin{equation} \label{eq:#1}%
	\csname #2\endcsname#3%
	\end{equation}
	\fi%
}
\newcommand{\EQgrp}[3][]{%
	\ifx&#1&%
\begin{subequations} \label{eq:#2}%
	\begin{empheq}{align}
	\csname #2\endcsname#3%
	\end{empheq}
\end{subequations}
	\else%
\begin{subequations} \label{eq:#1}%
	\begin{empheq}{align}
	\csname #2\endcsname#3%
	\end{empheq}
\end{subequations}
	\fi%
}

\newcommand{\acidx}[1]{%
	\ac{#1}%
	\edef\uaclevel{\acl{#1}}%
	\def\laclevel{\acs{#1}}%
}
\DeclareAcronym{FLRW}{
	short=FLRW,
	long=Friedmann-Lema\^{i}tre-Robertson-Walker,
	tag=acronym
}
\DeclareAcronym{DM}{
	short=DM,
	long=Dark Matter,
	tag=acronym
}
\DeclareAcronym{CDM}{
	short=CDM,
	long=Cold Dark Matter,
	tag=acronym
}
\DeclareAcronym{HDM}{
	short=HDM,
	long=Hot Dark Matter,
	tag=acronym
}
\DeclareAcronym{WDM}{
	short=WDM,
	long=Warm Dark Matter,
	tag=acronym
}
\DeclareAcronym{BAOs}{
	short=BAOs,
	long=Baryonic Acoustic Oscillations,
	tag=acronym
}
\DeclareAcronym{DE}{
	short=DE,
	long=Dark Energy,
	tag=acronym
}
\DeclareAcronym{LCDM}{
	short=$\Lambda$CDM,
	long=$\Lambda$ Cold Dark Matter,
	tag=acronym
}
\DeclareAcronym{LSFDM}{
	short=$\Lambda$SFDM,
	long=$\Lambda$ Scalar Field Dark Matter,
	tag=acronym
}
\DeclareAcronym{sCDM}{
	short=sCDM,
	long=Standard Cold Dark Matter,
	tag=acronym
}
\DeclareAcronym{SFDM}{
	short=SFDM,
	long=Scalar Field Dark Matter,
	tag=acronym
}
\DeclareAcronym{SF}{
	short=SF,
	long=scalar field,
	tag=acronym
}
\DeclareAcronym{BEC}{
	short=BEC,
	long=Bose Einstein Condensate,
	tag=acronym
}
\DeclareAcronym{CMB}{
	short=CMB,
	long=Cosmic Microwave Background,
	tag=acronym
}
\DeclareAcronym{GR}{
	short=GR,
	long=General Relativity,
	tag=acronym
}
\DeclareAcronym{SUSY}{
	short=SUSY,
	long=Super Symmetry,
	tag=acronym
}
\DeclareAcronym{SM}{
	short=SM,
	long=Standard Model of particle physics,
	tag=acronym
}
\DeclareAcronym{BBN}{
	short=BBN,
	long=big bang nucleo synthesis,
	tag=acronym
}
\DeclareAcronym{WIMP}{
	short=WIMP,
	long=Weakly Interacting Massive Particle,
	tag=acronym
}
\DeclareAcronym{SDSS}{
	short=SDSS,
	long=Sloan Digital Sky Survey,
	tag=acronym
}
\DeclareAcronym{SN1a}{
	short=SN Ia,
	long=Supernova(e) Type Ia,
	tag=acronym
}
\DeclareAcronym{ALPs}{
	short=ALPs,
	long=Axion-Like Particles,
	tag=acronym
}
\DeclareAcronym{ULA}{
	short=ULA,
	long=Ultra-Light Axion,
	tag=acronym
}
\DeclareAcronym{ULAs}{
	short=ULAs,
	long=Ultra-Light Axions,
	tag=acronym
}
\DeclareAcronym{FDM}{
	short=FDM,
	long=Fuzzy Dark Matter,
	tag=acronym
}
\DeclareAcronym{BECDM}{
	short=BECDM,
	long=Bose-Einstein-Condesate Dark Matter,
	tag=acronym
}
\DeclareAcronym{QCD}{
	short=QCD,
	long=Quantum Chromo Dynamics,
	tag=acronym
}
\DeclareAcronym{SI}{
	short=SI,
	long=Self-Interaction,
	tag=acronym
}
\DeclareAcronym{EOS}{
	short=EoS,
	long=Equation of State,
	tag=acronym
}
\DeclareAcronym{EOM}{
	short=EoM,
	long=Equation of Motion,
	tag=acronym
}
\DeclareAcronym{KGE}{
	short=KGE,
	long=Klein-Gordon equation,
	tag=acronym
}
\DeclareAcronym{IC}{
	short=ICs,
	long=Initial Conditions,
	tag=acronym
}
\DeclareAcronym{CLASS}{
	short=CLASS,
	long=Cosmic Linear Anisotropy Solving System,
	tag=acronym
}
\newcommand{\glo}[1]{%
	\acs{#1}%
}

\DeclareAcronym{LCDMmodel}
{
	short=$\Lambda$CDM model,
	long=dummy,
	tag=glossary
}

\DeclareAcronym{LSFDMmodel}
{
	short=$\Lambda$SFDM model,
	long=dummy,
	tag=glossary
}

\DeclareAcronym{flrwmetric}
{
	short=\ac{FLRW} metric,
	long=\textit{This metric is the most general type of a metric describing a homogeneous and isotropic spacetime},
	extra=\\\TBDgl,
	tag=glossary
}

\DeclareAcronym{friedmannequations}
{
	short=Friedmann equations,
	long=dummy,
	tag=glossary
}

\DeclareAcronym{friedmannequation}
{
	short=Friedmann equation,
	long=dummy,
	tag=glossary
}

\DeclareAcronym{scalarfield}
{
	short=scalar field,
	long=dummy,
	tag=glossary
}

\DeclareAcronym{inflaton}
{
	short=Inflaton,
	long=dummy,
	tag=glossary
}
\DeclareAcronym{inflation}
{
	short=inflation,
	long=dummy,
	tag=glossary
}
\DeclareAcronym{cosmologicalconstant}
{
	short=cosmological constant,
	long=dummy,
	tag=glossary
}

\DeclareAcronym{criticaldensity}
{
	short=critical density,
	long=dummy,
	tag=glossary
}
\DeclareAcronym{densitycontrast}
{
	short=density contrast,
	long=dummy,
	tag=glossary
}
\DeclareAcronym{primordialpowerspectrum}
{
	short=primordial power spectrum,
	long=dummy,
	tag=glossary
}

\DeclareAcronym{powerspectrum}
{
	short=power spectrum,
	long=dummy,
	tag=glossary
}
\DeclareAcronym{hubbleparameter}
{
	short=Hubble parameter,
	long=dummy,
	tag=glossary
}
\DeclareAcronym{conformaltime}
{
	short=conformal time,
	long=dummy,
	tag=glossary
}
\DeclareAcronym{kleingordonequation}
{
	short=Klein-Gordon equation,
	long=dummy,
	tag=glossary
}

\newcommand{\mnras}{MNRAS}             
\newcommand{\aap}{A\&A}                
\newcommand{\araa}{ARA\&A}             
\newcommand{\apjl}{ApJ}                
\newcommand{\apjs}{ApJS}               
\newcommand{\jcap}{J.~Cosmology Astropart. Phys.} 
\newcommand{\na}{New~Astron.}          
\newcommand{\nar}{New~Astron.~Rev.}    
\newcommand{\PaperLi}{LRS14}  
\newcommand{\PaperSDR}{SDR21}  

\newcommand{\lb}{\lambda_\text{deB}}
\makeatletter
\newcommand*{\hyperlinkcite}[1]{\hyper@link{cite}{cite.#1}}
\makeatother

\begin{document}

\preprint{APS/123-QED}

\title{Cosmological structure formation in complex scalar field dark matter versus real ultralight axions: a comparative study using CLASS}
\author{Horst Foidl}
\email{horst.foidl@outlook.com}
\author{Tanja Rindler-Daller}%
\email{tanja.rindler-daller@univie.ac.at}
\affiliation{%
	Institut f\"ur Astrophysik, Universit\"atssternwarte Wien,
	Fakult\"at f\"ur Geowissenschaften, Geographie und Astronomie,
	University of Vienna, T\"urkenschanzstr.17, A-1180 Vienna, Austria
}%


\date{\today}

\begin{abstract}
Scalar field dark matter (SFDM) has become a popular alternative to standard collisionless cold dark matter (CDM), because of its potential to resolve the small-scale problems that has plagued the latter for decades. Typically, SFDM consists of a single species of bosons of ultralight mass, $m \gtrsim 10^{-22}$ eV/c$^2$, in a state of Bose-Einstein condensation, hence also called BEC-DM. In this paper, we continue the study of $\Lambda$SFDM cosmologies, which differ from $\Lambda$CDM in that CDM is replaced by SFDM, by calculating the evolution of the background Universe, as well as the formation of linear perturbations, focusing on scalar modes. We consider models with complex scalar field, where we include a strongly repulsive, quartic self-interaction (SI), also called SFDM-TF, as well as complex-field models without SI, referred to as fuzzy dark matter (FDM). To this end, we modify the Boltzmann code CLASS, in order to incorporate the physics of complex SFDM which has as one of its characteristics that it leads to a non-standard, early expansion history, where complex SFDM (or FDM) dominates over all the other cosmic components, even over radiation, in the very early Universe, because its equation of state is maximally stiff. We calculate various observables, such as the temperature anisotropies of the cosmic microwave background, the matter power spectra and the unconditional Press-Schechter halo mass function for various models, and thereby expand previous findings in the literature that were limited either to the background, or to a semi-analytical approach to SFDM density perturbations neglecting the early stiff phase. We also compare our results of each, SFDM and FDM, with ultralight axion models (ULAs) without SI and modelled as real fields. Thereby, we characterize in detail the differences between $\Lambda$SFDM, $\Lambda$FDM and $\Lambda$ULA in terms of their background evolution and their linear structure growth. Our calculations confirm previous results of recent literature, implying that SFDM models with $\gtrsim$ kpc-size halo cores are disfavored, which questions the ability of SFDM to explain the small-scale problems on dwarf-galactic scales. Furthermore, we find that the gain in kinetic energy of SFDM due to the phase of the complex field leads to marked differences between SFDM/FDM versus ULAs. The mild falloff in the SFDM matter power spectrum toward high $k$ is explained by an evolution history of perturbations similar to that of CDM, although based on different physical effects, namely the rapidly shrinking Jeans mass for SFDM as opposed to the Meszaros effect for CDM. In addition, we find that the sharp cutoff in the matter power spectrum of ULAs is also followed by a mild falloff, albeit at very small power.
\end{abstract}


\maketitle

\newpage
\setcounter{tocdepth}{4}
\newpage

\section{Introduction}\label{sec:introduction}
The nature of the cosmological dark matter (DM) remains one of the biggest open problems in contemporary science. The standard, collisionless cold dark matter (CDM) paradigm assumes that DM is non-relativistic (``cold''), as of a very early time in the evolution of the Universe. The most popular CDM candidates have been weakly interacting, massive ($>$ GeV/c$^2$) particles (``WIMPs''), usually thermal relics that are believed to be superpositions of supersymmetric partners to known particles of the standard model of physics; a natural implication should supersymmetry be realized in Nature. 
Another CDM candidate is the QCD axion, a pseudo-scalar boson of small mass, $m \sim 10^{-5}$ eV/c$^2$, that is born cold from the outset and which has been proposed originally to resolve the so-called charge-parity problem of the strong nuclear force.

Either of these CDM candidates is in accordance with astronomical observations on large scales, e.g. with the temperature anisotropy of the cosmic microwave background radiation (CMB), the cosmic web of structure as revealed by galaxy surveys, as well as with big bang nucleosynthesis (BBN) which provides bounds on the allowed amount of non-relativistic and relativistic \textit{non-baryonic} matter in the Universe. However, this broad concordance is mostly a result of the non-relativistic behavior of these CDM candidates, while the determination of the microscopic nature of DM requires further detailed investigations. In fact, CDM candidates have been continuously probed by direct or indirect detection efforts. However, neither candidate has been detected so far, although constraints on allowed parameter spaces are being regularly updated; for very recent constraints see (ADMX collab.) \citet{admx} for the axion, and (PandaX-4T collab.) \citet{pandax-4t} for WIMPs.

In addition to this null-detections, it has been pointed out for more than 20 years that theoretical predictions for CDM halo models are in conflict with many observations on the scales of dwarf galaxies, despite the fact that, overally, the latter tend to be very much DM-dominated. However, CDM predicts DM densities in the central regions of galactic halos that are much higher than observations of velocity profiles, especially those of dwarf galaxies, reveal. This is the cusp-core problem. Other issues include the fact that CDM predicts generally more substructure than observed (``missing-satellites'' problem), and that subhalos massive enough to hold on to their baryons should have formed stars and be observable as satellites that could not have possibly been missed by now in observations of the Local Group, say; yet, they are not observed in the numbers predicted (too-big-to-fail problem).
The reader may consult reviews on these issues for more details, see \citet{Bullock2017}, \citet{Weinberg15}.
In conjunction with null-detections, these so-called small-scale structure problems motivate the study of alternative DM candidates, different from WIMPs or the QCD axion.
In this paper, we consider one of these alternatives, namely scalar field dark matter (SFDM), which is made of (ultra-)light bosons, all condensed into their ground state, described by a single scalar field, also known as ``Bose-Einstein-condensed dark matter (BEC-DM)''. Depending on the detailed particle model, it encompasses a broad family, and the length scale below which structure formation is suppressed, is typically much larger than that for CDM, which makes them interesting from the point of view of the mentioned small-scale challenges. Early works on SFDM models include e.g. \citet{Sin1994}, \citet{Lee1996},
\citet{Peebles2000},  \citet*{Lesgourgues2002}, \citet*{Arbey2003}.


If the bosons have no self-interaction (SI) and their mass $m$ is so small that their de Broglie wavelength $\lambda_\text{dB} = h/(mv)$ is of kpc-size, where $v$ is some characteristic velocity, SFDM is known e.g. as ``fuzzy dark matter (FDM)'' (\citet*{Hu2000}, \citet{Matos_Urena_Lopez_2000}), or ``$\psi$DM'' \citet{Schive_FDM}. In this case, $\lambda_\text{dB}$ is related to the scale below which structure formation is suppressed, as a result of quantum pressure due to the Heisenberg uncertainty principle.  
If the particles share the same Lagrangian than the QCD axion, but are ultra-light, they are known as ``ultra-light axion-like particles (ULAs)'' (\citet{2010PhRvD..81l3530A}, \citet{Marsh2012}). If the (ultra-)light bosons interact via a strong repulsive, quartic SI, when they are in the so-called Thomas-Fermi (TF) regime, SFDM has been also called ``SFDM-TF'' in \citet*{Dawoodbhoy2021}, though the general terms ``BEC-DM", ``BEC-CDM", or ``(super-)fluid DM'' have been used as well; some earlier works concerning this regime include e.g. \citet{Goodman2000}, \citet{Peebles2000}, \citet{Boehmer2007}, and  \citet{RindlerDaller2012}. 
In this case, the characteristic length scale below which structure formation is suppressed is related to the so-called TF radius, $R_{TF} \gg \lambda_\text{dB}$, which depends upon $m$ and the SI coupling strength in a characteristic combination. Now, either $\lambda_\text{dB}$ for FDM and ULAs, or $R_{TF}$ for SFDM-TF can be of kpc-size and, as a result, halos in these models will have near-constant density cores, instead of the cusps predicted by CDM, i.e. the ``cutoff'' scale of structure formation also affects halo structure, potentially resolving the cusp-core problem.
 
In terms of particle models, we briefly add that
\mbox{(ultra-)}light bosons are predicted by 
string theories and other extra-dimensional models, see e.g. \citet{1995PhRvL..75.2077F}, \citet{1997PhRvD..56.6391G}, \citet{Svrcek_2006}, or \citet{2016PDU....14...84F}. 

Our paper constitutes an important continuation of previous works with regard to SFDM models with underlying complex scalar field, as follows.
A fully self-consistent analysis of the background evolution of $\Lambda$SFDM universes has been carried out by \citet*{Li2014} (in the forthcoming abbreviated as \hyperlinkcite{Li2014}{\PaperLi}) and \citet*{Li2017}. In these papers, CDM was replaced by complex-field SFDM, while the other cosmic components, as well as the metric of the background Universe, were the same as for $\Lambda$CDM. \hyperlinkcite{Li2014}{\PaperLi} found that the complex field of SFDM gives rise to a distinctive expansion history, dictated by the equation of state\footnote{Actually, some statements apply to the time-averaged value for this EoS parameter $w$; this will become clear in later sections. } (EoS), $w = p/\rho$ of SFDM, where $p$ is the pressure and $\rho$ the energy density of the SFDM background. That EoS evolves smoothly from maximally stiff, $w \approx 1$, through an intermediate radiation-like behavior, $w \approx 1/3$, to a CDM-like, cold phase with $w \approx 0$. During both phases when $w \approx 1$ and $w \approx 0$, SFDM dominates over all the other cosmic components, while in the intermediate phase when SFDM is radiation-like, $w \approx 1/3$, it is very much subdominant to radiation. That phase only arises, if there is a repulsive SI included in the models for SFDM.
Thus, $\Lambda$SFDM models differ from $\Lambda$CDM in that we have a non-standard expansion history, in which SFDM behaves relativistically in the early Universe, dominating before radiation in its stiff phase, followed by radiation-domination during which time SFDM can still be relativistic, if $w \approx 1/3$. Finally, SFDM morphes into a fluid that resembles CDM, at which point it gives rise to the standard matter-dominated epoch in which halos and galaxies form. The fact that SFDM behaves relativistically in the early Universe, even dominating over radiation very early, implies that cosmological observables can be used to constrain the allowed parameter space of SFDM; these are notably the epoch of BBN, as well as the time of matter-radiation equality, which is probed by CMB observations. These constraints have been considered and determined in detail in \hyperlinkcite{Li2014}{\PaperLi}, and we refer the reader to this paper for many fundamentals concerning the modelling of $\Lambda$SFDM, as well as the resulting constraints. Moreover, it has been known from other literature that early stiff phases can have further cosmological implications. Notably, it can boost an ambient cosmic gravitational-wave background, which is predicted by standard inflationary models. Therefore, in a forthcoming paper, \citet{Li2017} have expanded upon the work of \hyperlinkcite{Li2014}{\PaperLi} by introducing the inflationary and reheating epochs before the stiff phase of SFDM and adding tensor modes on top of the $\Lambda$SFDM background Universe.
These tensor modes give rise to gravitational waves, once they enter the horizon, and it turns out that these modes are amplified if they enter during the stiff phase of SFDM.
Since these amplified modes add to the amount of relativistic degrees of freedom in the early Universe, e.g. during BBN, they constitute a means to constrain $\Lambda$SFDM even further, see \citet{Li2017}. On the other hand, such an amplified stochastic gravitational-wave background can be searched for using gravitational-wave observatories. In fact, detailed forecasts for possible signals measured with aLIGO (\citet{Abbott2020}) have been derived in \citet{Li2017}, for models that are still in accordance with cosmological constraints, such as BBN. 

SFDM models with strongly repulsive SI (``SFDM-TF'') have been also studied in the recent works by \citet*{Dawoodbhoy2021} and \citet*{Shapiro2021} (abbreviated \hyperlinkcite{Shapiro2021}{\PaperSDR} in the forthcoming). 
While \citet{Dawoodbhoy2021} focused on nonlinear halo formation, performing the first simulations of 1D spherical collapse of galactic SFDM-TF halos within a static background, \hyperlinkcite{Shapiro2021}{\PaperSDR} not only continued the study of nonlinear halo formation, but it also included an analysis of the linear growth of structure in SFDM-TF, i.e. it established the initial conditions for halo formation in an expanding background Universe. To this end, a semi-analytical approach has been used in \hyperlinkcite{Shapiro2021}{\PaperSDR}, where analytical approximations for the density perturbations of DM in the form of SFDM and of radiation have been applied in the respective eras, in which perturbation modes of interest for structure formation enter the horizon, namely during radiation-domination and matter-domination. In our paper here, we will be particularly interested in the results of this previous linear structure formation analysis which, however, is blind to the earlier stiff phase of SFDM; also the equations for the perturbations were not solved exactly. We direct the reader to the above paper references for the analysis of nonlinear structure formation in SFDM-TF, as well as for details on the analytic linear perturbation analysis.

As a result of this analysis, important findings have been reported in \hyperlinkcite{Shapiro2021}{\PaperSDR}, which concern small-scale structure in SFDM-TF. In essence, it has been shown that models that would resolve e.g. the cusp-core problem of CDM, having $R_{TF} \gtrsim 1$ kpc, imply a correspondingly large Jeans scale that suppresses structure and halo formation at earlier epochs, in the first place. However, that Jeans scale shrinks rapidly over cosmic time, i.e. modes that are initially suppressed can grow later.  Translating these linear findings to a Press-Schechter-type halo mass function, it has been found in \hyperlinkcite{Shapiro2021}{\PaperSDR} that, different from ULAs, the cutoff in structure happens already at higher halo mass scale, but the subsequent falloff is much shallower than in ULAs. (We note that \hyperlinkcite{Shapiro2021}{\PaperSDR} use the term ``FDM", but we want to avoid confusion with our nomenclature here.) As a result, and depending on the choice of parameters, SFDM-TF can have more small-scale structure than ULAs, but less than CDM. Consequently, there are nuances concerning the small-scale problems that have been exposed with that work.
Given the simplifications that have been assumed in \hyperlinkcite{Shapiro2021}{\PaperSDR}, another focus of our paper concerns the detailed comparison to these earlier results. 

In studying linear structure formation, we modify the Boltzmann code CLASS in order to incorporate complex-field SFDM models, with or without SI. As a result, we have to deal with the complexity of including the stiff phase, before radiation-domination, into CLASS, which has not been carried out before to our knowledge. Once modified, we were able to perform numerically accurate calculations with our amended CLASS version, in order to calculate linear structure formation in SFDM models. In particular, we are able to compare complex-field models with real-field models that have been studied in the literature more frequently. This way, our paper can be regarded as an in-depth comparison study, concerning $\Lambda$SFDM models of various properties.

This paper is organized as follows: in Sec. \ref{sec:LSFDMbasic}, we set the stage by presenting the fundamentals and main equations of motion that underlie $\Lambda$SFDM cosmologies. Sec. \ref{sec:class} discusses the main issues, concerning the implementation of $\Lambda$SFDM models with stiff phases into CLASS, which is directly related to the physics of complex SFDM models. In Sec. \ref{sec:resultsSFDM}, we present our comparison study of complex SFDM vs. real ULA models, which confirms and extends previous work of \hyperlinkcite{Li2014}{\PaperLi} and \hyperlinkcite{Shapiro2021}{\PaperSDR}.
Sec. \ref{sec:resultsFDM} concerns the same level of comparison but between complex FDM vs. real ULA models. Since both lack SI, the differences between complex and real fields are particularly exposed. Finally, in Sec. \ref{sec:implications} we present a discussion on the various implications for small-scale structure in light of the comparison study of previous sections. Sec. \ref{sec:conclusion} contains our main conclusions and summary.


\section{Basic Equations of the \boldmath$\Lambda$SFDM Model}\label{sec:LSFDMbasic}
\acuse{EOS} 
\acuse{SFDM} 
\acuse{SI} 
\acuse{ULAs} 
\acuse{CDM} 

In this section, we summarize a few basic equations that underlie $\Lambda$SFDM cosmologies in general, while we specify concrete models in the next section. In this paper, we will be concerned with the background evolution and the growth of density perturbations in the linear regime (``scalar modes''). These perturbations arise in the DM component -- here SFDM --, in baryons, in the radiation component, and in the respective gravitational potentials. Our models also include a cosmological constant (CC), $\Lambda$, which does not develop any perturbations.
In contrast to \citet{Li2017}, we do not consider tensor modes in this paper. Also, we disregard vector modes.

\subsection{Basic Equations for the Homogeneous Background Universe}\label{sec:basiceqbg}

$\Lambda$SFDM is a cosmological model that is based on the \glo{LCDMmodel}, but replaces standard CDM with SFDM. We use the same background metric than in $\Lambda$CDM, i.e. 
a spatially flat \glo{flrwmetric}, which obeys the cosmological principle of homogeneity and isotropy of the background Universe. It reads
\EQidx{EQflrwbg}{}
with the time-dependent scale factor $a$ that describes the expansion of the Universe, the metric tensor ${g}_{\mu \nu}$ of the background and $\syidx{SYMkron}$ the Kronecker delta. This metric enters the left-hand side of the Einstein field equations. In its general form, it is given by 
\EQidx{EQEinsteinL}{,}
with the Ricci tensor $R_{\mu \nu}$ and the Ricci scalar $R$ (here $g_{\mu \nu}$ denotes the general metric, not just the background).
The right-hand side contains the total energy-momentum tensor $T_{\mu \nu}$ that includes all cosmic components of interest that add to the energy density of the Universe, i.e. SFDM, baryons and radiation. As is customary, we will shift $\Lambda$ to the right-hand side, incorporating it into $T_{\mu \nu}$ by defining an effective energy density $\rho_{\Lambda} = \Lambda c^2/(8\pi G)$ of the CC.   

Now, the $0-0$ component of the Einstein equation, applied to the background Universe with metric in \eqref{eq:EQflrwbg}, is the \glo{friedmannequation}
\EQidx{EQfriedmannLSFDM}{,}
with the time-dependent background energy densities for radiation ($\syidx{SYMed}_{r}$), baryons ($\syidx{SYMed}_{b}$) and SFDM ($\syidx{SYMed}_{SFDM}$), as well as the cosmological constant ($\rho_{\Lambda}$).
$H(t)$ is the \glo{hubbleparameter} defined as
\EQidx{EQH}{,}
where the dot denotes the derivative with respect to cosmic time $t$.
Its present-day value, the Hubble constant, is denoted as $H_0$. The \glo{criticaldensity}, defining a flat Universe, is given by
\EQidx{EQcritdenst}{.}
At the present-day, it is
\EQidx{EQcritdens}{.}
Given the symmetries of $T_{\mu \nu}$, we have the energy conservation equation
\EQidx{EQecons}{,}
where $\syidx{SYMed}$ and $\syidx{SYMpr}$ stand for the background energy density and pressure, respectively, of any cosmic component. They only depend upon time $t$, or scale factor $a$, but not on spatial coordinates. The energy density and pressure are related by the EoS
\EQidx{EQeos}{,}
where $w$ is often called the EoS parameter, which can also change with time.
Furthermore, it is convenient to introduce the so-called density parameters 
\EQidx{EQdenspar}{,}
which are nothing but the background energy densities, relative to the critical density \eqref{eq:EQcritdenst}. 
The Friedmann equation for the spatially flat $\Lambda$SFDM background Universe can be alternatively written as a closure condition, using the present-day values,
\EQidx{EQclosureLCDM}{.} 

The background evolution of the standard cosmic components is the same as in $\Lambda$CDM,
\EQgrp{EQdensitiesLCDM}{,} 
with \eqref{subeqn:EQdensitiesLCDMr} for radiation (its EoS parameter in (\ref{eq:EQeos}) is $w_r=1/3$); \eqref{subeqn:EQdensitiesLCDMm} for baryonic matter ($w_b=0$), and \eqref{subeqn:EQdensitiesLCDML} for the \glo{cosmologicalconstant} $\Lambda$ ($w_{\Lambda}=-1$). 
Now, the most complicated component is SFDM which is defined by the complex scalar field (SF)
\EQidx{EQcSF}{.}
In contrast to standard CDM where $w=0$, the EoS of SFDM is not a constant, but varies with time depending on the choice of SFDM particle parameters via its Lagrangian. To calculate the EoS, the equation of motion (EoM) of the SF, the \glo{kleingordonequation}, has to be solved. To derive the \glo{kleingordonequation}, the Euler-Lagrange equation in its general form, as commonly used in field formalism
\EQidx{EQele}{}
is applied, where $D$ is the covariant derivative and $\star$ denotes the complex conjugate (of course, real scalar fields are included by requiring $\varphi = \varphi^{\star}$). Using the general form of the Lagrangian density for a \acidx{SF}
\EQidx{EQlagrangianSF}{} 
gives the general form of the \glo{kleingordonequation}
\EQgrp{EQkggeneral}{,}
where $g$ is the determinant of the metric tensor $\syidx{SYMg}^{\mu\nu}$.
Using the FLRW metric (\ref{eq:EQflrwbg}) in \eqref{eq:EQkggeneral}, we recover the \glo{kleingordonequation} in the familiar form, applied to the homogeneous and isotropic background SF of SFDM,
\EQidx{EQkgbg}{.}
The metric couples the \glo{kleingordonequation} to the Einstein equations in \eqref{eq:EQEinsteinL}. Thus, in order to determine the EoS for SFDM, we need to solve the Einstein-Klein-Gordon system of equations, using the energy-momentum tensor for SFDM, which can be generally derived from 
\EQidx{EQemtsf}{.}

Applying the Lagrangian density \eqref{eq:EQlagrangianSF} and the FLRW metric \eqref{eq:EQflrwbg} to \eqref{eq:EQemtsf}, the energy-momentum tensor for the SFDM background field is simplified and becomes diagonal \eqref{subeqn:EQemtbgT}, and takes the form of a perfect fluid,
\EQgrp{EQemtbgSFDM}{.}

Equations \eqref{subeqn:EQemtbgde} and \eqref{subeqn:EQemtbgpr} express the SFDM energy density $\syidx{SYMed}_{SFDM}$ and pressure $\syidx{SYMpr}_{SFDM}$ and are clearly different from those of pressure-less standard CDM. With these two equations, the EoS of the \glo{scalarfield} can be expressed as
\EQidx{EQeosgeneral}{.}
This equation closes the set of equations, necessary to describe the evolution of the $\Lambda$SFDM background Universe: the \glo{friedmannequation} \eqref{eq:EQfriedmannLSFDM}; the energy conservation equation \eqref{eq:EQecons} along with equations \eqref{eq:EQdensitiesLCDM}; the \glo{kleingordonequation} \eqref{eq:EQkgbg}, and the EoS for \acidx{SFDM} \eqref{eq:EQeosgeneral}.\par

\subsection{Basic Equations for Linear Structure Formation}\label{sec:SFDMtheory}

In this subsection, we state the basic equations describing the evolution of density perturbations in the linear regime (``scalar modes''). From here on, we use the overhead bar to denote background quantities. The \glo{densitycontrast} $\syidx{SYMdenscontrast}$ of any cosmic component, defined as
\EQgrp{EQdenscontrast}{,}
is strictly below unity in the linear regime. We apply small perturbations
\EQgrp{EQgrperturbations}{}
to the metric \eqref{subeqn:EQgrperturbationsm} and energy-momentum tensor  \eqref{subeqn:EQgrperturbationst} of the background Universe. The resulting tensors are both symmetric, containing four scalar components.
For the metric perturbations $\delta\syidx{SYMg}_{\mu\nu}$ these tensor components represent: the generalized gravitational potential $\syidx{SYMmpnewpot}$, the local distortion $\syidx{SYMmpnewlapse}$ of the scale factor $a(t)$, the potential $b$, such that $\delta\syidx{SYMg}_{0j} = \partial_i b$ and the potential $\mu$, such that $\delta\syidx{SYMg}_{ij} = (\partial_i \partial_j - \frac{1}{3} \syidx{SYMkron}\Delta)\mu$, where $\Delta$ denotes the Laplace operator.
For the perturbations of the energy-momentum tensor $\delta\syidx{SYMemt}_{\mu\nu}$ these tensor components represent: the density perturbations $\delta\syidx{SYMed}$, the pressure perturbations $\delta\syidx{SYMpr}$, the velocity divergence $\syidx{SYMemtvelocity}$ and the anisotropic stress $\syidx{SYMemtstress}$.
Thus, we have to solve for the unknown ``hydrodynamical'' perturbation variables,\\
~\\
\centerline{  \{$\syidx{SYMdenscontrast}, \delta\syidx{SYMpr}, \syidx{SYMemtvelocity}, \syidx{SYMemtstress}$\}  }\\
~\\
for each cosmic component (except for $\Lambda$). In terms of radiation, the evolution of these quantities determines -- among other things -- the temperature fluctuations of the CMB photons, which involves solving the collisional Boltzmann equation. The evolution of the perturbations of the baryons results e.g. in baryonic acoustic oscillations (BAOs). These topics are described in detail in many textbooks covering the theory of structure formation to which we refer the reader, e.g. \citet{Weinberg2008,Coles2002,Mukhanov2005,Dodelson2003,Peebles1993}.
However, we outline how scalar quantities are determined for SFDM, as follows.\par
To find the equations describing the evolution of the density perturbations of the \acidx{SF}, we use a similar approach used to derive the EoM for the \acidx{SF} in the background Universe by applying the Euler-Lagrange equation in its general form in \eqref{eq:EQele}.
Also, general-relativistic perturbation theory conveniently works with the \glo{conformaltime} $\syidx{SYMct}$, defined as
\EQidx{EQct}{,}
since this definition provides symmetry between space and time coordinates. The line element of the metric transforms accordingly as
\EQidx{EQflrwbgct}{,}
and the \glo{hubbleparameter} \eqref{eq:EQH} transforms to
\EQidx{EQHct}{,}
where the prime denotes the derivative with respect to $\tau$.
To get the \glo{kleingordonequation} for the perturbations, we apply small perturbations to the metric \eqref{subeqn:EQgrperturbationsm} and use the transformed line element \eqref{eq:EQflrwbgct}, giving the perturbed line element in conformal synchronous gauge
\EQidx{EQflrwperturbed}{,}
where $h_{ij}$ denotes the metric perturbations. \citet{Ma1995} derived the line element of the perturbed metric among other equations used in linear structure formation in Newtonian and synchronous gauge. A general derivation of equations used in the context of scalar fields can be found e.g. in \citet{Ratra1991} or \citet{Perrotta1999}. As we want to calculate the evolution of individual perturbation modes, we need to apply perturbations to the background \acidx{SF}, $\bar{\varphi}$, where we denote the perturbations with $\phi$, and transform the resulting perturbed \glo{kleingordonequation} to Fourier space (see \citet{Matos2002,Caldwell1998,Ferreira1998}), thus 
\EQgrp{EQkgperturbation}{.}
Finally, the perturbed density and pressure (or density and pressure ``contrast'') of SFDM can be written in terms of its field variables
\EQgrp{EQemtperturbation}{.}

\subsection{Cosmological Parameters}

For this work, 
we choose the standard minimal set of cosmological parameters, as determined by the \citet{Collaboration2020} shown in Table \ref{tab:planck2018}.%
\begin{table}[H] 
	\caption{Cosmological parameters from Planck 2018\label{tab:planck2018}}
	\begin{ruledtabular}
		\begin{tabular}{lrl}
			Parameter        & Value            & Comment \\
			\hline
			H$_{0}$          & $67.556$         & \\
			T$_{CMB}$~[K]    & $2.7255$         & \\
			N$_{ur}$         & $3.046$          & \\
			$\Omega_{\gamma}$ & $5.41867 \times$10$^{-05}$  & derived from T$_{CMB}$\\
			$\Omega_{\nu}$    & $3.74847 \times$10$^{-05}$  & derived from N$_{ur}$\\
			$\Omega_{b}$      & $0.0482754$      & \\
			$\Omega_{SFDM}$   & $0.263771$       & \\
			$\Omega_{\Lambda}$& $0.687762$       & \\
			$\Omega_{k}$      & $0$              & \\
			$\tau_{reio}$     & $0.0925$         & \\
			A$_{s}$          & $2.3 \times$10$^{-9}$       & \\
			n$_{s}$          & $0.9619$            & adiabatic ICs\\
		\end{tabular}
	\end{ruledtabular}
\end{table}
These values served as our initial conditions used in the CLASS input file, for all computations carried out in this paper, to investigate various $\Lambda$SFDM universes. 
Our results and comparison will be presented in sections \ref{sec:resultsSFDM} and \ref{sec:resultsFDM}. 

\section{Numerical Implementation of the \boldmath$\Lambda$SFDM Model Into CLASS} \label{sec:class}

The software \acidx{CLASS} is an accurate Boltzmann code, that is designed, not only to provide a user-friendly way to perform cosmological computations, but also to provide a flexible coding environment for implementing customized cosmological models. These concepts, including an overview of coding conventions, can be found in \citet{Lesgourgues2011}. The code is publicly available at \href{http://class-code.net/}{http://class-code.net/}, where updated versions, with improved functionality and models, are provided on a regular basis. The version used in this work, the up-to-date version, when we started the implementation of the \acidx{SFDM} model, is version 2.9 (21.01.2020). This version uses the Planck 2018 cosmological parameters from \citet{Collaboration2020} as the default parameter set. Additionally, CLASS provides different sets of precision configuration files, to reflect varying requirements on precision needed in the results and available computation time. The precision configuration offering the highest accuracy is proofed to be in conformance with the Planck results within a $0.01\%$ level. This allows an indirect comparison to Planck observational data, by using the $\Lambda$CDM reference configuration of \acidx{CLASS}.\par
The modular concept of \acidx{CLASS} and the coding conventions make it possible to enhance the existing code with alternative particle species, without the risk of compromising existing functionality. Therefore, we decided to use \acidx{CLASS} in order to study $\Lambda$SFDM models. We calculate the cosmic evolution of the background Universe (which confirms earlier results in the literature), but foremost the linear growth of structure, via the computation of the power spectra for the matter density perturbations and the temperature anisotropies of the CMB, where we compare our new results with some existing ones in the literature. Apart from some minor modifications, connected to the addition of the \acidx{SFDM} related input parameters, the majority of modifications had to be carried out in the background module and the perturbation module of CLASS. We use a synchronous gauge in all calculations, for which we have to add a non-zero, but very subdominant, cold dark matter component, $\Omega_{0,cdm} = 10^{-4}$, following the choice of \citet{UrenaLopez2016}\footnote{They used a value of $10^{-6}$. As a consequence of the results of our tests with the intended range of our model parameters, we had to modify the parameter to $10^{-4}$, to guarantee stability in the equation solver of CLASS}.\par

\subsection{Background Module}

As described in the Introduction, we are particularly interested in models of SFDM as a complex field with repulsive SI, as studied by \hyperlinkcite{Li2014}{\PaperLi} and \citet{Li2017}. There are a number of motivations to consider a complex, rather than a real scalar field. The $U(1)$ symmetry of a complex field induces a conserved particle number, which is of special interest in cosmology. However, we refer the reader to the above papers for more details on the motivation and underlying formalism.

We want to complement those studies by considering here SFDM density perturbations and their power spectra. For this purpose, it will be sufficient to focus on the older fiducial model used in \hyperlinkcite{Li2014}{\PaperLi}, never mind that it has been updated in \citet{Li2017}, because the underlying parameters are not very different. Also, we want to reassess the models studied by \hyperlinkcite{Shapiro2021}{\PaperSDR}, which already reflect tighter constraints. All these models will be compared, in turn, with other scalar field cosmologies with model parameters of interest from previous literature. 

In this section, we present the Lagrangian of our models of interest; for brevity we omit here the subscript ``SFDM", as well as the overhead bar over field quantities.

In the background module we use physical units, transforming the SF $\varphi$ into $\syidx{SYMsfbec}$ such that
\EQidx{EQtransBECtoSF}{.}
The \acidx{SFDM} model Lagrangian density is then
\EQidx{EQlagrangianBEC}{}
for the complex SF
\EQidx{EQcSFBEC}{ }
with its oscillation frequency
\EQidx{EQcSFomega}{. }
As in previous works, we choose a SF potential with quadratic mass term and quartic self-interaction (SI),
\EQidx{EQpotentialBEC}{,} 
where we assume that the SI coupling strength is constant and non-negative, $\lambda \geq 0$.
In case of FDM models, we will neglect SI in the last term of \eqref{eq:EQpotentialBEC}. For ULAs, we also neglect SI\footnote{This way, we follow most of the literature that neglects the attractive SI ($\lambda < 0$) of ULAs, but we stress that the complete Lagrangian of ULAs includes SI-terms, like for the QCD axion.}, and we adopt a real field, $\psi = \psi^{\star}$. Thus, FDM and ULAs are special cases of our SFDM Lagrangian.
 Even if not indicated every time, we stress that for the rest of this paper it is understood that \textit{we use the term ``SFDM'' to describe models with complex field and strongly repulsive SI (also called SFDM-TF); ``FDM'' refers to models with complex field but without SI, while ``ULAs'' finally refer to real fields without SI.}

\begin{figure} [!htbp]
	{\includegraphics[width=\PW\columnwidth]{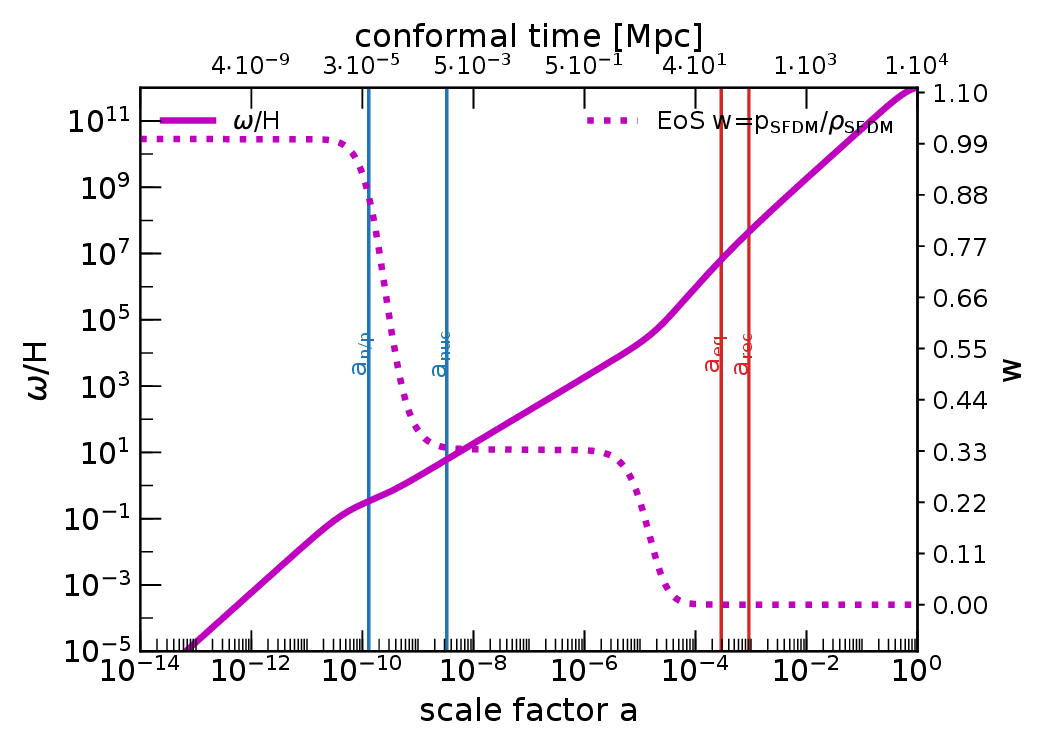}}		
	\caption[Evolution of the oscillation of the SF]
	{\textbf{Evolution of the oscillation of the SF}. During the evolution of the Universe, the oscillation frequency $\omega$ of the SFDM scalar field increases with time, while the expansion rate $H$ decreases. Both have units of $[1/s]$. The ratio $\omega/H$ (left axis) expresses the oscillation of the SF relative to the expansion rate, indicated by the magenta solid line. The dotted magenta line displays the evolution of the EoS (right axis) of SFDM from $w \approx 1$ to $w \approx 1/3$, to $w \approx 0$. The vertical lines in blue bracket the epoch of BBN, between neutron-proton freeze-out $a_{n/p}$ and nuclei production $a_{nuc}$, while the vertical lines in red (at higher scale factor) indicate the time of matter-radiation equality $a_{eq}$, followed by recombination $a_{rec}$.
	}
	\label{fig:SFDM-oHw}
\end{figure}

Inserting the above equations into \eqref{eq:EQkgbg} (the \glo{kleingordonequation}) and \eqref{eq:EQemtbgSFDM} (the contributions to the energy-momentum tensor) results in (see also \hyperlinkcite{Li2014}{\PaperLi} equations (17)-(19)):
\EQgrp{EQkgbgsfdm}{,}
where \eqref{subeqn:EQkgbgsfdm} represents the \glo{kleingordonequation} for the SFDM background, its energy density \eqref{subeqn:EQkgbgsfdmed} and its pressure \eqref{subeqn:EQkgbgsfdmpr}. Thus, the EoS $w(t) = \syidx{SYMprbg}/\syidx{SYMedbg}$ for \acidx{SFDM} is
\EQidx{EQeossfdm}{.}
Again, the evolution of the background Universe is given by the following set of equations: the \glo{friedmannequation} \eqref{eq:EQfriedmannLSFDM}, the energy conservation equation \eqref{eq:EQecons}, the \glo{kleingordonequation} \eqref{eq:EQkgbgsfdm} and the \acidx{EOS} for \acidx{SFDM} \eqref{eq:EQeossfdm}.
The oscillation frequency \eqref{eq:EQcSFomega} of the SF increases immensely with time, which poses difficulties in the numerical treatment of the equations.
In Fig.~\ref{fig:SFDM-oHw},  we show the evolution of $\omega/H$, anticipating some of the results for the fiducial \acidx{SFDM} model discussed in section \ref{sec:resultsSFDM}.

Initially, the expansion rate dominates over the oscillation frequency by some orders of magnitude, but for realistic SFDM models this dominance has to cease before BBN is over. As in \hyperlinkcite{Li2014}{\PaperLi}, we bracket the epoch of BBN between the time of neutron/proton freeze-out at $a_{n/p}$ ($T_{n/p} = 1.293$ MeV, the difference between the neutron and the proton mass) and the time of nuclei production $a_{nuc}$ (around $T_{nuc} \approx 0.07$ MeV).

After BBN, the oscillation frequency dominates over the expansion rate, reaching some orders of magnitude which indicates a very fast oscillating \acidx{SF}. Additionally, the impact of the \acidx{EOS} on the expansion rate, can be clearly seen in the change of slope of the curve for $\omega/H$. In order to deal with the numerical difficulties, we use the approach put forward in \hyperlinkcite{Li2014}{\PaperLi} (see details there), and adapt it to the needs of \acidx{CLASS}, as follows. When the oscillation frequency exceeds a certain threshold and the field enters the fast oscillation regime, the exact integration of equations \eqref{eq:EQkgbgsfdm} is replaced by equations describing the evolution of the time-averaged background density and pressure, averaged over a sufficient number of oscillation periods. Using the approximation \eqref{subeqn:EQkgbgsfdmavgapprox} gives the equations \eqref{subeqn:EQkgbgsfdmavgde} for the time-averaged density and \eqref{subeqn:EQkgbgsfdmavgpr} for the time-averaged pressure:
\EQgrp{EQkgbgsfdmavg}{.}
The \acidx{EOS} can be evaluated using equation (25) in \hyperlinkcite{Li2014}{\PaperLi},
\EQidx{EQeossfdmavg}{.}
This is a numerically very robust method to determine the evolution of density and pressure in the fast oscillation regime of the field. In the earlier ``slow oscillation regime", the exact integration of the equations is applied. Unfortunately, there is no way to determine the \acidx{IC} for the field at a scale factor $a=10^{-14}$, where \acidx{CLASS} starts integration, by default. Therefore, the following approach was implemented: the matching point between the slow and fast oscillation regimes is determined by applying the numerical procedure developed by \hyperlinkcite{Li2014}{\PaperLi} and customizing the parameter shooting mechanism provided by \acidx{CLASS}. This gives the accurate \acidx{IC} for a backward integration of the exact equations until the starting point of CLASS is reached, at which point the \acidx{IC} for the regular integration process of \acidx{CLASS} are used.\par
During this integration, not only the quantities describing the evolution of the background Universe are computed, but also all \textit{field-related} background quantities, appearing in the perturbation equations \eqref{eq:EQkgperturbation} and \eqref{eq:EQemtperturbation}. In contrast to the physical units we use in the background module, in the perturbation module, described below, we stick to CLASS conventions and use pure field quantities and natural units, normalized to the reduced Planck mass. All quantities computed in the background module, relevant for the perturbation module, were transformed using
\eqref{eq:EQtransBECtoSF}, converted to natural units and normalized to CLASS conventions.

\subsection{Perturbation Module}

The oscillation behavior of the field impacts the approach used for the integration process applied in the perturbation module. We use the matching point, determined during the calculation of the background Universe, when the field enters the fast oscillation regime. At this point, we change from the exact integration of the perturbed \glo{kleingordonequation} \eqref{eq:EQkgperturbation} and using the contributions to the perturbations of the energy-momentum tensor given by \eqref{eq:EQemtperturbation}, to the application of a fluid approximation of the fast oscillating \acidx{SF} (equations  \eqref{eq:EQfldperturbation} and \eqref{eq:EQfldcontributions} below). \acidx{CLASS} already provides a fluid approximation to model fluid dark energy (details are found in \citet{Lesgourgues2011a}). This implementation uses a generalized form of the fluid equations for \acidx{SF} perturbations, presented in \citet{Hu1998} and equivalently in \citet{Hlozek2015}. The implementation of the \acidx{EOS} of the dark energy (DE) in CLASS is based on the CLP parametrization of \citet{Chevallier2001,Linder2003}. This approach has been modified to the \acidx{EOS} of SFDM, giving equations \eqref{eq:EQfldperturbation} and \eqref{eq:EQfldcontributions},
\EQgrp{EQfldperturbation}{,}
where the prime denotes the derivative with respect to \glo{conformaltime} $\tau$, $c_{s}^2$ denotes the sound speed squared of the fluid, and $c_{ad}^2$ is the adiabatic sound speed squared. The quantities with the subscript ``apx'' denote the approximated quantities for the \acidx{SF}. The integration of \eqref{eq:EQfldperturbation} determines the contributions of SFDM to the perturbations of the energy-momentum tensor, given by the following equations,
\EQgrp{EQfldcontributions}{.}
So, in the ``fast oscillation regime", instead of integrating the perturbed \glo{kleingordonequation} \eqref{eq:EQkgperturbation}, equation  \eqref{eq:EQfldperturbation} is integrated to get the quantities of the fluid approximation for the field. From these quantities, equation \eqref{eq:EQfldcontributions} determines the contributions of SFDM to the perturbations of the energy-momentum tensor, instead of using equations \eqref{eq:EQemtperturbation} in the slow oscillation regime of the \acidx{SF}.

\subsection{Computation of Power Spectra}

The evolution of the (dark) matter density perturbations is initiated by primordial fluctuations created in the \glo{inflation} era. The statistical description of these perturbations is based on the \glo{primordialpowerspectrum} $P_p(k)$. The evolution of perturbations from the initial time up to the time of interest, can then be accounted for by the so-called transfer function $T(k, t)$. The matter \glo{powerspectrum} of the density perturbations $P(k,t)$ is obtained via
\EQidx{EQpowspectrum}{,}
where the \glo{primordialpowerspectrum}, assumed to be Gaussian and nearly scale-invariant, is described by a power law 
\EQidx{EQpowspectrprimordial}{.}
The \glo{powerspectrum} $P(k, t)$ expresses the variance of the density perturbations. As the computation of the evolution of the density perturbations already takes place in Fourier space, the accumulation of the perturbations into the transfer function $T(k, t)$ can be performed easily.\par
Deriving the \glo{powerspectrum} of the temperature fluctuations of the CMB photons is much more complicated and involves solving the collisional Boltzmann equation, which is described in detail in many textbooks covering the theory of structure formation like e.g. \citet{Weinberg2008,Coles2002,Mukhanov2005,Dodelson2003,Peebles1993}. \acidx{CLASS} uses a very efficient way to do this, using a methodology introduced with the code CAMBFAST in \citet{Seljak1996}, called \emph{line-of-sight integration}. \acidx{CLASS} offers the computation of both spectra, matter power spectrum $P(k, t)$ and the power spectrum of the spherical temperature fluctuations in the CMB, $C_l(t)$, without the need of extensive code adaptation. Only the contributions to the corresponding source functions  $S(k, t)$ for the density perturbations and the temperature fluctuations need to be modified, in order to reflect the characteristics of the \acidx{SFDM} model.

\section{Results for SFDM and Comparison to ULAs}\label{sec:resultsSFDM}

\subsection{Model Parameters}

According to the Lagrangian in \eqref{eq:EQlagrangianBEC} and \eqref{eq:EQpotentialBEC}, SFDM has two free model parameters: $m$, the mass of the particle and $\lambda$, the coupling strength of the quartic 2-particle SI. On the other hand, FDM and ULAs have only the mass $m$ as a free parameter. Theoretical models allow a huge range of these parameters, so we rely greatly on astrophysical constraints.
As mentioned above, one purpose of this paper is to complement the work of \hyperlinkcite{Li2014}{\PaperLi} and \citet{Li2017}, by calculating here the power spectra of linear structure formation of complex field $\Lambda$SFDM models. We focus in particular on the fiducial model of \hyperlinkcite{Li2014}{\PaperLi}, whose parameters can be found in the first line in Table \ref{tab:SFDM}. Its $m$ and $\lambda$ were chosen such, that the model fulfills cosmological constraints, as well as resulting in galactic core radii of size $\sim 1$ kpc. This latter choice was motivated by the small-scale problems of CDM, as described in the Introduction. These problems lend also motivation to pick parameters for FDM and ULAs that produce kpc-size cores. Indeed, for ULAs this means to choose masses\footnote{Just a few years ago, a fiducial choice of $m \sim 10^{-23}$ eV/c$^2$ has been common for ULAs or FDM, also because that way, the wave-like quantum nature is more exposed on galactic scales. However, recent constraints from Local Group dwarf galaxies have disfavored this extreme low-mass range, see e.g. \citet{Nadler2021}.} around or above $m \gtrsim 10^{-22}$~eV/c$^2$. 
 
As mentioned in the Introduction, the minimal length scale which is relevant for galactic cores in SFDM-TF is given by the TF radius, 
\EQidx{EQRTF}{,}
where $G$ denotes the gravitational constant; for details see \hyperlinkcite{Li2014}{\PaperLi}. It depends upon the ratio $\lambda/m^2$, i.e. by fixing a TF radius of interest, say $R_{TF} \sim 1$ kpc in order to address the small-scale problems, we have fixed that ratio. So, this leaves still a freedom in the choice of $m$, but there are other constraints that affect the allowed values of $m$, and these constraints have been derived in \hyperlinkcite{Li2014}{\PaperLi}. The \acidx{SFDM} model parameters that we study in our paper are shown in Table \ref{tab:SFDM}. As just said, the first line includes the fiducial model parameters from \hyperlinkcite{Li2014}{\PaperLi}, which was also considered in \hyperlinkcite{Shapiro2021}{\PaperSDR}. The other models are further reference cases from the study of \hyperlinkcite{Shapiro2021}{\PaperSDR}, because another aim of our work here is to compare our results using CLASS with those presented in \hyperlinkcite{Shapiro2021}{\PaperSDR}. As discussed in the Introduction, in order for linear perturbation modes to collapse later in the cosmic history into nonlinear halos that exist today, e.g. Milky Way hosts, \hyperlinkcite{Shapiro2021}{\PaperSDR} find that $R_{TF}$ should be of sub-kpc size. In other words, only small enough $R_{TF}$ are favored, and the suppression of structure with $R_{TF} \sim 1$ kpc seems too strong. Therefore, we include the reference models of \hyperlinkcite{Shapiro2021}{\PaperSDR} with their small $R_{TF}$ in our study, as well; see Table \ref{tab:SFDM}.     
Column ``$m_{22}$'' gives the mass in units of $10^{-22}$~eV/c$^2$. Column ``$R_{TF}$'' gives the TF radius. In addition, in this section we compare our SFDM models to two ULA models, one of which has the same mass as the fiducial SFDM case, while the second model has a smaller mass (equivalent to a larger $\lambda_{deB}$), a popular model in previous works, e.g. \citet{Schive_FDM}, \citet{Robles_SFDM_smallscale}, \hyperlinkcite{Shapiro2021}{\PaperSDR}.

\begin{table}[!htbp] 
	\caption{Model parameters for complex SFDM and real ULAs \label{tab:SFDM}}
	\begin{ruledtabular}
		\begin{tabular}{rrlr}
			m [eV/c$^2$]          &  m$_{22}$  & $\syidx{SYMlambdasi}/(m c^2)^2$ [eV$^{-1}$ cm$^3$] & $R_{TF}$\footnote{Actually, these values differ very marginally from those used in \hyperlinkcite{Shapiro2021}{\PaperSDR}, because we use the fiducial value of $\syidx{SYMlambdasi}/m^2$ from \hyperlinkcite{Li2014}{\PaperLi} to ``scale down'' to higher-mass models. $R_{TF}$ is calculated from \eqref{eq:EQRTF}.}\\
			\hline
			$3.0 \times 10^{-21}$ & 30    & $2 \times 10^{-18}$       & 1.1 kpc\\
			$5.0 \times 10^{-20}$ & 500   & $2 \times 10^{-20}$       & 110 pc\\
			$5.0 \times 10^{-19}$ & 5000  & $2 \times 10^{-22}$       & 11 pc\\		
			$5.0 \times 10^{-18}$ & 50000 & $2 \times 10^{-24}$       & 1 pc\\
			\hline				
			$3.0 \times 10^{-21}$ & 30   & --                         & --\\	
			$8.0 \times 10^{-23}$ & 0.8  & --                         & --\\			
		\end{tabular}
	\end{ruledtabular}
\end{table}

\subsection{Background Evolution}\label{sec:SFDMbackground}

In this subsection, we basically reproduce some results of \hyperlinkcite{Li2014}{\PaperLi} in order to confirm that our CLASS modification works correctly. Also, we include here results concerning the new model parameters of Table \ref{tab:SFDM}.

\begin{figure} [!htbp]
	\centering	
	\includegraphics[width=\PW\columnwidth]{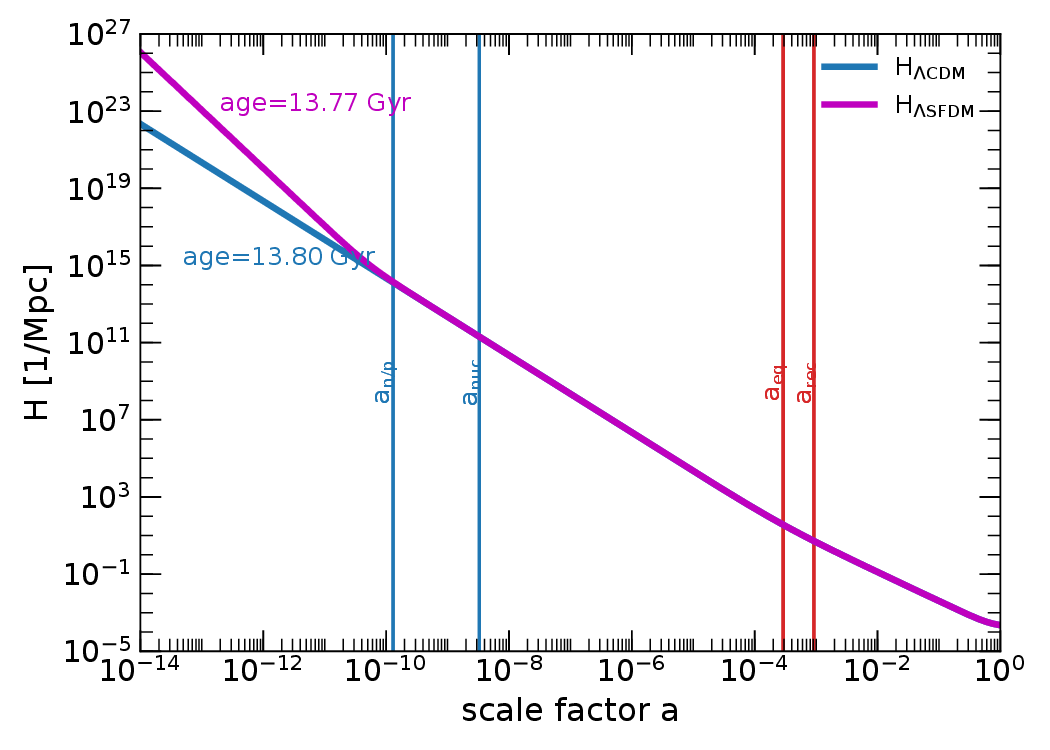}		
	\caption[Expansion rate of the fiducial $\Lambda$SFDM model ($m_{22}=30$, $R_{TF}=1.1$~kpc) and for $\Lambda$CDM]
	{\textbf{Expansion rate of the fiducial $\Lambda$SFDM model of \hyperlinkcite{Li2014}{\PaperLi} (\boldmath{$m_{22}=30$, $R_{TF}=1.1$}~\textnormal{kpc}), compared to $\Lambda$CDM}, over cosmic history. Vertical lines are the same as in Fig. \ref{fig:SFDM-oHw}, which showed quantities of the same fiducial model (see also Figure 2 in \hyperlinkcite{Li2014}{\PaperLi}). 
	}
	\label{fig:SFDM-H}
\end{figure}

First, we confirm the different expansion history of $\Lambda$SFDM, compared to $\Lambda$CDM, seen in Fig.~\ref{fig:SFDM-H} for the same fiducial model as in \hyperlinkcite{Li2014}{\PaperLi}. In $\Lambda$SFDM, we see an initial faster decrease of the expansion rate, compared to $\Lambda$CDM. This is due to the evolution of the \acidx{EOS}, which is explained shortly. The result is a slightly younger Universe, $13.77$~Gyr, for $\Lambda$SFDM, compared to $13.80$~Gyr for $\Lambda$CDM.\par

\begin{figure*} [!htbp]
	{\includegraphics[width=0.49\textwidth]{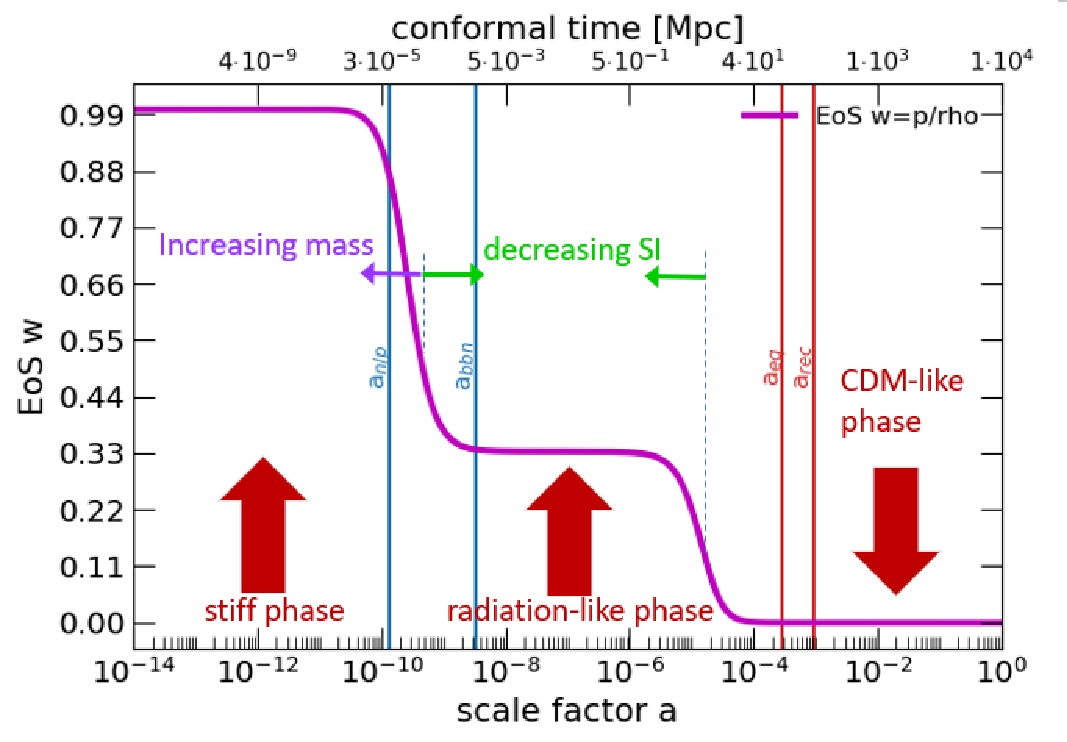}}
	{\includegraphics[width=0.49\textwidth]{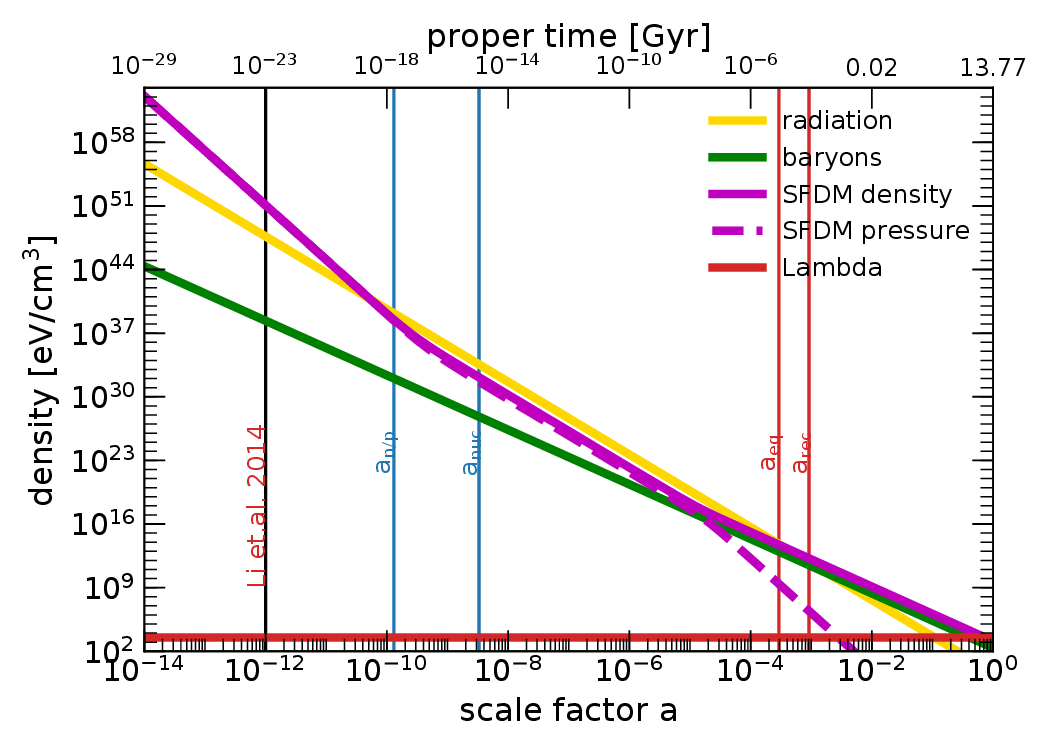}}	
	\caption[Evolution of the EoS of the SF]
	{\textbf{Evolution of the EoS of SFDM}. The left-hand panel displays the evolution of the \acidx{EOS} of \acidx{SFDM} for the fiducial model of \hyperlinkcite{Li2014}{\PaperLi} ($m_{22}=30$, $R_{TF}=1.1$~\textnormal{kpc}), which evolves in three phases. The initial stiff phase ($w = 1$) ends shortly before BBN is over, with the transition to the radiation-like ($\langle w \rangle = 1/3$) phase. Shortly before $a_{eq}$, the \acidx{EOS} enters the CDM-like phase ($\langle w \rangle = 0$). Note that a decrease of the value of SI (green arrows) leads to a shorter radiation-like phase, while a shift to higher masses (purple arrow) leads to an earlier transition from the stiff phase to the radiation-like phase.
	We thus confirm the results of \hyperlinkcite{Li2014}{\PaperLi}, using CLASS, so this figure may be compared to Figure 1 in \hyperlinkcite{Li2014}{\PaperLi}.
	The right-hand panel displays the evolution of all background energy densities, as well as the SFDM background pressure. The solid lines display their evolution for radiation (yellow), baryons (green), SFDM (magenta) and the cosmological constant $\Lambda$ (red). The magenta dashed line displays the pressure of SFDM. Note that it does not go to zero abruptly.
	The left-most vertical line ``Li \textit{et al.} 2014'' indicates the starting redshift of the calculation in \hyperlinkcite{Li2014}{\PaperLi}.
	 Vertical lines are the same as in Fig. \ref{fig:SFDM-oHw}: the vertical lines in blue bracket the epoch of BBN, between neutron-proton freeze-out $a_{n/p}$ and nuclei production $a_{nuc}$, while the vertical lines in red (at higher scale factor) indicate the time of matter-radiation equality $a_{eq}$, followed by recombination $a_{rec}$.  
	}
	\label{fig:SFDM-EOS-rho}
\end{figure*}

The evolution of the \acidx{SF} is characterized by its \acidx{EOS}, which, in contrast to standard CDM, is a function of time. Figure \ref{fig:SFDM-EOS-rho} shows the evolution of the \acidx{EOS} (left-hand panel) and the energy densities (right-hand panel) for the fiducial $\Lambda$\acidx{SFDM} model of \hyperlinkcite{Li2014}{\PaperLi} (first line of Table \ref{tab:SFDM}). Our CLASS calculations confirm their results. Note that we show the evolution of all cosmic components in the right-hand panel, while the left-hand panel only exhibits the various impacts of the SFDM parameters onto the form of its EoS.
As discussed before, the ratio $\omega/H$, which expresses the oscillation of the SF relative to the expansion rate, gives rise to different EoS. Initially, the expansion rate dominated over the oscillation frequency by some orders of magnitude until $\sim a_{n/p}$. Up to this point, there has been only one half of an oscillation of the \acidx{SF} and the \acidx{EOS} of SFDM is stiff with $w \approx 1$. Once the oscillation frequency dominates over the expansion rate, the \acidx{EOS} of SFDM drops to that of a radiation-like state, with an average EoS parameter $\langle w \rangle \approx 1/3$. The model is tuned such that, after $a_{eq}$ SFDM is in the CDM-like phase with $\langle w \rangle \approx 0$. By then, $\omega$ has increased over $H$ by $\sim 12$ orders of magnitude, thus indicating a very fast oscillating scalar field.\par

The oscillation frequency depends upon both \acidx{SFDM} model parameters, i.e. it determines the time of the drop from the stiff phase to the radiation-like phase, as well as the transition from the radiation-like to the CDM-like phase; see equation (B6) in \hyperlinkcite{Li2014}{\PaperLi} for this latter case. Increasing the mass leads to faster oscillation, thus an earlier drop (i.e. the transition of the \acidx{EOS} shifts to the left). Decreasing the SI lowers the oscillation frequency, resulting in the opposite impact on the transition from the stiff phase to the radiation-like phase. Equation \eqref{eq:EQeossfdmavg} is used to determine the characteristics of the transition from the radiation-like phase to the CDM-like phase. We find that the value at the ``edge'' at $\langle w\rangle = 1/6$ is a good point to characterize the time of the relatively sharp transition from the radiation-like phase to the CDM-like phase of SFDM and equation \eqref{eq:EQeossfdmavg} can be rewritten to give us the corresponding density $\bar{\rho}_{1/6}$ at that time which reads  
\EQidx{EQrhow}{,}
with $\syidx{SYMlambdasit} = \syidx{SYMlambdasi}/(m c^2)^2$, which is proportional to $R_{TF}$ introduced in equation \eqref{eq:EQRTF}. Thus, lowering the SI strength $\syidx{SYMlambdasit}$ shifts this point to a higher density (i.e. to the left). The evolution of background density and pressure is directly related to the evolution of the \acidx{EOS}, as can be clearly seen in the right-hand panel of Fig.~\ref{fig:SFDM-EOS-rho}. Initially, the energy density of SFDM dominates over radiation by $\sim 7$ orders of magnitude, but it evolves proportional to $a^{-6}$, whereas radiation evolves proportional to  $a^{-4}$. The energy density of SFDM drops below that for radiation just before $a_{nuc}$. At the end of the stiff phase, the EoS of SFDM drops to $1/3$, thus SFDM behaves radiation-like and evolves proportional to $a^{-4}$. Just before $a_{eq}$, SFDM becomes CDM-like, evolving proportional to $a^{-3}$. The transition between these EoS is fast, but smooth, which can be also seen by the dashed magenta line for the SFDM time-averaged pressure in the right-hand panel of Fig.~\ref{fig:SFDM-EOS-rho}.\par

\begin{figure*} [!htbp]
	\centering	
	\includegraphics[width=0.49\textwidth]{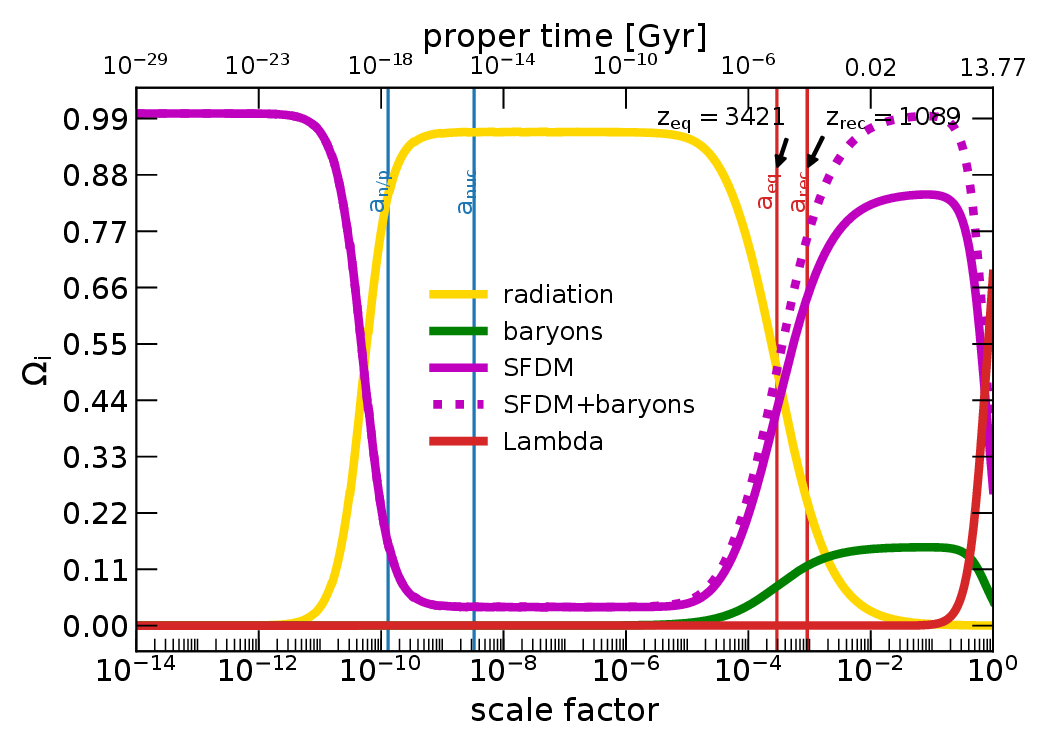}
	\includegraphics[width=0.49\textwidth]{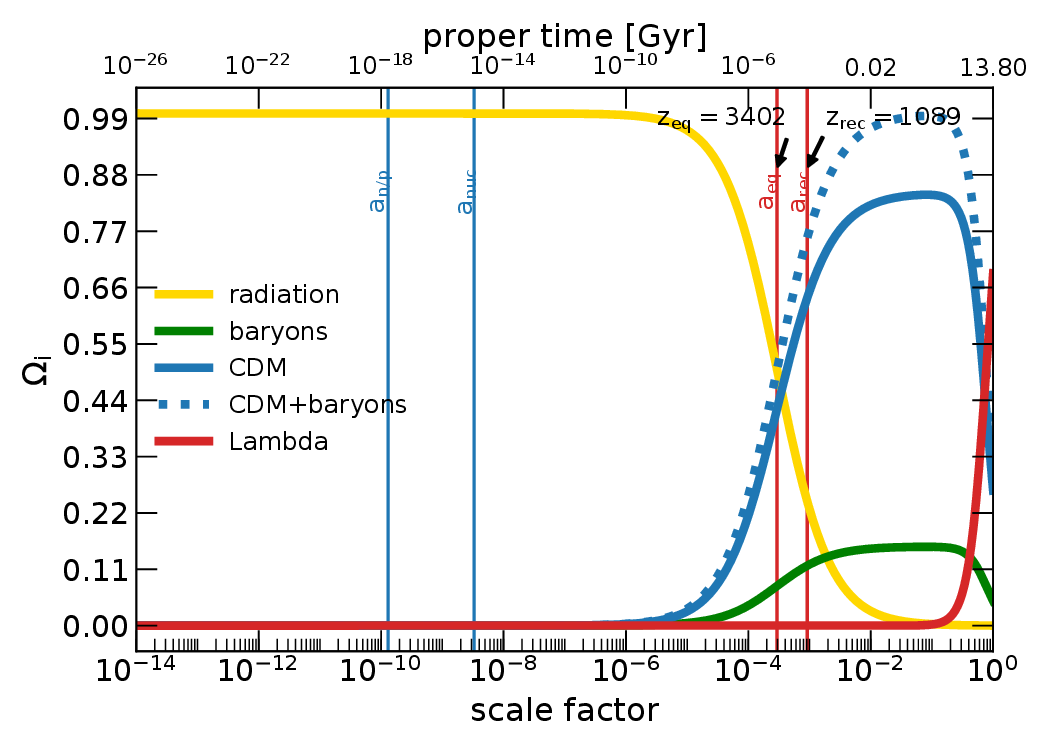}		
	\caption[Evolution of density parameters in $\Lambda$SFDM model]
	{\textbf{Evolution of density parameters in $\Lambda$SFDM (left panel) versus $\Lambda$CDM (right panel)}. The density parameters $\Omega_i$ express the energy densities $\rho_i$ as fractions of the critical density. The solid lines display their evolution for radiation (yellow), baryons (green) and the cosmological constant $\Lambda$ (red) in both panels. In the left-hand panel the magenta solid line displays the evolution for SFDM, while the magenta dotted line displays the evolution for the entire matter content (SFDM + baryons) in $\Lambda$SFDM. This plot may be compared to Figure 3 in \hyperlinkcite{Li2014}{\PaperLi}. Note the plateau during the radiation-dominated era, in which SFDM behaves radiation-like due to the presence of SI. In the right-hand panel, the blue solid line displays the evolution for CDM and the blue dotted line displays the evolution for the entire matter content (CDM + baryons) in $\Lambda$CDM. Same vertical lines as in Fig.3.
	}
	\label{fig:SFDM-CDM-Omega}
\end{figure*}

Figure \ref{fig:SFDM-CDM-Omega} displays the evolution of the density parameters in the \glo{LSFDMmodel} compared to their evolution in the \glo{LCDMmodel}. In contrast to $\Lambda$CDM, where the evolution starts with a radiation-dominated era, the evolution of $\Lambda$SFDM starts with an era, where the energy density of the Universe is dominated by SFDM. During this era, SFDM evolves in its stiff phase, where $w \approx 1$. As the density drops rapidly, radiation becomes the dominant constituent of the Universe. The SFDM density parameter drops to a plateau, whose height is determined by the strength of the \acidx{SI}, during which SFDM is radiation-like. As the energy density of radiation drops, SFDM becomes dominant again, giving rise to the standard matter-dominated epoch. In the fiducial $\Lambda$SFDM model, matter-radiation equality takes place at $z=3421$, slightly earlier\footnote{In {\PaperLi}, the fiducial model was picked such that its $z_{eq}$ was in accordance with the up-to-date $\Lambda$CDM parameters as measured by an earlier Planck data release, while CLASS used here comes with a newer set of slightly different cosmological parameters for $\Lambda$CDM.} than in the $\Lambda$CDM model for which $z=3402$, compared to $z_{eq} = 3407 \pm 31$ in column 4 (``TT,TE,EE+lowE"; 1$\sigma$ confidence interval) of Table 2 in \citet{Collaboration2020}. So, the values for both of our models are within the 1$\sigma$ confidence interval with respect to the Planck value. The evolution of the density parameters after matter-radiation equality shows no significant differences between the fiducial \glo{LSFDMmodel} and $\Lambda$CDM.\par

\begin{figure*} [!htbp]
	{\includegraphics[width=0.49\textwidth]{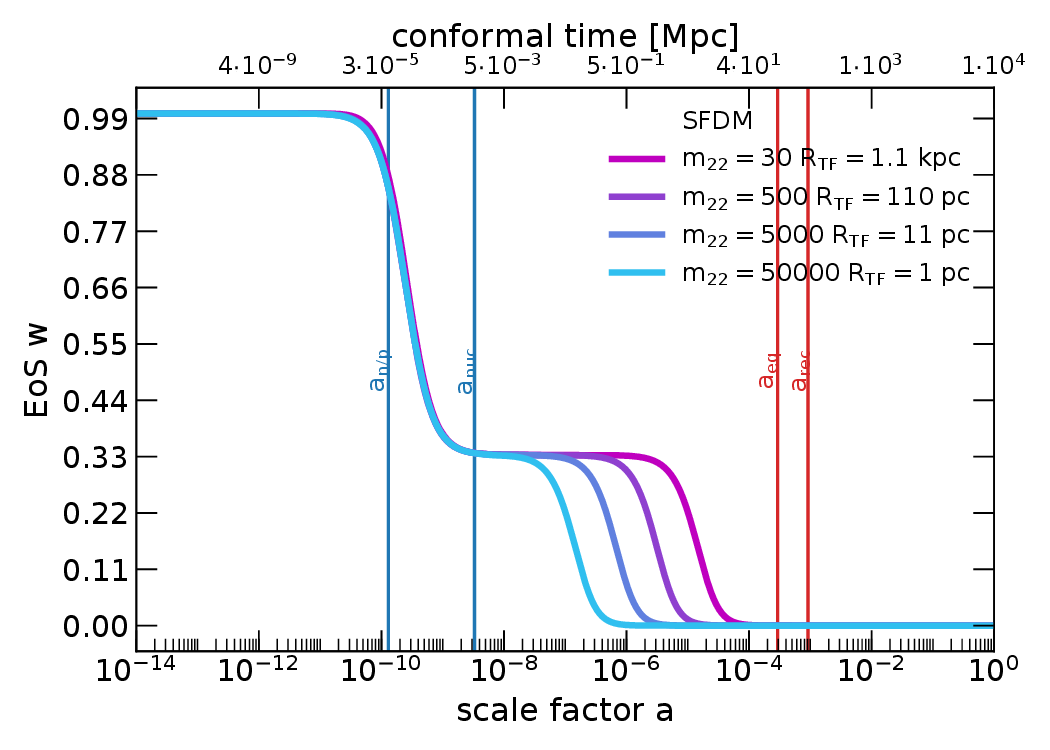}}
	{\includegraphics[width=0.49\textwidth]{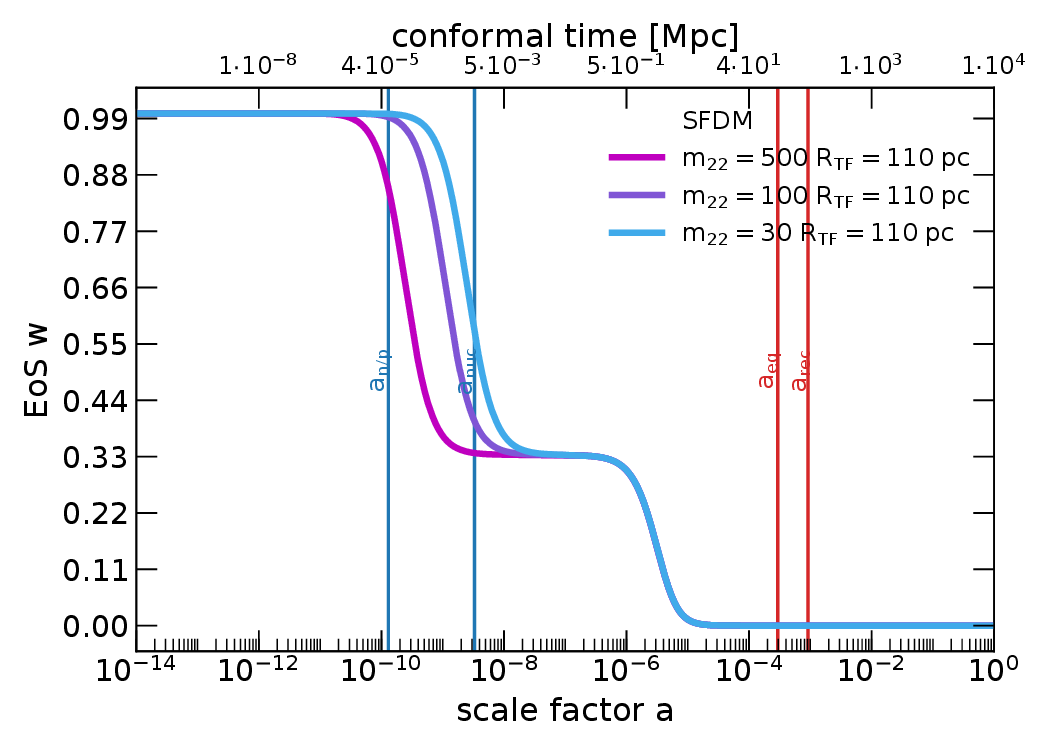}}		
	\caption[Evolution of the EoS for $\Lambda$SFDM models]
	{\textbf{Evolution of the EoS for \glo{LSFDMmodel}s with varying \boldmath{R$_{TF}$} and $m$, according to table II}. The left-hand panel displays the evolution of the \acidx{EOS} for the SFDM reference models of \hyperlinkcite{Shapiro2021}{\PaperSDR} with varying $R_{TF}$ and all of them meet the cosmological constraints set by \hyperlinkcite{Li2014}{\PaperLi}, because for all models the stiff phase ends before BBN is over. The transition to the CDM-like phase depends upon $R_{TF}$. The right-hand panel displays the evolution of the \acidx{EOS} for fixed $R_{TF} = 110$ pc, but for different masses as an illustration (note that the cases $m_{22} = 30, 100$ violate the constraints of \hyperlinkcite{Li2014}{\PaperLi}). Same vertical lines as in Fig.3.
	}
	\label{fig:SFDM-multi-SFDM}
\end{figure*}

Figures \ref{fig:SFDM-oHw} and \ref{fig:SFDM-CDM-Omega} essentially reproduced the results and findings of \hyperlinkcite{Li2014}{\PaperLi} with respect to the cosmology of a fiducial $\Lambda$SFDM model, confirming those results now with CLASS, and convincing ourselves that our code implementation works as intended. As pointed out earlier, this is especially important, given the unconventional stiff phase of SFDM, i.e. having to deal with a non-radiation-dominated background stiffer than radiation, in the early Universe. This difficulty presented a major challenge to our modification of CLASS, but now we are able to extend the analysis of \hyperlinkcite{Li2014}{\PaperLi} by calculating not only the background evolution, but also perturbation spectra for any $\Lambda$SFDM model parameters. 

Since the work of \hyperlinkcite{Li2014}{\PaperLi}, more SFDM models have been studied in the literature. In particular, the TF regime of SFDM has been analyzed in detail in the recent works by \citet{Dawoodbhoy2021} and \hyperlinkcite{Shapiro2021}{\PaperSDR}. For the sake of our analysis here, we only highlight the important findings in \hyperlinkcite{Shapiro2021}{\PaperSDR}, which pertain to the linear regime of structure growth. 
\hyperlinkcite{Shapiro2021}{\PaperSDR} investigate linear structure formation in the \glo{LSFDMmodel} by applying analytical approximations for the density perturbations in the radiation-dominated and matter(SFDM)-dominated epochs of the Universe. Their models are ``blind'' to the stiff phase of SFDM, however.
The most important finding of \hyperlinkcite{Shapiro2021}{\PaperSDR} concerns the novel constraints on SFDM parameters. It has been found there that, in order for linear SFDM perturbations to collapse into halos later in the history, the TF radius (related to the parameter combination $\lambda/m^2$) should be smaller than kpc-size. The reason stems from the scale-factor dependence of the corresponding (comoving) Jeans scale (which acts as a filtering scale) in the TF regime of SFDM, which also depends upon that parameter combination: this scale shrinks much faster as a function of scale factor than that for other models, such as ULAs, or FDM. On the other hand, there is an important nuance, because this same Jeans filtering scale can later ``recover''. As a result, the Jeans scale is responsible for an early initial cutoff in the power spectrum, while the subsequent decline thereafter is much shallower than in other models, like ULAs and FDM. 

Our task here will be to test the findings of the semi-analytic approach of \hyperlinkcite{Shapiro2021}{\PaperSDR} using CLASS, where we can also deal with the stiff phase and quantify its impact (if there is any) which has been neglected in \hyperlinkcite{Shapiro2021}{\PaperSDR}. Thereby, we will find new surprising results, whose explanations we will lay out in due course.  

\hyperlinkcite{Shapiro2021}{\PaperSDR} use a number of model parameters to compare their results to previous literature, especially ULAs. To aid the comparison with our results from CLASS simulations we use the same model parameters as summarized in Table \ref{tab:SFDM}, also for ULAs. Remember that ULAs have neither a stiff phase nor SI, so to include them in our comparison serves mainly as a way to expose the differences between complex and real scalar field dark matter. We will compare our results with those of \citet{UrenaLopez2016}, who investigated structure formation in real-field ULA models without SI. Instead of implementing ULAs directly into our version of CLASS (which was optimized for an early stiff phase), we compute ULA models using an amended version of CLASS provided by these authors, which is publicly available \footnote{available at \href{https://github.com/lurena-lopez/class.FreeSF}{https://github.com/lurena-lopez/class.FreeSF}}.\par

Let us first turn to the background evolution of the models in question.
The left-hand panel in Fig.~\ref{fig:SFDM-multi-SFDM} displays the evolution of the \acidx{EOS} of the SFDM reference models of table II with varying $R_{TF}$ as indicated in the legend. The mass of the individual model is chosen such that the cosmological constraints by \hyperlinkcite{Li2014}{\PaperLi} are met. Therefore, the stiff phase ends for all models at the same time, but the duration of the radiation-like phase is determined by equation \eqref{eq:EQrhow}, i.e. it depends upon $R_{TF}$, hence resulting in the transition to the CDM-like phase at different times. The right-hand panel displays the evolution for the reference model with $R_{TF} = 110$ pc and two alternative masses to demonstrate the dependence of the \acidx{EOS} from $m$. It can be clearly seen that the time of the transition from radiation-like phase to CDM-like phase is only influenced indirectly by $m$ via the normalized strength of the \acidx{SI}, $\syidx{SYMlambdasit}$ in equation \eqref{eq:EQrhow}, but the end of the stiff phase is directly linked to $m$: the higher $m$, the faster the \acidx{SF} oscillates, which results in an earlier end of the stiff phase. This is in accordance with the findings of \hyperlinkcite{Li2014}{\PaperLi}.

\par
\begin{figure} [!htbp]
	\centering	
	\includegraphics[width=\PW\columnwidth]{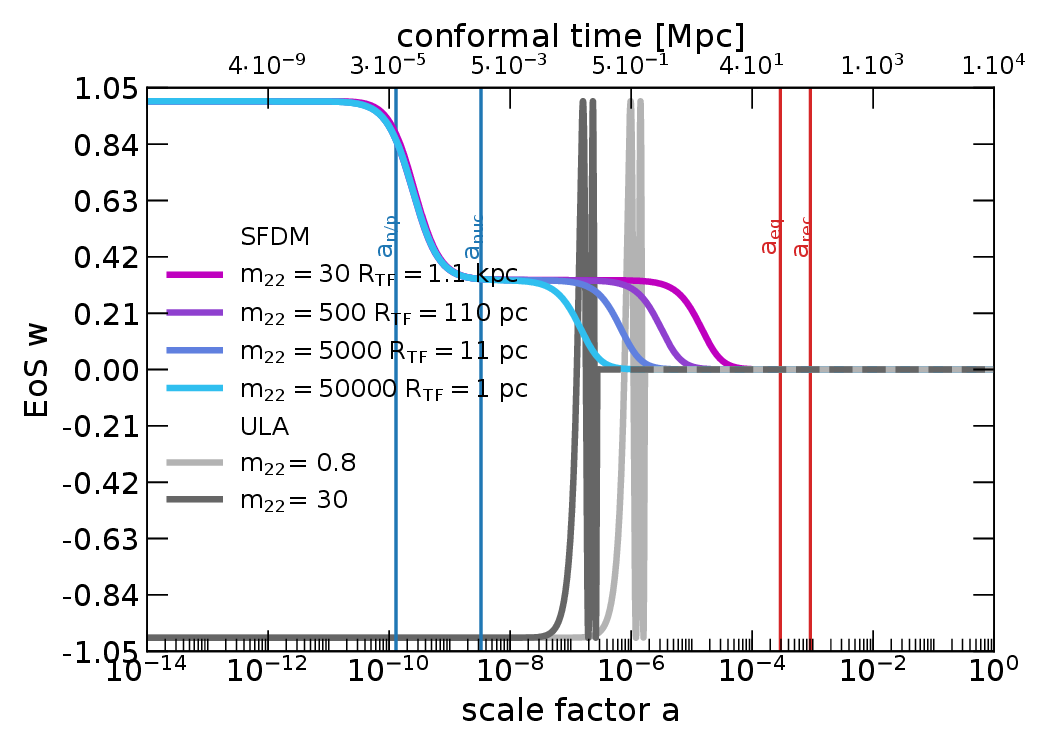}		
	\caption[\acidx{EOS} for multiple SFDM and ULA models]
	{\textbf{\acidx{EOS} for multiple SFDM and ULA models}. The colored lines display the \acidx{EOS} of the SFDM reference models also shown in Fig.~\ref{fig:SFDM-multi-SFDM}. The gray lines display the \acidx{EOS} of two ULA models. All model parameters are from table II. For ULAs, the evolution starts in a CC-like phase ($w=-1$), which enters an oscillatory regime, where the \acidx{EOS} oscillates between $-1$ and $+1$. The oscillation frequency increases rapidly with $a$, which prohibits reasonable plotting. Therefore, the oscillations have been suppressed in the plot at some time, indicated by the dashed lines where $\langle w \rangle = 0$. Vertical lines are the same as in Fig. 3. 
	}
	\label{fig:SFDM-w-multi-ULA}
\end{figure}

Comparing the SFDM models to ULAs enables a qualitative insight into the characteristics of complex versus real \acidx{SF}s.
For this purpose, we plot again the EoS parameter of SFDM (left-hand panel of Fig.~\ref{fig:SFDM-multi-SFDM}), along with that of the two ULA models in table II, in Fig.~\ref{fig:SFDM-w-multi-ULA}.
First, we look at equation \eqref{eq:EQeosgeneral}; the two terms in numerator and denominator correspond to the kinetic energy and the potential energy of the \acidx{SF}. 
SFDM starts in the early Universe with $w=1$, as indicated by the colored lines in Fig.~\ref{fig:SFDM-w-multi-ULA}, whereas ULAs (gray lines in Fig.~\ref{fig:SFDM-w-multi-ULA}) start with $w=-1$. This immediately indicates different physical properties for complex and real \acidx{SF}s. For complex \acidx{SF}s the kinetic energy is the dominating term, resulting in $w=1$, leading to the conclusion that complex fields gain kinetic energy from the phase of the field. Real scalar fields like ULAs do not gain kinetic energy, resulting in the dominant potential term in \eqref{eq:EQeosgeneral}, which results in $w=-1$, i.e. a CC-like behavior of ULAs in that early phase. Consequently, SFDM dominates in the early Universe over all other cosmic components, while ULAs never dominate in the early Universe. Therefore, they are not subject to constraints from BBN, or $z_{eq}$, in terms of their impact on the expansion history.

Now, this early phase, characterized by $w=+1$ for SFDM and $w=-1$ for \acidx{ULAs}, respectively, ends as soon as the oscillation frequency $\omega$ of the field exceeds the expansion rate $H$, and the field thereafter enters the oscillatory regime. This can be best seen for ULAs: for them, $\omega$ increases very rapidly, too. However, the EoS of ULAs oscillates between $w=-1$ and $w=+1$. These oscillations are suppressed in the plots after the third oscillation, and a dashed line at $w=0$ is plotted, which corresponds to the average value of the \acidx{EOS} in the oscillatory regime (i.e. the CDM-like phase). The fast transition to the oscillatory regime also indicates a very fast transition from the CC-like phase to the CDM-like phase of ULAs. In contrast to that, SFDM shows a soft transition from stiff to radiation-like to CDM-like phase and moreover, its \acidx{EOS} effectively converges to $w=0$. In fact, this is reflected in the mild decrease of pressure as seen in the right-hand panel of Fig.~\ref{fig:SFDM-EOS-rho}, and distinguishes complex from real scalar field models. Thus, the EoS parameter $w$ of the complex scalar field oscillates with declining amplitude around the averaged value, as a result of time-averaging density \eqref{subeqn:EQkgbgsfdmavgde} and pressure \eqref{subeqn:EQkgbgsfdmavgpr}. In contrast, the exact amplitude of the $w$ of ULAs never converges, but keeps oscillating between the two extremes given by $w=-1$ and $w=+1$, though the averaged amplitude is zero, see also \citet{Matos2002,Magana2012a}. This behavior is attributed to the different nature of complex and real scalar fields and has nothing to do with the presence of \acidx{SI}. Indeed, we confirm the same difference between complex FDM and ULAs discussed below; see Fig.~\ref{fig:FDM-UL-rho}. 

We note in passing that the overall evolution of density parameters $\Omega_i$ vs $a$ for $\Lambda$-ULA models looks the same as for $\Lambda$CDM shown in the right-hand panel of Fig.\ref{fig:SFDM-CDM-Omega} (see also Fig.\ref{fig:FDM-UL-Omega}), never mind the different EoS between ULAs and CDM.  \par

\subsection{Evolution of Density Perturbations}

\begin{figure} [!htbp]
	{\includegraphics[width=\PW\columnwidth]{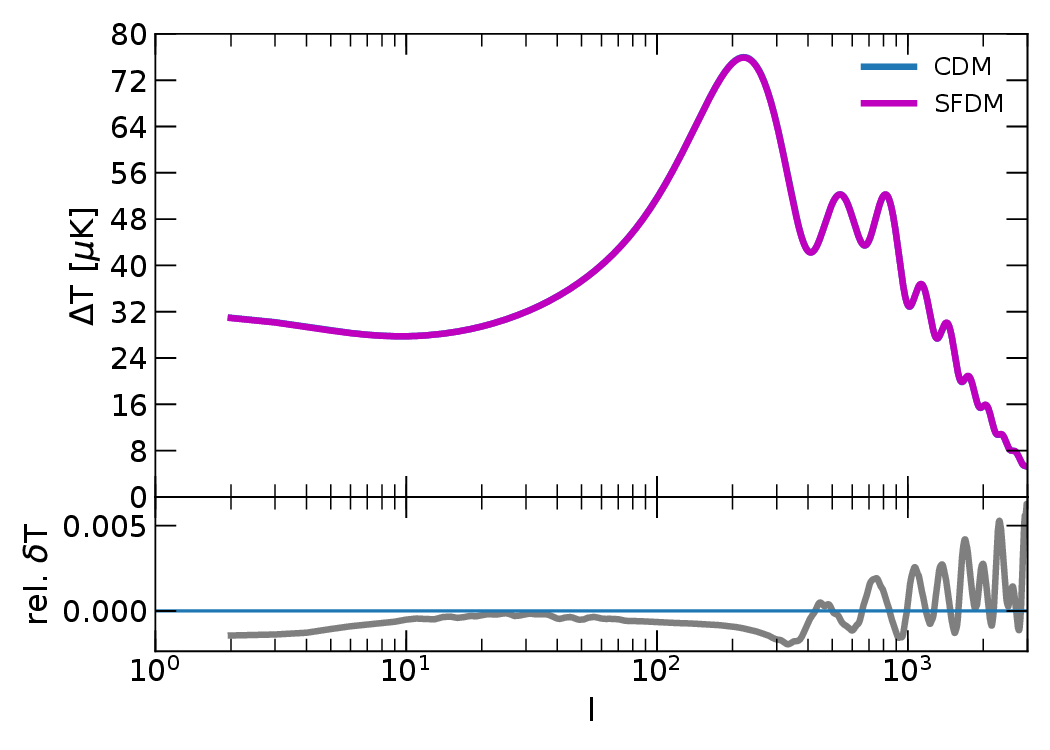}}			
	\caption[Temperature anisotropies in the CMB for SFDM]
	{\textbf{Temperature anisotropies in the CMB calculated for the fiducial $\Lambda$SFDM model of \hyperlinkcite{Li2014}{\PaperLi} ($m_{22} = 30, R_{TF} = 1.1$ kpc), compared to $\Lambda$CDM}. The top part of the panel displays the temperature anisotropies in the CMB for the fiducial \glo{LSFDMmodel} (magenta) of \hyperlinkcite{Li2014}{\PaperLi} vs. that for $\Lambda$CDM (cyan); they lie basically on top of each other. The bottom part of the panel displays the relative temperature differences between the models; the fiducial \glo{LSFDMmodel} shows no significant differences to $\Lambda$CDM.
	}
	\label{fig:SFDM-Temp}
\end{figure}

\begin{figure*} [!htbp]
	{\includegraphics[width=0.49\textwidth]{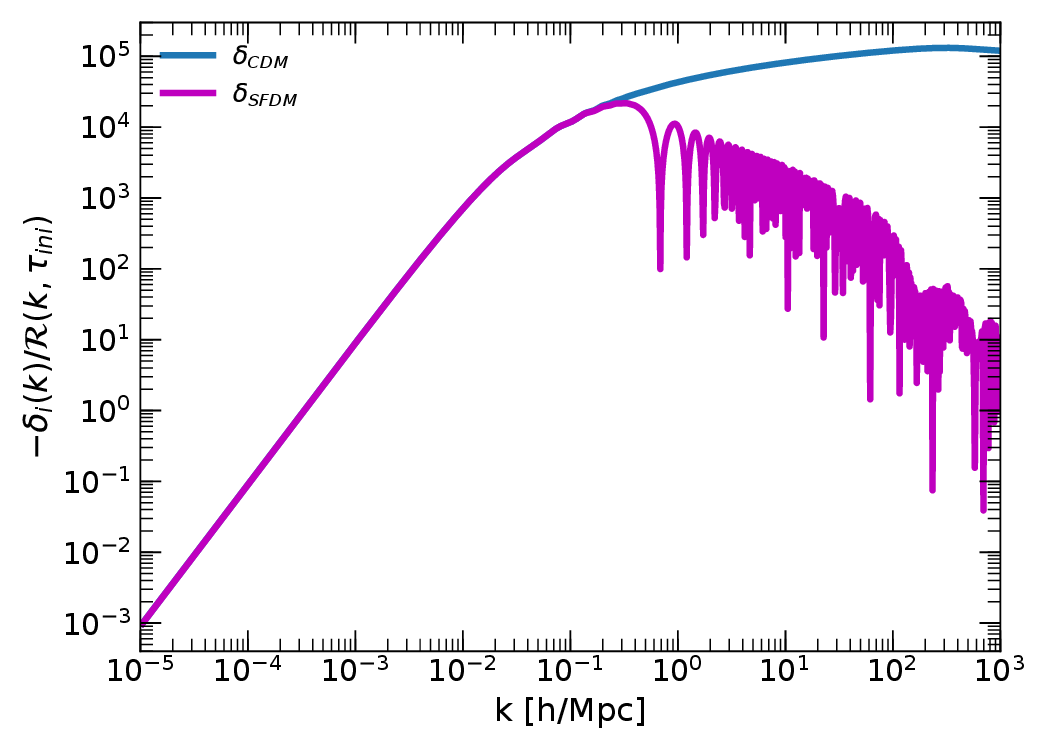}}		
	{\includegraphics[width=0.49\textwidth]{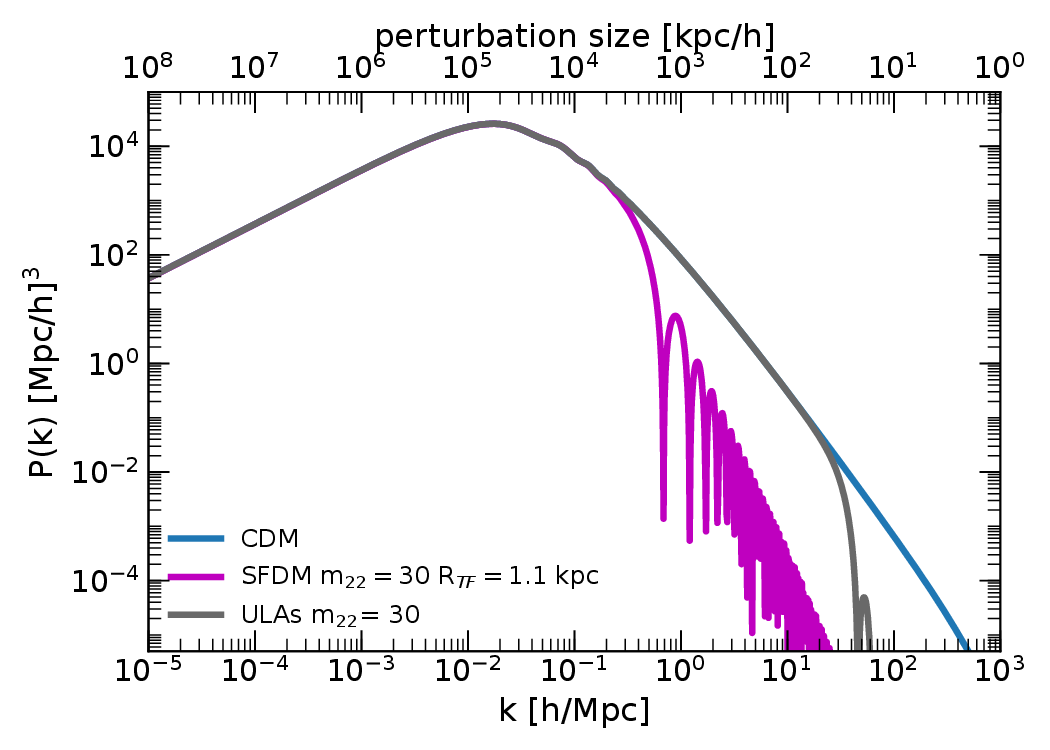}}	
	\caption[Transfer functions and matter power spectra of $\Lambda$SFDM and $\Lambda$CDM models]
	{\textbf{Transfer functions and matter power spectra of the fiducial \glo{LSFDMmodel} and of $\Lambda$CDM at \boldmath{$z=0$}}. The left-hand panel displays the transfer functions calculated for the fiducial \glo{LSFDMmodel} of \hyperlinkcite{Li2014}{\PaperLi} (solid magenta line) and for $\Lambda$CDM (solid cyan line). The scaling is according to the CLASS convention ($\mathcal{R}(k,\tau_{ini})=1$). The right-hand panel displays the matter power spectra of the same \glo{LSFDMmodel} and $\Lambda$CDM. In addition, the matter power spectrum for one ULA model of same mass (gray solid line) is included, as well.
	}
	\label{fig:SFDM-mtk-mpk}
\end{figure*}

In this subsection, we turn to the perturbation spectra. 
First, we display the power spectra of the spherical anisotropies in the CMB temperature, calculated for the fiducial $\Lambda$SFDM model of \hyperlinkcite{Li2014}{\PaperLi} along with that for $\Lambda$CDM in Fig.~\ref{fig:SFDM-Temp}. The lower part displays the relative deviation of the $\Delta T$ between both models. There are no significant differences. However, as we will see below, there are differences in the matter power spectra between SFDM and CDM. This is explained by the fact that the CMB temperature is determined only by the energy density component $T_{00}$ of the energy momentum tensor, whereas the density perturbations also depend upon the diagonal elements $T_{ii}, i=1,..,3$, which include the pressure (see e.g. \citet{Dodelson2003}), and this pressure differs between SFDM and CDM, being higher for SFDM. Hence, as the evolution of density perturbations unfolds, small-scale structures are clearly suppressed in SFDM compared to CDM, which can be seen in the left-hand panel of Fig.~\ref{fig:SFDM-mtk-mpk}.

It shows the transfer functions of the density perturbations for SFDM and CDM in the CLASS convention, normalized to $\mathcal{R}(k,\tau_{ini})=1$, where $\mathcal{R}(k,\tau_{ini})$ denotes the spatial curvature perturbation. There is a clear cutoff in the transfer function for SFDM with oscillations in k-space. In addition to these oscillations, the envelope of the transfer function clearly displays some wiggles toward small structures (above $k \sim 10^2$), which are explained shortly. The right-hand panel in Fig.~\ref{fig:SFDM-mtk-mpk} displays the corresponding matter power spectra for CDM, SFDM and ULAs. SFDM and ULAs show both a clear cutoff in their power spectrum. The cutoff for the fiducial SFDM model takes place for structures of size $\sim 2 \times 10^4$~kpc/h, and the falloff toward smaller spatial scales (high $k$) is mild, which is discussed in detail below. On the other hand, for ULAs of the same particle mass, the cutoff happens at $\sim 2 \times 10^2$~kpc/h with a very steep falloff toward high $k$.\par

\begin{figure} [!htbp]
	{\includegraphics[width=\PW\columnwidth]{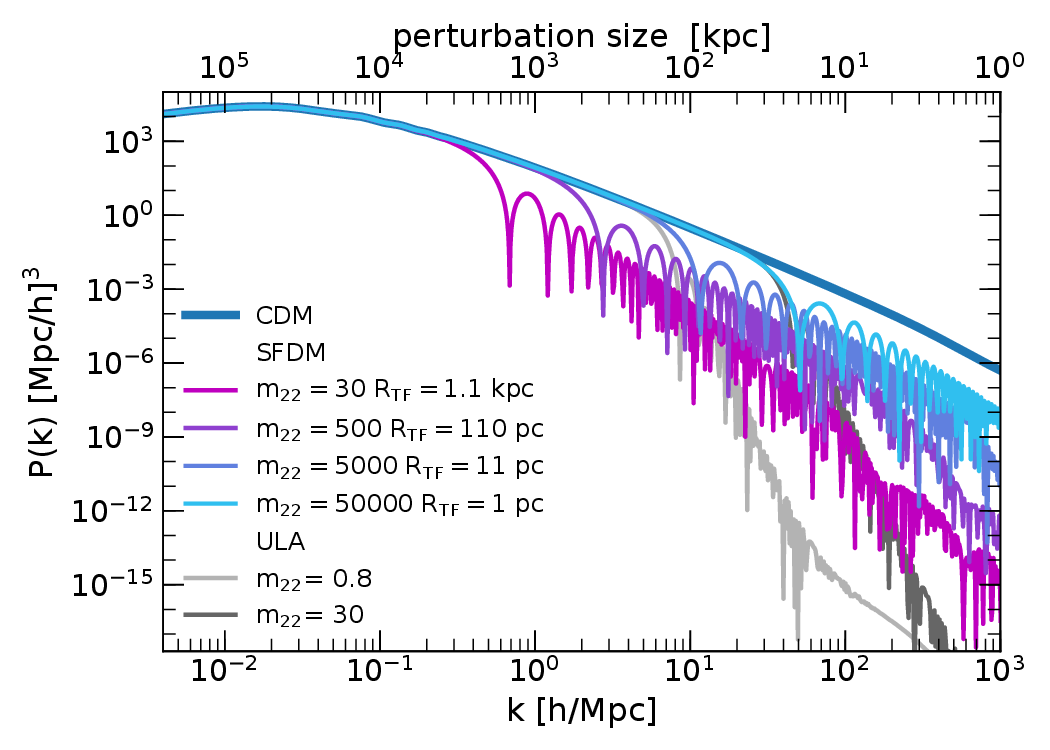}}		
	\caption[Matter Power Spectra of multiple SFDM models and ULA models]
	{\textbf{Matter power spectra of multiple SFDM and ULA models according to Table \ref{tab:SFDM}, all at \boldmath{$z=0$}}. The thick dark blue solid line displays the matter power spectrum of CDM for comparison. The colored solid lines display the matter power spectra of SFDM models with varying SI strength (via $R_{TF}$) and masses. The gray lines show two different ULA models. This plot may be compared to Figure 7 (top panel) in \hyperlinkcite{Shapiro2021}{\PaperSDR}, but note our expanded $y$-axis range here, down to very small power.
	}
	\label{fig:mPk-multi-SFDM}
\end{figure}

As mentioned earlier, one task of our work consists in checking and comparing to the findings of \hyperlinkcite{Shapiro2021}{\PaperSDR} with respect to i) the strong suppression of power in the matter power spectrum that occurs for $\lambda/m^2 \sim R_{TF} \gtrsim 1$ kpc, and ii) the subsequent falloff of the spectra which is milder than in other DM models.
Therefore, we compute with CLASS the matter power spectra of the reference models used and analyzed in \hyperlinkcite{Shapiro2021}{\PaperSDR}, which are displayed in Fig.~\ref{fig:mPk-multi-SFDM}. In fact, we confirm that the cutoff in the power spectra takes place at increasingly smaller $k$ with increasing $R_{TF}$. We also confirm that the falloff of the spectra toward higher $k$ is milder for SFDM models, compared to ULAs. However, we show an expanded range of power on the y-axis of Fig.~\ref{fig:mPk-multi-SFDM}, in order to include the features seen in the transfer function of Fig.~\ref{fig:SFDM-mtk-mpk}. This way, we can notice two things: first, the SFDM models show the same wiggles in the envelope of the power spectra toward high $k$, as seen in the transfer functions. Second and interestingly, for ULA models the initial very steep cutoff \textit{also} flattens at some point but only at very small power, which can be explained as follows.

In contrast to CDM, the evolution of the density perturbations in SFDM depends not only on the time of horizon entry, but has additional features: the strength of \acidx{SI} parameterized by $\lambda/m^2$ (or $R_{TF}$), and the implied Jeans mass by which we have to distinguish if the enclosed mass of a perturbation mode that enters the horizon is sub-Jeans or super-Jeans. It has been shown in \hyperlinkcite{Shapiro2021}{\PaperSDR} that, in the CDM-like phase of SFDM, the Jeans scale due to SI shrinks rapidly with time ($M_J \propto ~a^{-3}$). 
More precisely, the evolution of individual modes can be characterized as follows (see \hyperlinkcite{Shapiro2021}{\PaperSDR} and \citet{Suarez2015}): sub-Jeans modes entering the horizon in the radiation-like phase of SFDM perform an acoustic oscillation with a constant amplitude. If they transition to CDM-like behavior before matter-radiation equality $a_{eq}$, oscillation continues but the amplitude grows $\propto a^{1/4}$, until they eventually get super-Jeans. Super-Jeans modes display an evolution like standard CDM without any oscillation.\par

Now, complex FDM and real ULAs have characteristic Jeans scales as well, never mind that they do not go through a radiation-like phase. However, in contrast to complex FDM and SFDM models, for ULAs the Jeans mass $M_J$ shrinks $\propto a^{-3/4}$ (e.g. \citet{Hu2000}). As a result, an eventual transition of an individual perturbation mode from sub-Jeans to super-Jeans is deferred compared to the other models and therefore a stronger suppression at small spatial scales occurs (even after the time of recombination, as seen in the top row right-hand panel of Fig.~\ref{fig:FDM-Tk-tau}, discussed in detail in the next section). This fact, in combination with the constant amplitude of the oscillations provides an explanation for the very steep cutoff in the power spectra of the ULA models.\par
Now, the ensuing mild falloff of the envelopes of the matter power spectra for the SFDM models has to do with the length of the suppression of growth of the amplitudes of the oscillations from the time they enter the horizon, until they get super-Jeans. This phenomenon also occurs in CDM, and therefore the slopes of the power spectra are similar.
Finally, the ULA model with m$_{22}=0.8$ has no radiation-like phase in its EoS, but its modes remain sub-Jeans until $\tau \simeq 10^2$, long after $a_{eq}$. So, growth of these modes is suppressed for a very long time. The duration of the suppression is determined by the time of horizon entry, as in CDM. Again, it results in a shallow slope, similar than for CDM.\par

\begin{figure*} [!htbp]
	{\includegraphics[width=0.49\textwidth]{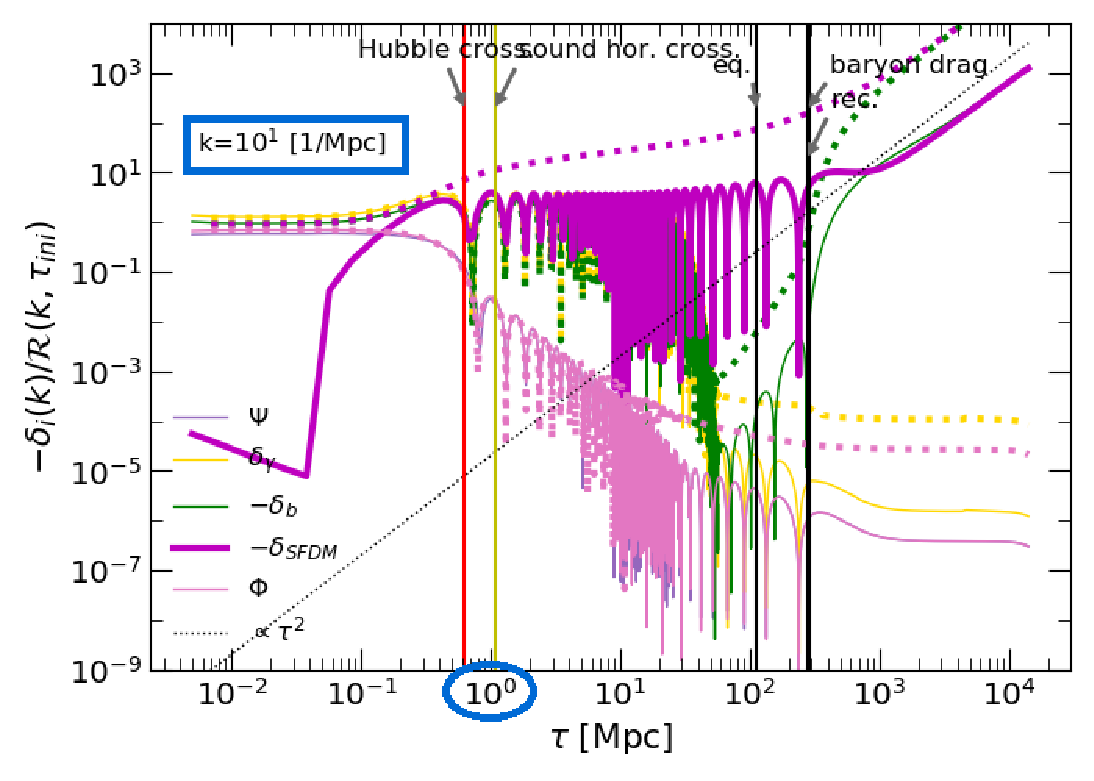}}
	{\includegraphics[width=0.49\textwidth]{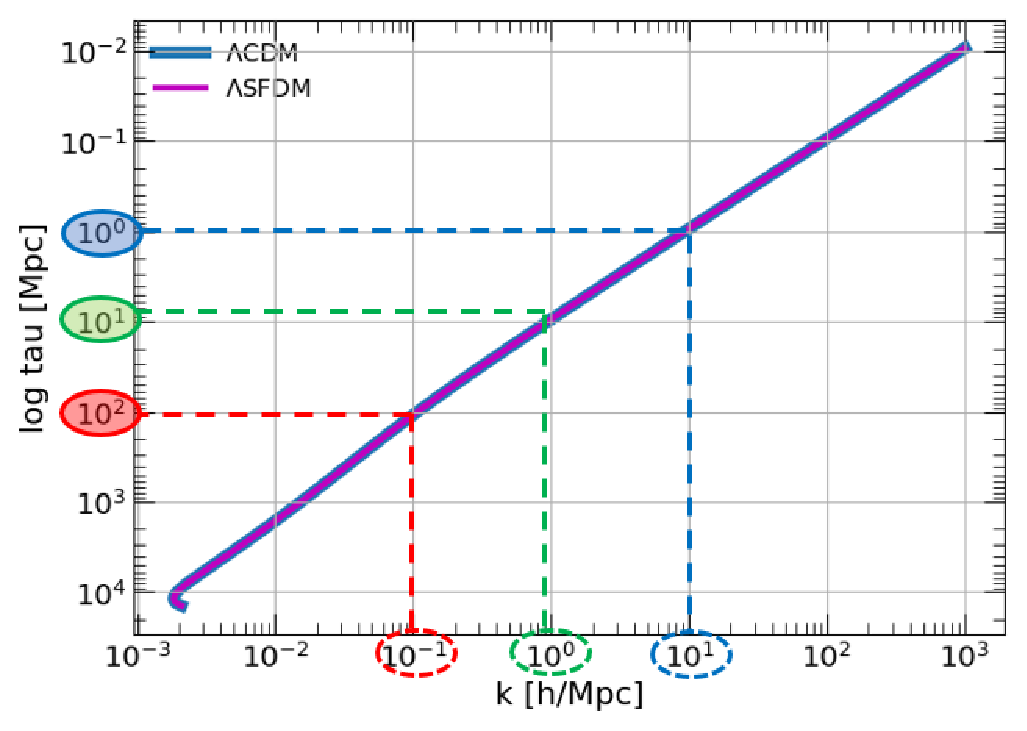}}	
	{\includegraphics[width=0.49\textwidth]{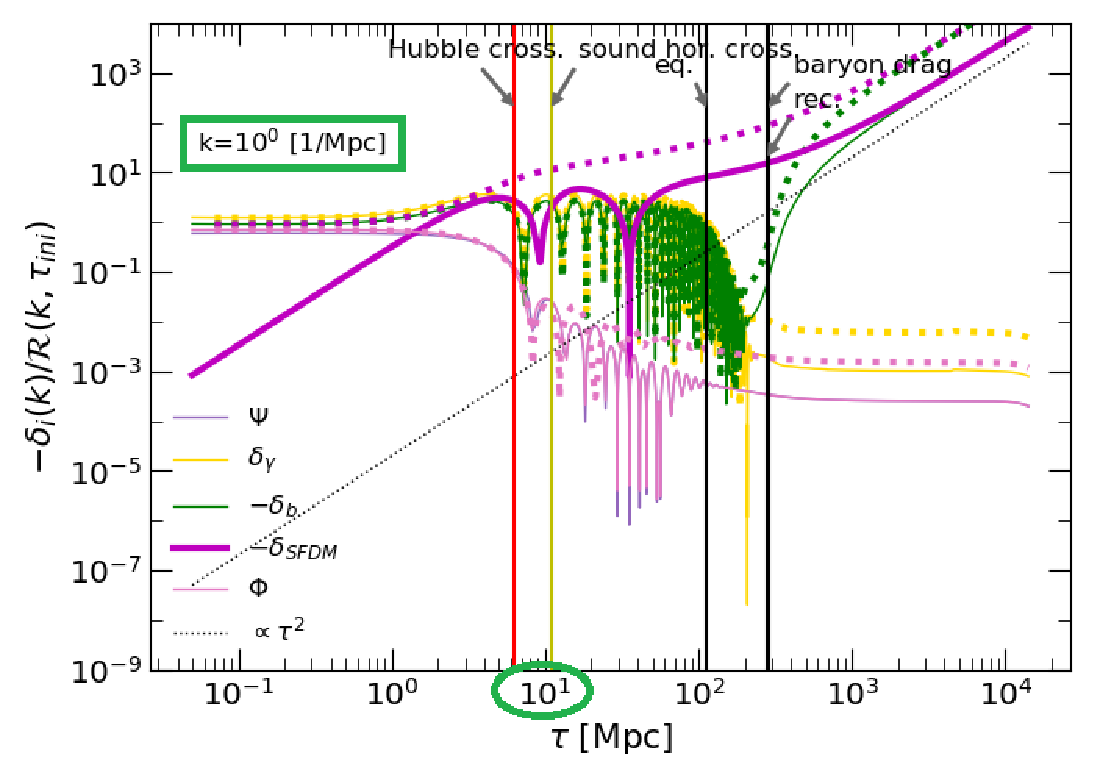}}
	{\includegraphics[width=0.49\textwidth]{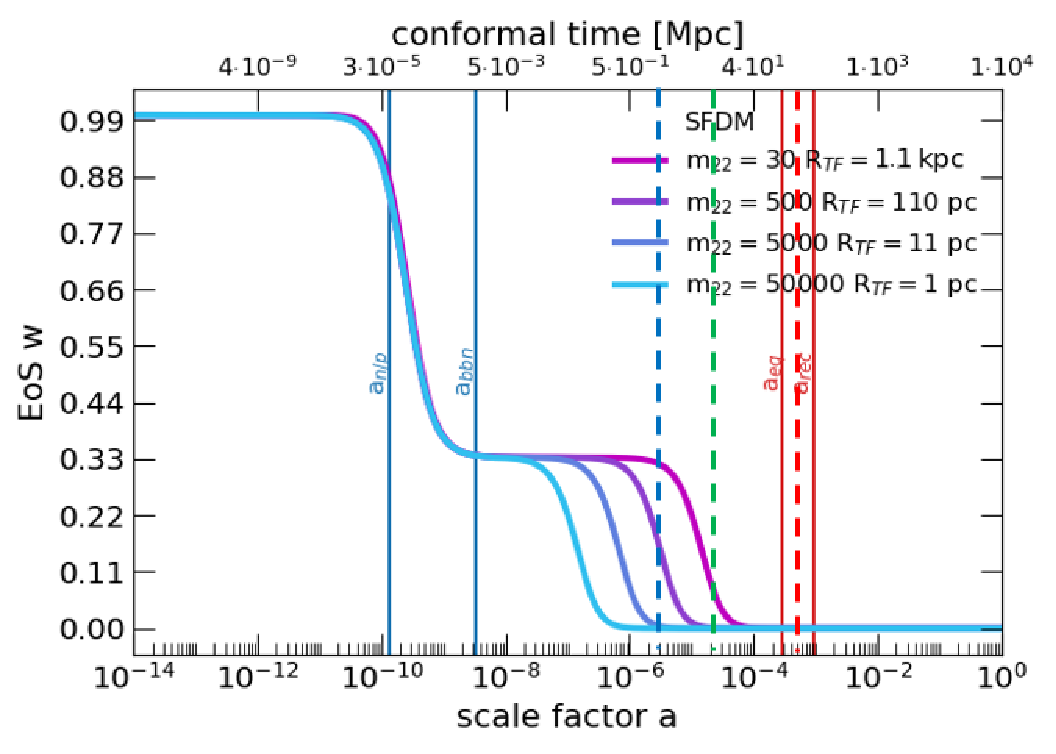}}		
	{\includegraphics[width=0.49\textwidth]{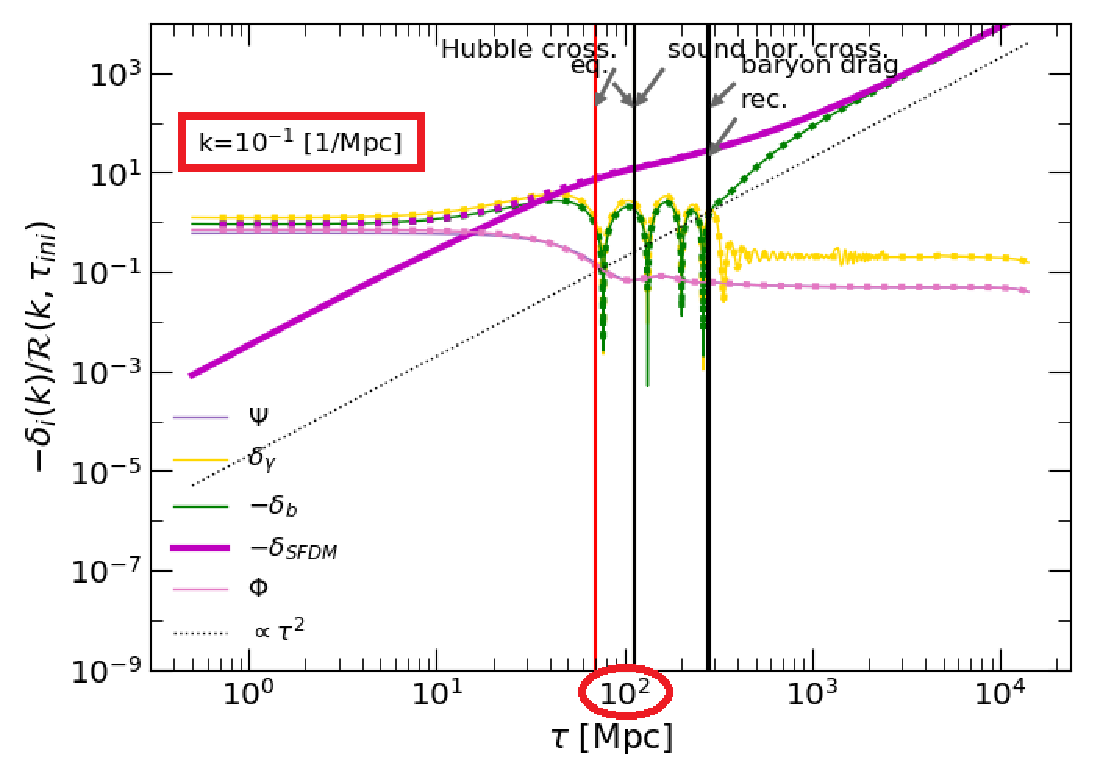}}
	{\includegraphics[width=0.49\textwidth]{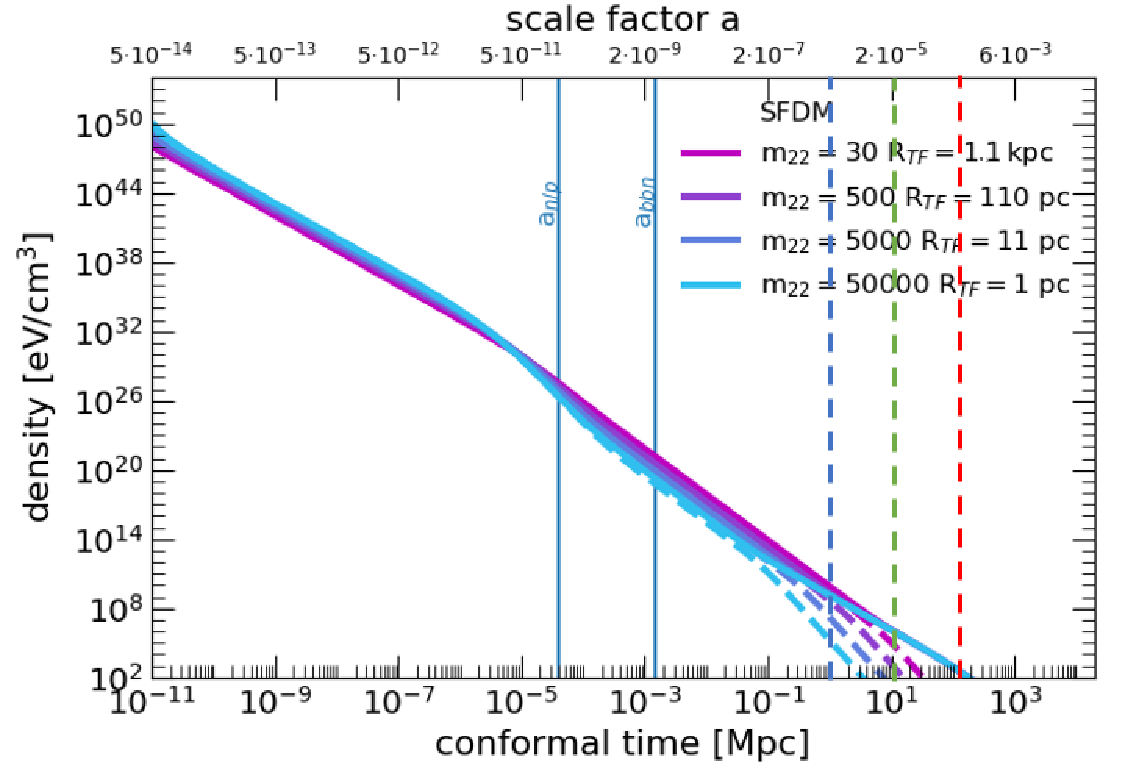}}							
	\caption[Explanation of the cutoff and shallow drop in the SFDM's matter power spectrum]
	{\textbf{Illustration of the cutoff and subsequent shallow falloff in the matter power spectrum of SFDM.} These plots illustrate the position of the cutoff in the matter power spectrum, as well as the subsequent shallow falloff of the matter power spectrum seen in Fig.~\ref{fig:SFDM-mtk-mpk} and Fig.~\ref{fig:mPk-multi-SFDM}, first found in \hyperlinkcite{Shapiro2021}{\PaperSDR}.
	The left-hand panels illustrate the evolution, as a function of conformal time $\tau$, of three individual perturbation modes of SFDM ($k = 10, 1, 0.1$ Mpc$^{-1}$ from top to bottom in thick magenta lines), along with those of the other cosmic components (baryons in green and photons in yellow) and the perturbed metric potentials $\Psi$ and $\Phi$ (pink thin and thick lines).
	The dotted curves in the same color indicate the corresponding evolution of perturbations in the CDM model (CDM itself is the dotted magenta curve).
	Some benchmark epochs are indicated with vertical lines of different color: Hubble crossing (red), sound horizon crossing (green), matter-radiation equality (``eq", thin black) and time of recombination and baryon drag (thick black). The top right-hand panel shows the respective time of the modes' horizon entry. The middle and bottom right-hand panels show the evolution of the EoS and energy density of SFDM, respectively, but now include the time of horizon-entry of the selected modes, as an illustration.
	See main text for more explanations. 
	}
	\label{fig:SFDM-Tk-tau}
\end{figure*}

In Fig. \ref{fig:SFDM-Tk-tau}, we illustrate the points regarding the cutoff in the matter power spectra of SFDM models, and their ensuing shallow falloff. The left-hand panels display the evolution, as a function of conformal time $\tau$, of three characteristic modes of perturbations of the \textit{fiducial} SFDM model (R$_{TF}=1.1$~kpc, $m_{22}=30$), in the order of their entry into the horizon, from top to bottom. The top right-hand panel displays the respective time of horizon crossing, while the middle and bottom right-hand panels show the \acidx{EOS} and the evolution of density and pressure of SFDM, respectively, with these horizon entry times indicated, for better illustration. The three characteristic modes are $k=10^{1}$~1/Mpc (blue), $k=10^{0}$~1/Mpc (green) and $k=10^{-1}$~1/Mpc (red). The perturbation with the smallest size (blue) is the first one to enter the horizon at $\tau=10^{0}$~Mpc, followed by the others at $\tau=10^{1}$~Mpc (green) and $\tau=10^{2}$~Mpc (red). Solid lines indicate all components of $\Lambda$SFDM, while dotted lines display the corresponding quantities for $\Lambda$CDM which we include in the plots for comparison's sake. The time of horizon entry of our chosen SFDM perturbation modes is also indicated in the middle right-hand panel by the colored vertical dashed lines. Here it is seen that, upon horizon entry of the ``blue'' and ``green'' perturbations, SFDM still has significant pressure, whereas by the time the ``red'' one enters, SFDM has already morphed into the pressure-less, CDM-like phase. The evolution of the SFDM perturbations (solid magenta line) can be explained as follows. The ``blue'' perturbation enters the horizon just at the end of the radiation-like phase of SFDM, and it is sub-Jeans, which explains the acoustic oscillation and the constant amplitude. Shortly after that, the \acidx{EOS} drops to CDM-like, resulting in a growth of that perturbation amplitude like $a^{1/4}$. The transition to super-Jeans coincides with matter-radiation equality (``eq''), the oscillation ceases and the subsequent evolution is the same as for a CDM perturbation (top left). The ``green'' mode enters the horizon just at the transition to pressure-less, CDM-like \acidx{EOS} of SFDM. The result is an acoustic oscillation and a growth of the amplitude. For this mode, the transition to super-Jeans coincides with $a_{eq}$ and the successive evolution is the same as for a CDM perturbation (middle left). Finally, the ``red'' mode enters the horizon during the CDM-like phase of SFDM, and it is super-Jeans. It therefore displays a CDM-like evolution throughout (bottom left). From this we can conclude, that the cutoff occurs for the minimum value of $k$, where the mode is still sub-Jeans at the time of horizon crossing. This induces acoustic oscillations of the density perturbations of SFDM, where higher pressure results in a higher frequency of the oscillation. 

The ensuing mild falloff of the envelopes of the matter power spectra for SFDM is determined by the time span during which the growth of the amplitudes of the oscillations is suppressed, beginning with the time a perturbation mode enters the horizon. Since in SFDM models, the growth of density amplitudes is limited to modes that are super-Jeans, and that the growth is only $\propto a^{1/4}$ for CDM-like modes before they get super-Jeans, the beginning of the growth of structure is nearly at the same time, at roughly $a_{eq}$, for all perturbation modes, as seen in the left panels of Fig.~\ref{fig:SFDM-Tk-tau}. This is analogous in standard CDM, where growth of structure is suppressed until $a_{eq}$, due to the \emph{Meszaros effect}\footnote{The Meszaros effect is not governed by CDM's Jeans mass and suppresses the formation of CDM structures until the sharp end at $a_{eq}$. The fact that our SFDM models get super-Jeans around $a_{eq}$ is a coincidence related to our model parameters. So we like to note, that although there is a similar behavior of CDM and SFDM, with respect to the suppression of structures toward high k, both are based on different physical processes.} (\citet{Meszaros1974}). Therefore, the slopes of the envelopes that result for the power spectra are similar in CDM and SFDM. After all, the expansion histories of the fiducial \glo{LSFDMmodel} and $\Lambda$CDM are similar, except for the initial stiff phase. Therefore, the time of horizon crossing of equal-sized perturbation modes does not differ, either. In both models, the common point in time for the end of suppression of the growth of structure, is matter-radiation equality $a_{eq}$. So, the quantity that impacts the suppression, in both models, is the nearly identical time of horizon entry of equal-sized perturbation modes, resulting in nearly identical slopes in the power spectrum. The wiggles seen in the SFDM transfer function toward high $k$ are explained by the small variations in the point in time when the density perturbations of SFDM start growing in super-Jeans mode, caused by the ``oscillatory'' effects governing also the ``overshooting'' of the transfer functions explained in section \ref{sec:resultsFDMC}.

Now, the transition to a flatter power spectrum can be also seen for ULAs in Fig.~\ref{fig:mPk-multi-SFDM}, where the initially very steep falloff of the power spectrum flattens to nearly the same degree as seen for SFDM. The reason, that this effect occurs only at very small powers for ULAs, is the evolution of the Jeans mass in that model, which shifts the transition from sub-Jeans to super-Jeans modes to much later times, compared to SFDM.
\\
Moreover, we point to the offset seen at low $\tau$ in the SFDM perturbation modes, away from the initial amplitudes of the other components, which is a marked difference to CDM, see Fig. \ref{fig:SFDM-Tk-tau}, solid magenta lines in the left-hand panels. We believe that this is a clear impact of the stiff phase, as follows. Relativistic perturbation theory predicts that the spatial curvature perturbation $\mathcal{R}(k,\tau_{ini})=const$, and even that $\mathcal{R}(k,\tau_{ini})=\mathcal{R}$, i.e. it has the same value for all perturbations, which is determined by the amplitude of primordial perturbations $A_{s}$. Since $A_{s}$ is not directly accessible to observations, it is calibrated using other quantities, among them $\sigma_8$ (\citet{Collaboration2020}). However, these predictions apply to a radiation-dominated Universe, and their validity may not be guaranteed for the very early stiff phase of SFDM. Nevertheless, the model parameters that we used guarantee that the Universe is radiation-dominated by the time BBN is over. Consequently, we can see that the SFDM perturbation modes converge to those predicted at horizon entry. In particular, they converge in amplitude to the respective CDM modes which are the dotted magenta lines in each panel.
\\
We also recognize that the evolution of the two metric potentials $\Psi$ and $\Phi$ is so close that their respective curves lie almost on top of each other, implying that the common assumption -- especially in analytic work -- of setting them equal is well justified.

\par
\begin{figure} [!htbp]
	{\includegraphics[width=\PW\columnwidth]{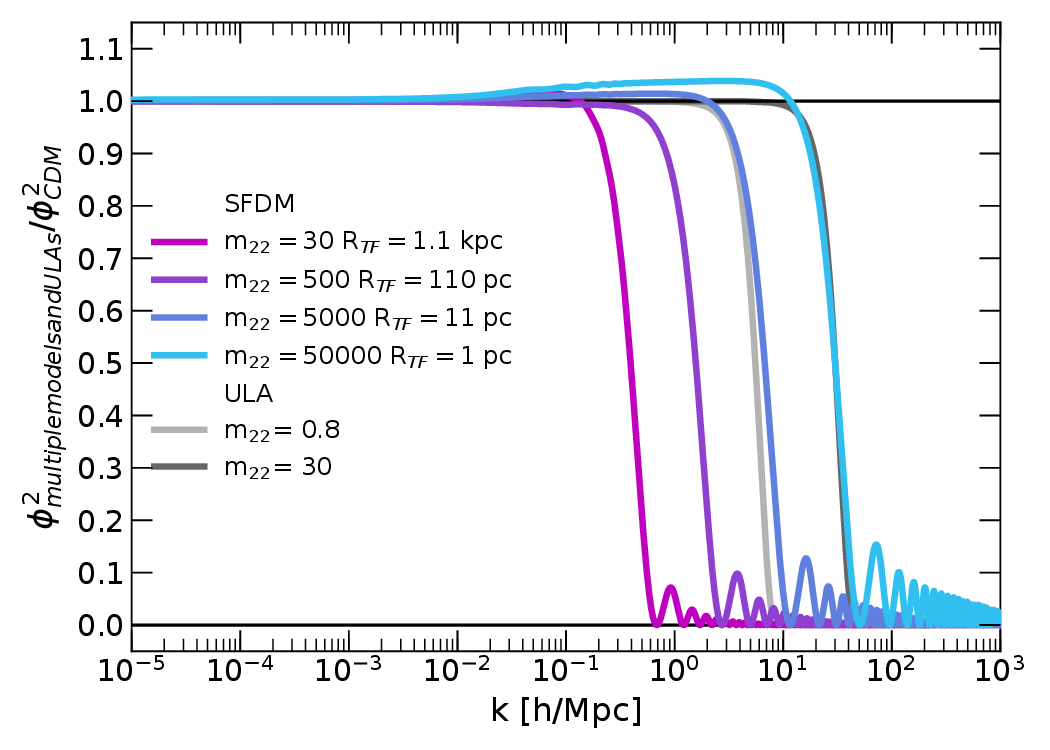}}			
	\caption[Transfer functions of multiple SFDM and ULA models relative to the CDM model, all at \boldmath{$z=0$}]
	{\textbf{Transfer functions of multiple SFDM and ULA models (according to Table \ref{tab:SFDM}) relative to CDM, all at \boldmath{$z=0$}}. The colored solid magenta lines display the transfer functions of different SFDM models with varying SI strength (via $R_{TF}$) and $m$ which all match cosmological constraints, according to \hyperlinkcite{Li2014}{\PaperLi}, i.e. the stiff phase ends before BBN is over. The gray lines display two different ULA models (the ULA case $m_{22}=30$ and the SFDM case with $m_{22}=50000$ overlap to a great extent). For some models, ``overshooting'' happens which is explained in Sec. \ref{sec:resultsFDM}. This plot may be compared to Figure 6 (top panel) in \hyperlinkcite{Shapiro2021}{\PaperSDR}.
	}
	\label{fig:mTk-multi-SFDM}
\end{figure}

Now, we additionally computed the transfer functions of the same models as used in \hyperlinkcite{Shapiro2021}{\PaperSDR} and normalized them by the transfer function of CDM, which are displayed in Fig.~\ref{fig:mTk-multi-SFDM}. Our results are comparable to those of \hyperlinkcite{Shapiro2021}{\PaperSDR}, but with a difference. For $k \gtrsim 10^{-2}$~h/Mpc, some of the transfer functions exhibit an ``overshooting'' compared to the CDM transfer function which, however, has no impact on the matter power spectra and CMB temperature anisotropies. The same effect appears for complex FDM, and since it is a simpler model to analyze, we defer the discussion of this feature to the next section \ref{sec:resultsFDM}. Moreover, our ULA model spectra do not show the spiky features seen in Figure 6 of \hyperlinkcite{Shapiro2021}{\PaperSDR}. This difference does not stem from the different code implementation -- axionCAMB (\citet{Hlozek2015}) there vs. CLASS here --, but it is rather caused by different $k$-samplings, used in CDM vs. ULA-model calculations. We chose an identical sampling rate for all models, resulting in smooth curves.\par

Finally, we elaborate on the fact that acoustic oscillations of the density perturbation modes of SFDM are induced as discussed above, where higher pressure in the EoS of SFDM results in a higher frequency of these oscillations. This phenomenon is akin to the well-known baryonic acoustic oscillations (``BAOs'') in the baryonic component, so we might call them ``SFDM or scalar field acoustic oscillations", or ``SAOs''. BAOs trace the subsequent formation of galaxies, and they are thereby seen in the large-scale structure revealed by galaxy surveys as characteristic BAO ``rings", which provide even a standard ruler for cosmological distance measurements. We could speculate that the same occurs for SAOs, which might reveal themselves as ``rings'' in the overall dark matter large-scale structure. However, since they are not visible in the baryonic component in the form of galaxies, it would be very hard to detect them. Also, the question arises whether these SAOs might create, or amplify ambient gravitational waves, so as to produce ring-like features in an otherwise stochastic gravitational-wave background. These questions are beyond the scope of this work, but might be of interest for future studies.

\section{Results for Complex Fuzzy Dark Matter and Comparison to ULAs}\label{sec:resultsFDM}

\subsection{Model Parameters}

In this section, we perform a comparison between complex-field FDM (i.e. SFDM without \acidx{SI}) and ULAs, which are implemented by a real \acidx{SF} as previously. Table \ref{tab:FDM} lists the model parameters of our runs with CLASS. In addition to the reference models, used to compare our results to other works, we include a complex FDM model in the third line, which is compatible with all cosmological constraints.
\begin{table} [!htbp] 
	\caption{Model parameters for complex FDM and real ULAs \label{tab:FDM}}
	\begin{ruledtabular}
		\begin{tabular}{rrlr}
			m [eV/c$^2$]          &  m$_{22}$  & $\syidx{SYMlambdasi}/(m c^2)^2$ [eV$^{-1}$ cm$^3$] & $R_{TF}$\\
			\hline
			$3.0 \times 10^{-21}$ & 30        & --   & --\\			
			$8.0 \times 10^{-23}$ & 0.8       & --   & --\\
			$3.0 \times 10^{-17}$ & $3\cdot 10^5$     & --   & --\\
			\hline				
			$3.0 \times 10^{-21}$ & 30        & --   & --\\	
			$8.0 \times 10^{-23}$ & 0.8       & --   & --\\					
		\end{tabular}
	\end{ruledtabular}
\end{table}

\subsection{Background Evolution}
\begin{figure*} [!htbp]
	{\includegraphics[width=0.49\textwidth]{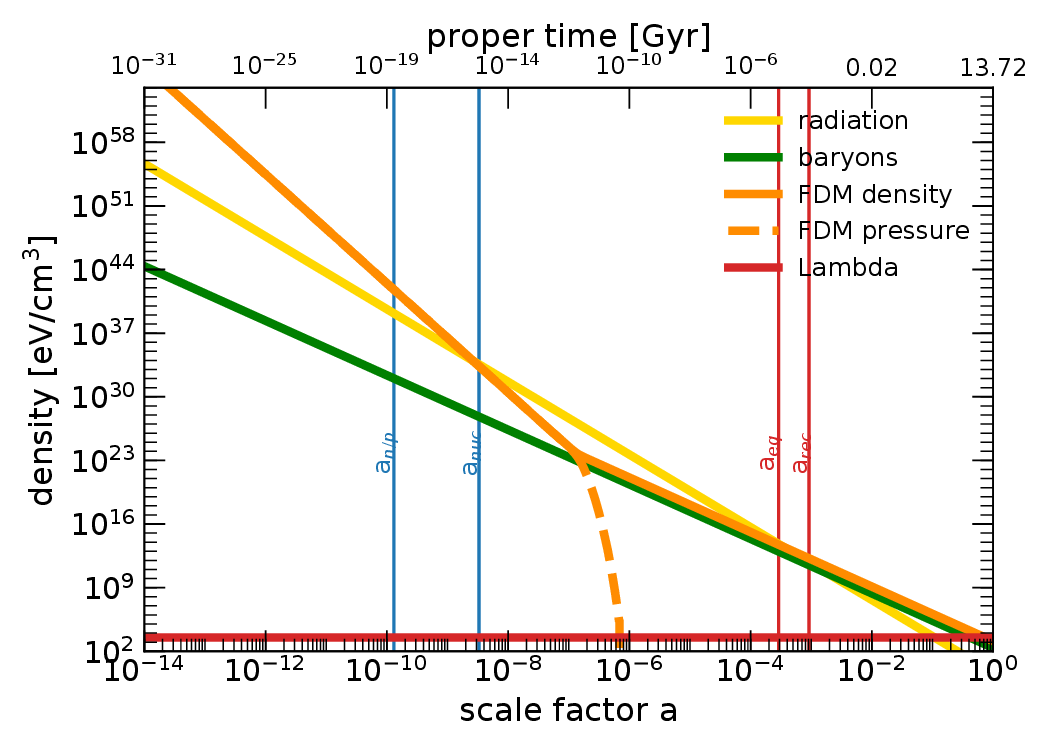}}	
	{\includegraphics[width=0.49\textwidth]{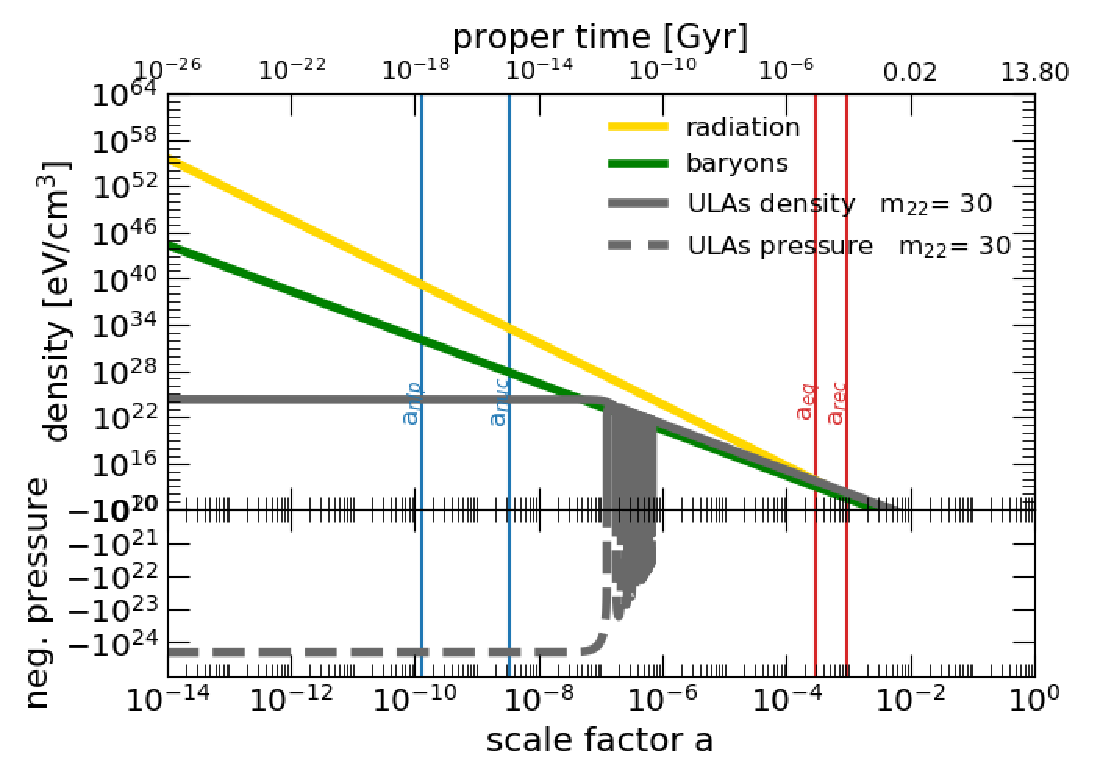}}		
	\caption[Evolution of energy densities for complex FDM and ULAs]
	{\textbf{Evolution of energy densities for complex FDM and real ULAs for the same mass $m_{22} = 30$}. 
	The left-hand panel shows the evolution of energy densities in the complex $\Lambda$FDM Universe: radiation (solid yellow), baryons (solid green) and FDM (solid orange). The pressure for FDM is indicated by the dashed orange line. The evolution of FDM starts in the stiff phase, with its density proportional to $a^{-6}$ in the slow-oscillation regime. Once it enters the fast-oscillation regime, the pressure shows a sharp drop and the FDM's \acidx{EOS} transforms to $\langle w\rangle \approx 0$, rendering it CDM-like. There is no radiation-like phase, different to SFDM in Fig.\ref{fig:SFDM-EOS-rho}.
	The right-hand panel displays the evolution of energy densities in the $\Lambda$-ULA Universe:  radiation (solid yellow), baryons (solid green) and ULAs (solid gray). The pressure for ULAs is indicated by the dashed gray line. The characteristic property of real ULAs is the negative pressure in the phase of slow oscillation of the \acidx{SF}. This results in a constant energy density, up to the time when the field oscillation dominates over the expansion rate $H$. In this phase, the \acidx{EOS} exhibits a very fast oscillation between the values $w=-1$ and $w=+1$ with an average value $w=0$, which makes ULAs behave as CDM. Once the oscillation frequency exceeds a certain threshold, we suppress the oscillations and use the averaged value of the \acidx{EOS} in the plot, instead. Also, in order to extend to the negative $y$-axis to display the ULAs on this log-log plot, we suppress a little stripe around the $y=0$-line; as a result $\Lambda$ is suppressed and not visible on the right-hand panel. 
	}
	\label{fig:FDM-UL-rho}
\end{figure*}

We start the discussion of complex FDM vs. ULAs, by comparing the evolution of the energy densities, which is shown in Fig.~\ref{fig:FDM-UL-rho}, where the left-hand panel displays the evolution of the densities for complex FDM (orange lines) and the right-hand panel those for ULAs (gray lines), both for the fiducial model parameter $m_{22}=30$. For complex FDM, the energy density evolves proportional to $a^{-6}$ in the stiff phase, which lasts until after $a_{nuc}$. Therefore, this model does not meet the cosmological constraints set by \hyperlinkcite{Li2014}{\PaperLi}. Without \acidx{SI}, there is no radiation-like phase, instead the stiff phase is immediately followed by the CDM-like phase of FDM, where pressure (dashed orange line) drops to low values. The energy density evolves in this phase proportional to $a^{-3}$, as expected. In the right-hand panel, we see a completely different evolution for ULAs that has been well known in the literature: the evolution starts in the CC-like phase with $w=-1$ as discussed in section \ref{sec:resultsSFDM}, i.e. starting out with a negative pressure, resulting in a constant energy density. At the end of the CC-like phase, the \acidx{SF} starts oscillating and the \acidx{EOS} performs oscillations between the values $w=-1$ and $w=+1$, resulting in an oscillation of the pressure (dashed gray line). The average value of $w=0$ causes the density to evolve proportional to $a^{-3}$, as expected. The oscillations are again suppressed in the plot. As explained in the last section, the marked difference in the background evolution of complex FDM ($w=1$) versus real ULAs ($w=-1$) in the early Universe has to do with the gain in kinetic energy for the former, due to the phase of the complex SF.
Furthermore, the transition to the pressure-less, CDM-like phase takes place within a single oscillation of the \acidx{SF}, which is very much faster compared to the drop of pressure for complex FDM. The milder drop in pressure, compared to ULAs, was seen already for SFDM in Fig. \ref{fig:SFDM-EOS-rho}. However, there is also a difference between FDM and SFDM, in that the pressure decreases more slowly for the latter. This is expected, since SI provides an additional source of pressure that is not available to FDM.

\begin{figure*} [!htbp]
	\centering
	\includegraphics[width=0.49\textwidth]{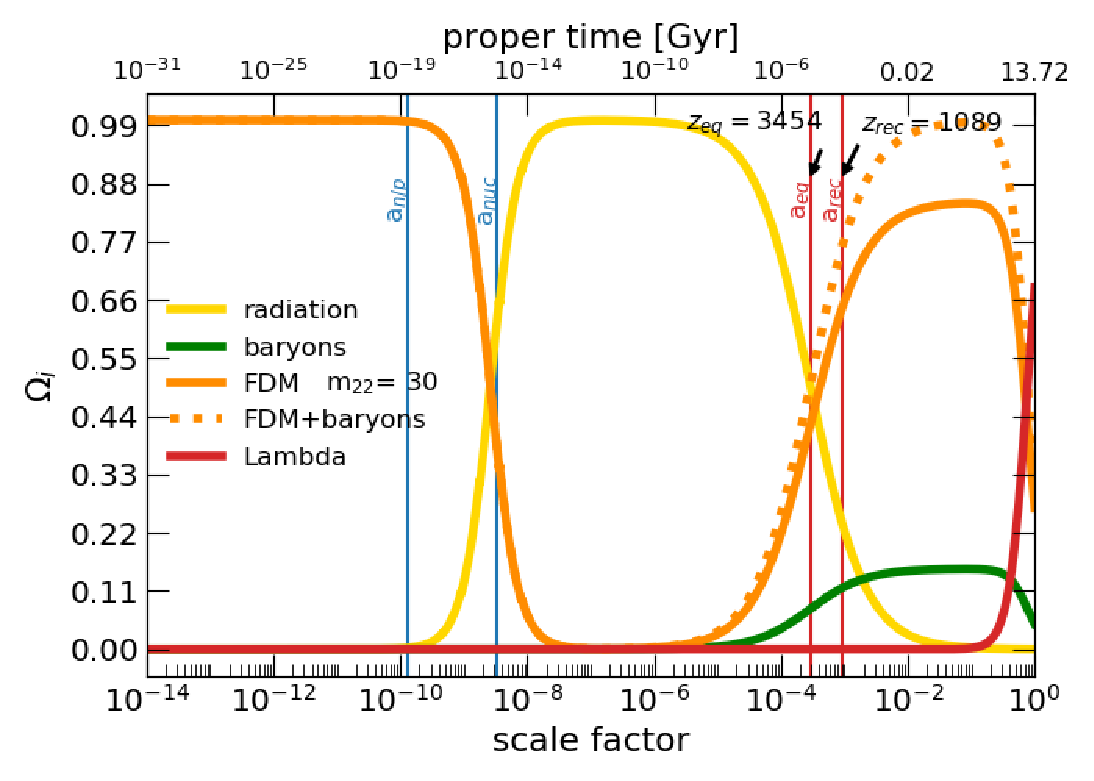}
	\includegraphics[width=0.49\textwidth]{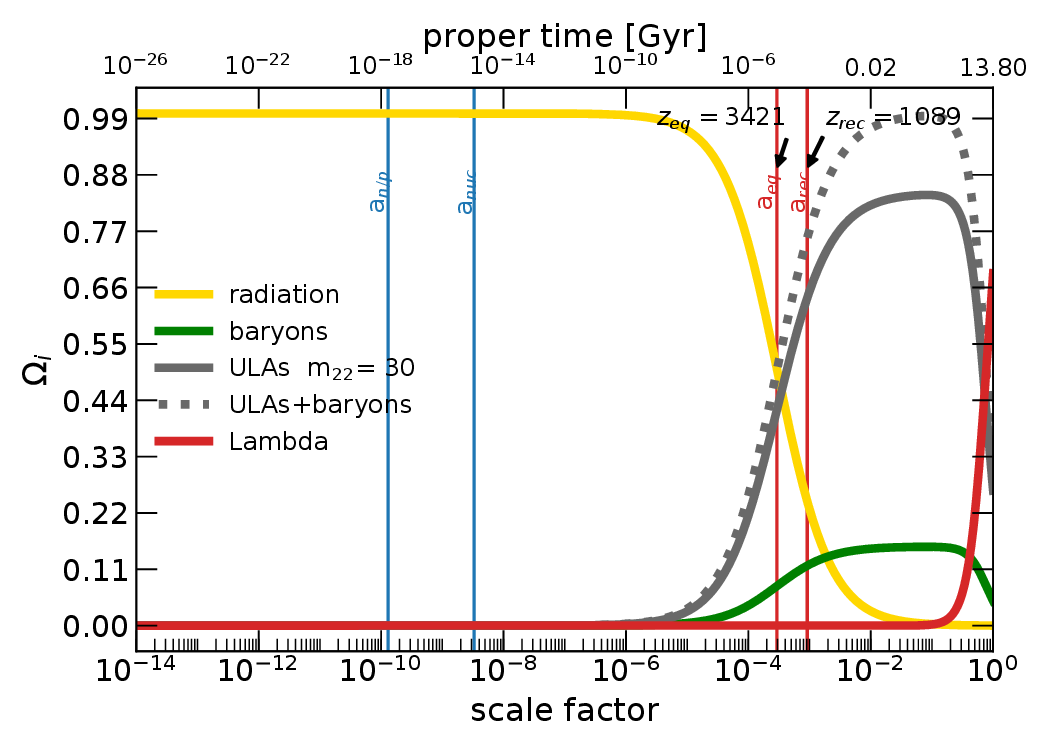}			
	\caption[Evolution of density parameters for complex FDM and ULAs]
	{\textbf{Evolution of density parameters for complex FDM and real ULAs for the same mass $m_{22} = 30$}. The solid lines display their evolution for radiation (yellow), baryons (green), complex FDM on the left-hand panel (orange)/ULAs on the right-hand panel (gray) and the cosmological constant $\Lambda$ (red). The dotted lines displays the evolution for the entire matter (ULAs/FDM + baryons). Vertical lines at benchmark epochs are the same as in Fig.\ref{fig:SFDM-EOS-rho}. The $\Lambda$-ULA model shows an evolution that is identical to that known from the \glo{LCDMmodel} (see rhp of Fig.\ref{fig:SFDM-CDM-Omega}), starting the evolution with a radiation-dominated epoch. In contrast, complex FDM dominates the early Universe. During this era, FDM evolves in the stiff phase, where its \acidx{EOS} equals unity. This is followed immediately by the CDM-like phase, as the absence of \acidx{SI} removed the plateau typical for the radiation-like phase. The times of $a_{eq}$ and $a_{rec}$ show no significant difference to the values known from $\Lambda$CDM, also the evolution of the density parameters in later epochs is the same.
	}
	\label{fig:FDM-UL-Omega}
\end{figure*}

\begin{figure} [!htbp]
	{\includegraphics[width=\PW\columnwidth]{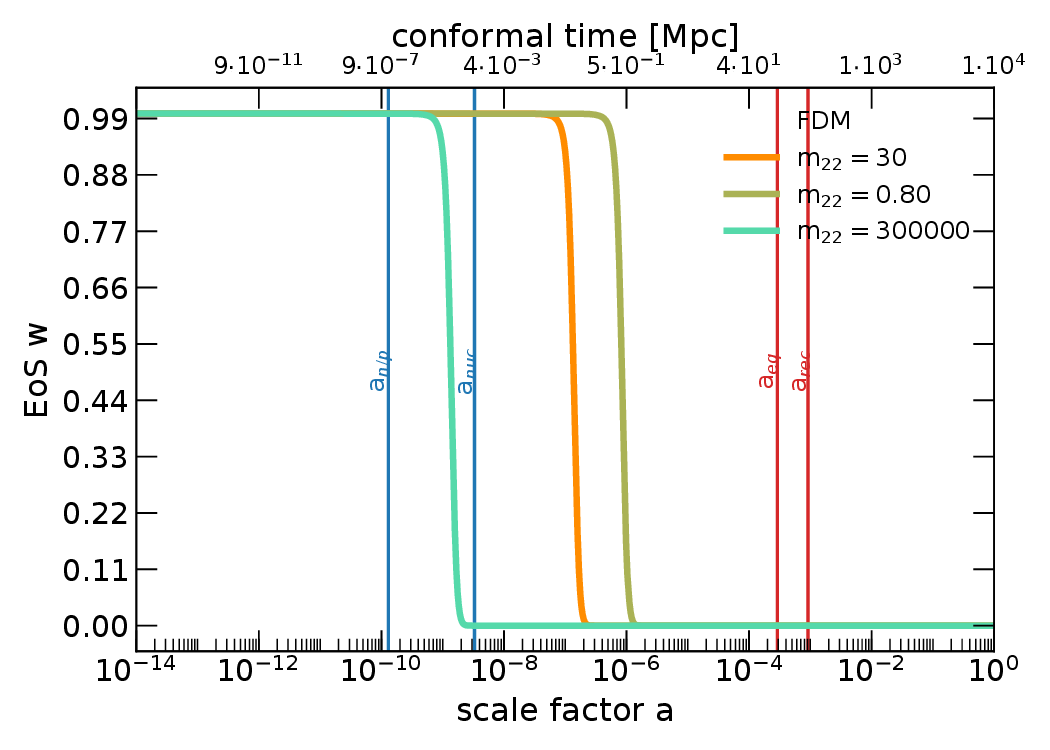}}		
	\caption[Equation of state of complex \acidx{FDM} models of different mass]
	{\textbf{EoS of complex FDM models of different mass}. The \acidx{EOS} parameter $w$ is shown for several complex FDM models: same mass than fiducial SFDM case of \hyperlinkcite{Li2014}{\PaperLi} ($m_{22} = 30$) (solid orange line); a model with smaller mass $m_{22} = 0.8$ (green line); and a high-mass case, $m_{22} = 3\cdot 10^{5}$ (turquoise line), that is compatible with cosmological constraints set by \hyperlinkcite{Li2014}{\PaperLi}. Since in FDM the oscillation frequency of the \acidx{SF} is only determined by the particle mass, the stiff phase ($w=1$) ends significantly later than in corresponding SFDM models, unless the FDM mass is chosen high. The absence of \acidx{SI} removes the plateau of the radiation-like phase ($\langle w \rangle =1/3$). Therefore, the CDM-like phase ($\langle w \rangle =0$) of FDM immediately follows its stiff phase.
	}
	\label{fig:FDM-EOS}
\end{figure}

We now compare the evolution of the density parameters for both models displayed in Fig.~\ref{fig:FDM-UL-Omega}, where the left-hand panel displays the evolution of complex $\Lambda$FDM and the right-hand panel that of $\Lambda$-ULA. Like for SFDM, FDM dominates in the early Universe, before radiation-domination. This is in contrast to ULAs which never dominate prior to radiation. Around $a_{nuc}$ radiation starts to dominate over FDM, given our choice of mass $m_{22}=30$. But since there is no SI, there is now no radiation-like phase, and hence no plateau. By contrast, the expansion history of $\Lambda$-ULA shows no significant differences to $\Lambda$CDM (see right-hand panel in Fig.\ref{fig:SFDM-CDM-Omega}). Now, the redshifts of matter-radiation equality are slightly shifted compared to CDM: $z_{eq}=3454$ for complex FDM, which is outside the 1$\sigma$ confidence interval of the reference value $z_{eq} = 3407 \pm 31$ (see section \ref{sec:SFDMbackground}), which is caused by the long stiff phase in this model (see Fig.~\ref{fig:FDM-UL-rho} \& \ref{fig:FDM-EOS} and explanations). On the other hand, $z_{eq}=3421$ for ULAs, which is within the 1$\sigma$ confidence interval.\par

Figure \ref{fig:FDM-EOS} displays the evolution of the \acidx{EOS} for complex FDM models with the same masses as the ULA reference models of \hyperlinkcite{Shapiro2021}{\PaperSDR} for comparison's sake. However, they do not meet the cosmological constraints set by \hyperlinkcite{Li2014}{\PaperLi}. Therefore, we added an additional model, with mass $m_{22} = 3\cdot 10^5$, which meets these criteria. Comparing the results to those shown in Fig.~\ref{fig:SFDM-multi-SFDM}, we can see that the transition from the stiff phase to the CDM-like phase (as there is no radiation-like phase for FDM) is steeper and occurs more rapidly than the EoS changes of SFDM.\par

\subsection{Evolution of Density Perturbations}\label{sec:resultsFDMC}

\begin{figure*} [!htbp]
	\includegraphics[width=0.49\textwidth]{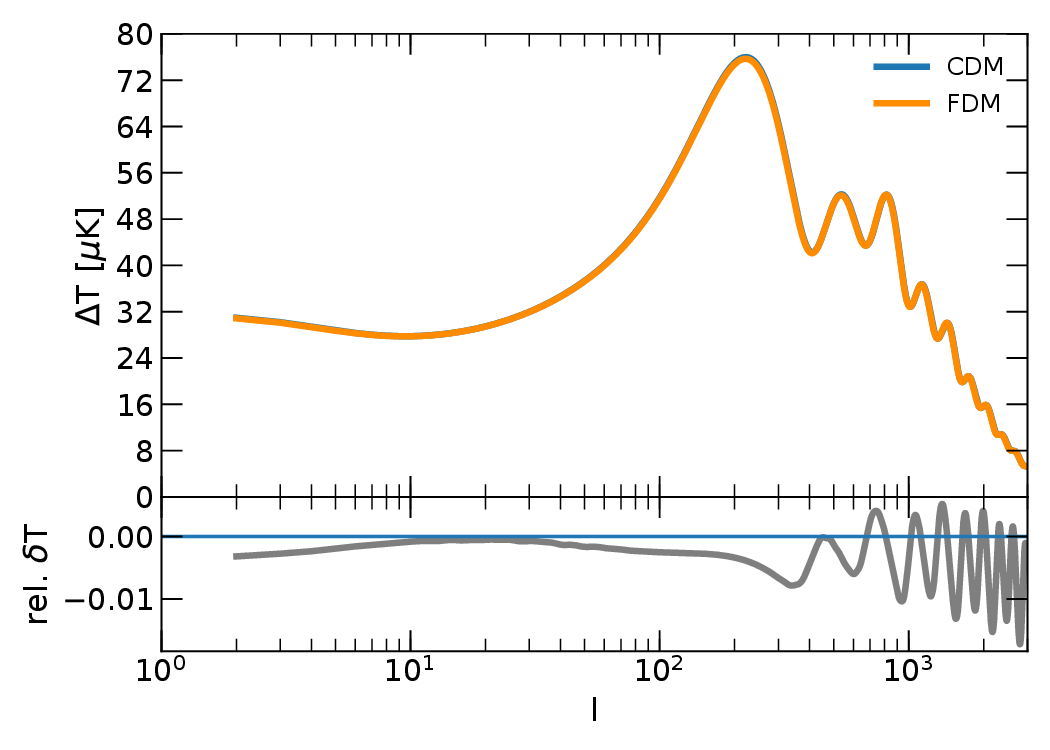}
	\includegraphics[width=0.49\textwidth]{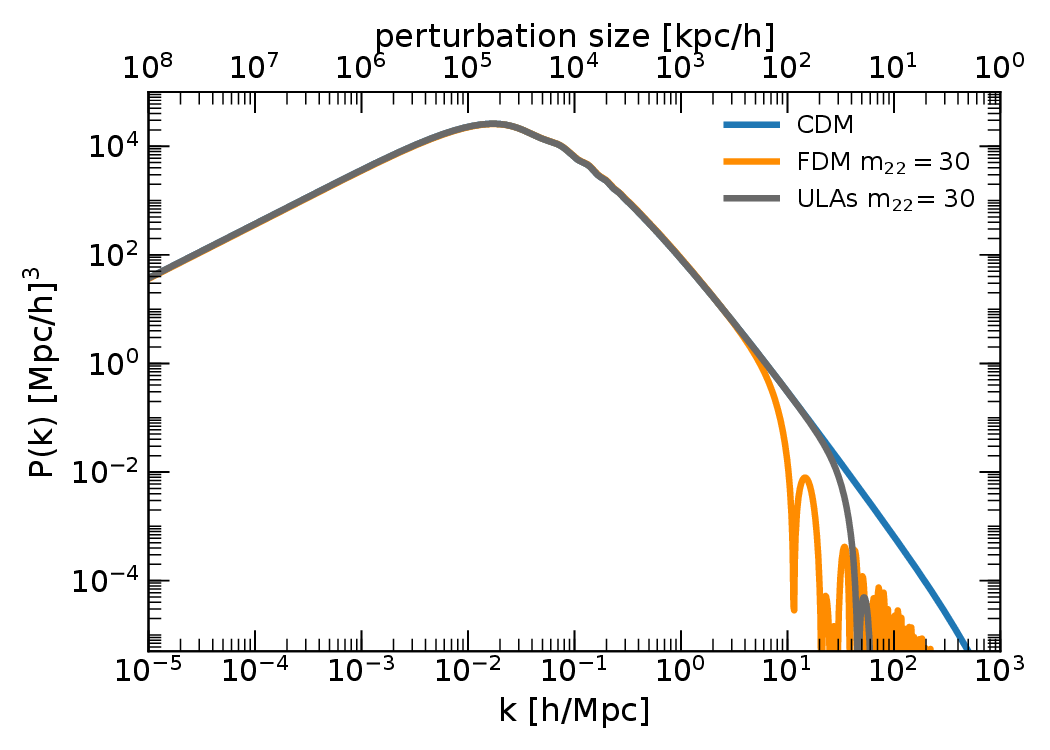}	
	\caption[Temperature anisotropies in the CMB and matter power spectra for complex FDM and CDM]
	{\textbf{Temperature anisotropies in the CMB and matter power spectra for complex $\Lambda$FDM and $\Lambda$CDM}. The left-hand panel displays the temperature anisotropies in the CMB for complex FDM (solid orange line) and CDM (solid blue line) vs. the size of the anisotropies. The particle mass for FDM is $m_{22}=30$, identical to that of the fiducial SFDM model shown in Fig.\ref{fig:SFDM-Temp}. No significant differences to CDM occur. The right-hand panel displays the matter power spectra for complex FDM (solid orange line), ULAs (solid gray line) and CDM (solid blue line) vs. wave number. We pick the same particle mass for FDM and ULAs, i.e. $m_{22}=30$. Both complex FDM and ULAs show a suppression of structure compared to CDM, but with different characteristics. In contrast to ULAs, which show a sharp cutoff in the power spectrum, the respective cutoff for complex FDM occurs at smaller $k$, along  with a subsequent shallower falloff toward high $k$. This behavior is similar to SFDM, see Fig.~\ref{fig:SFDM-mtk-mpk}, but note the cutoff at lower $k$ there due to SI ($R_{TF}=1.1$ kpc), while SI is absent here.
	}
	\label{fig:UL-Temp-mPKs}
\end{figure*}

Again, we start with a discussion of the CMB and matter power spectra, before we turn to the details of the perturbation evolution in each model.

The left-hand panel of Fig.~\ref{fig:UL-Temp-mPKs} displays the spherical temperature anisotropies of complex FDM with $m_{22} = 30$ (solid orange line) and CDM (solid blue line). There are no significant differences between both models as indicated by the relative differences shown in the bottom part of the panel. The right-hand panel of Fig.~\ref{fig:UL-Temp-mPKs} shows the matter power spectra of complex FDM (solid orange line) and ULAs (solid gray line) for the same mass, as well as CDM (solid blue line). For complex FDM, the cutoff in the power spectrum occurs at smaller wave number $k$ than for ULAs, although both models have the same particle mass. Similar to the results shown in Fig.~\ref{fig:SFDM-mtk-mpk}, ULAs display the very sharp cutoff in the power spectrum, whereas the matter power spectrum of complex FDM shows a shallower falloff toward high $k$, quite similar to the behavior seen in the previous section for SFDM, and reported first in \hyperlinkcite{Shapiro2021}{\PaperSDR}. 
We might explain this similarity as follows. In SFDM, we have two sources of pressure, the one resulting from the kinetic energy of the complex SF, and the one due to SI. In FDM, only the first source is available. Yet, that source produces a similar result for the overall appearance of the power spectrum.  
However, SI plays an important role in terms of where the cutoff happens. As can be seen from Fig.~\ref{fig:SFDM-mtk-mpk}, the cutoff for SFDM with $R_{TF}=1.1$ kpc happens at significantly smaller $k$ than for the FDM model here, despite the fact that both models have a complex SF with the same mass, i.e. the additional pressure of SFDM due to SI makes the cutoff shift to smaller $k$.

\begin{figure*} [!htbp]
	{\includegraphics[width=0.49\textwidth]{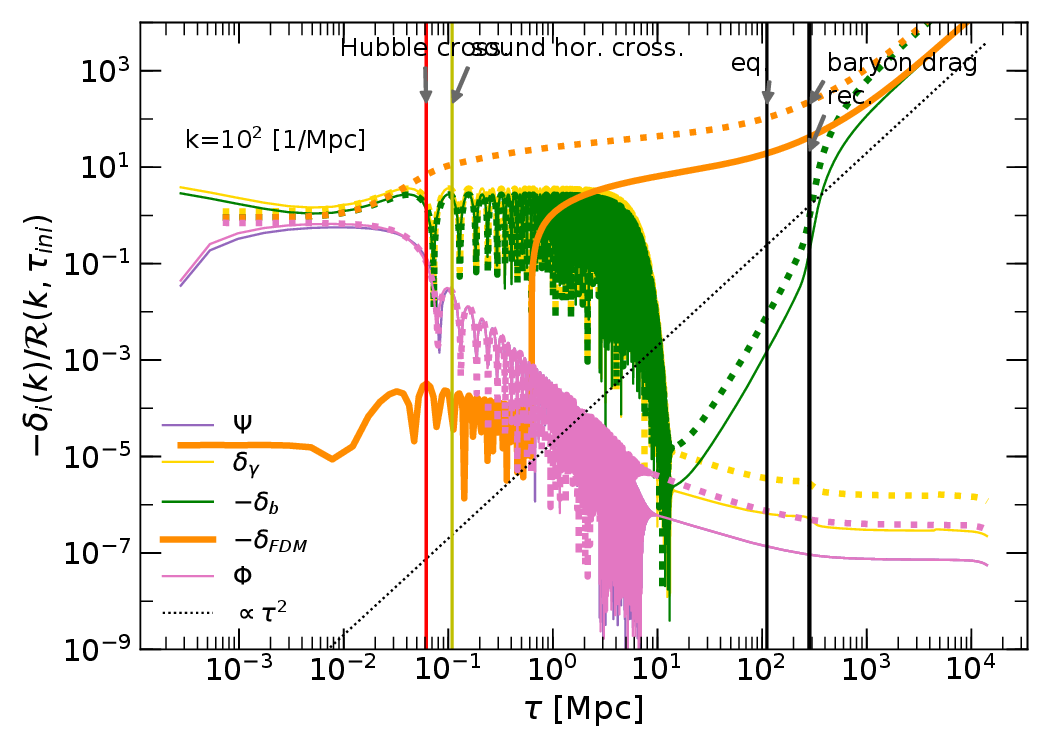}}
	{\includegraphics[width=0.49\textwidth]{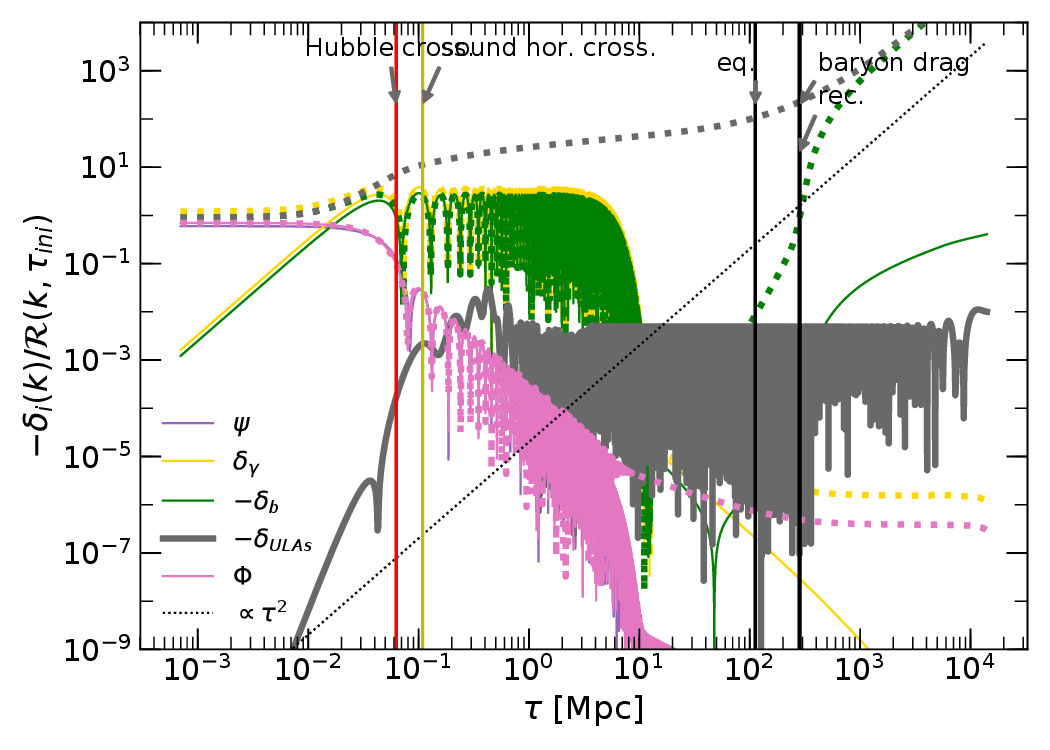}}
	{\includegraphics[width=0.49\textwidth]{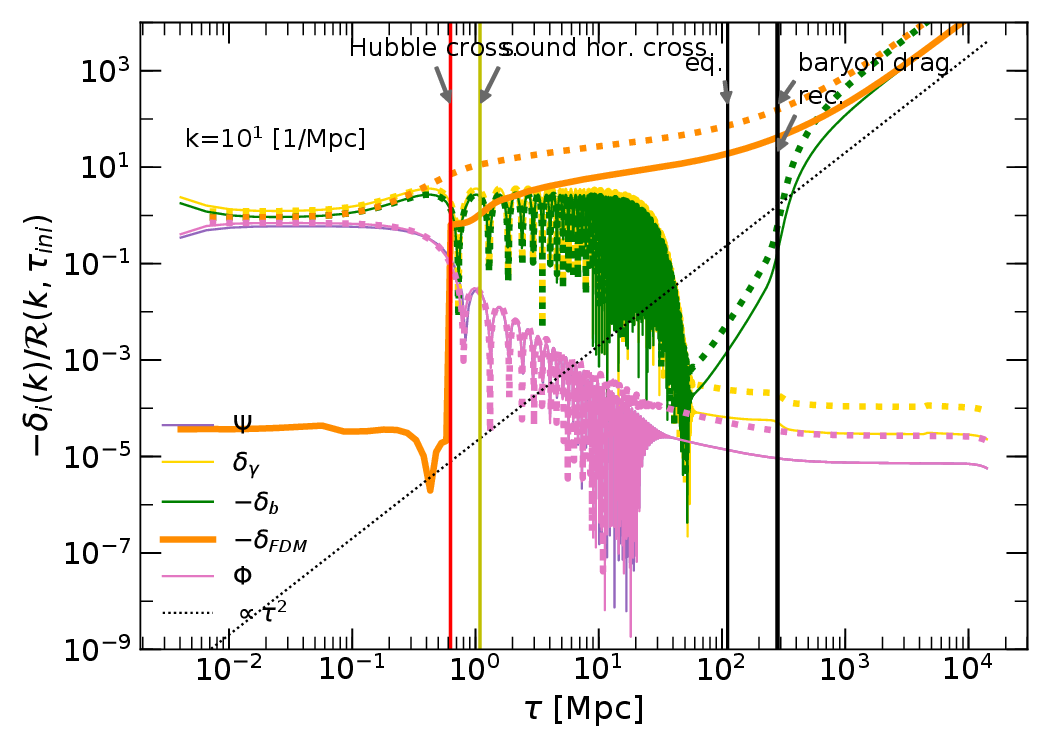}}
	{\includegraphics[width=0.49\textwidth]{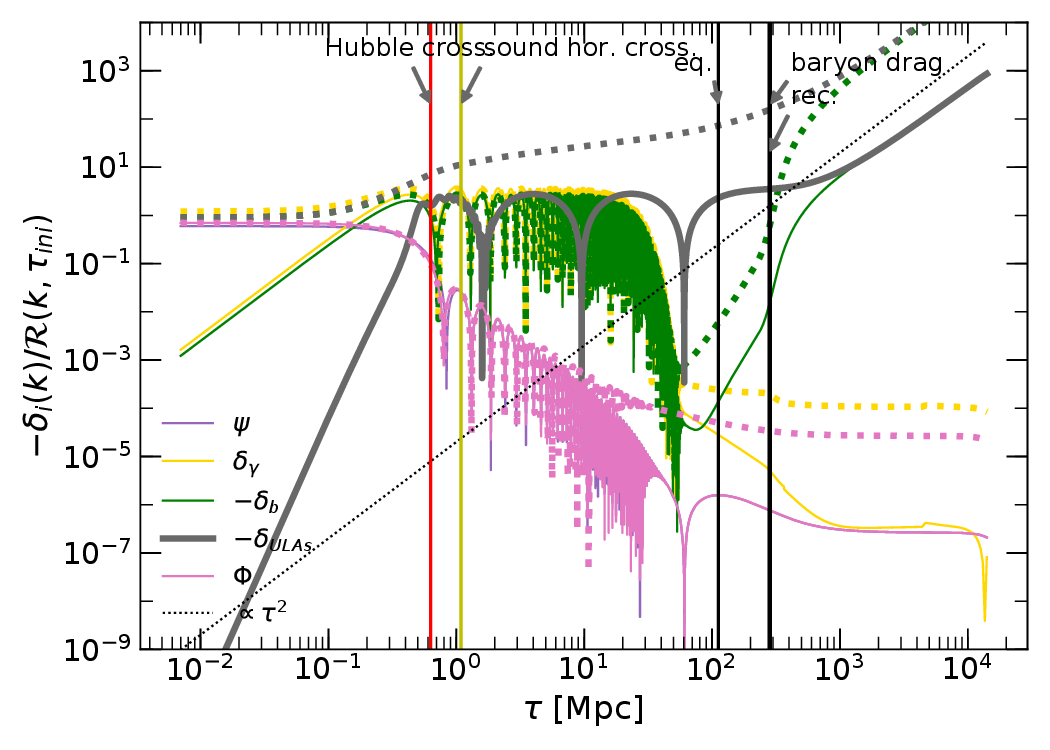}}
	{\includegraphics[width=0.49\textwidth]{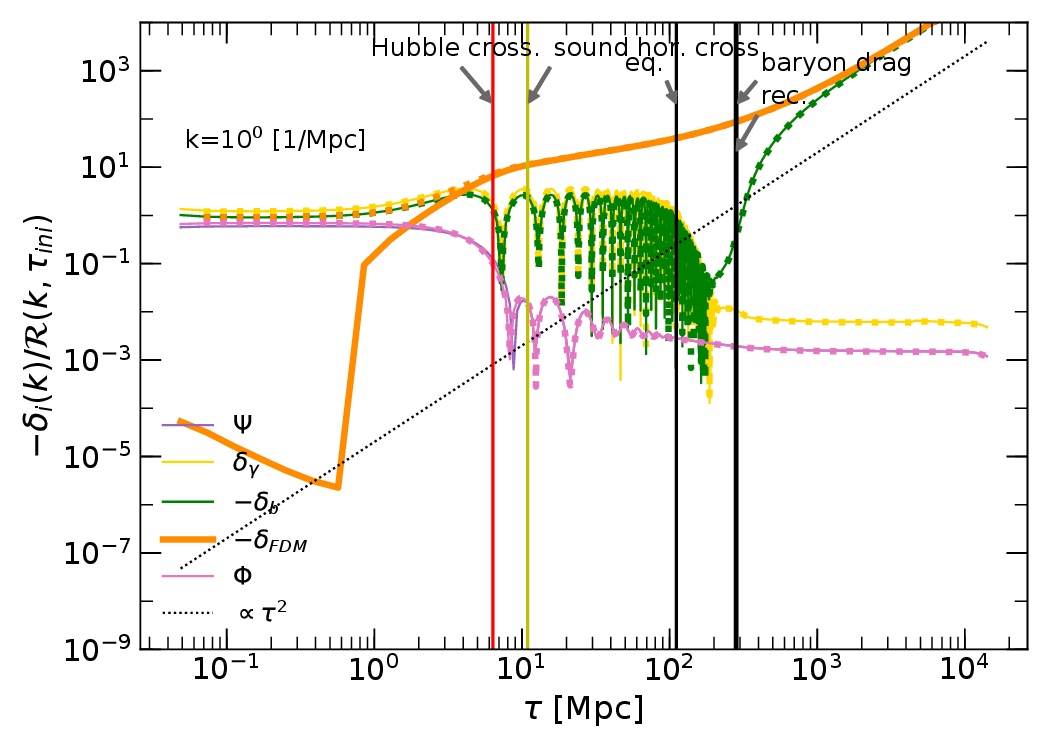}}
	{\includegraphics[width=0.49\textwidth]{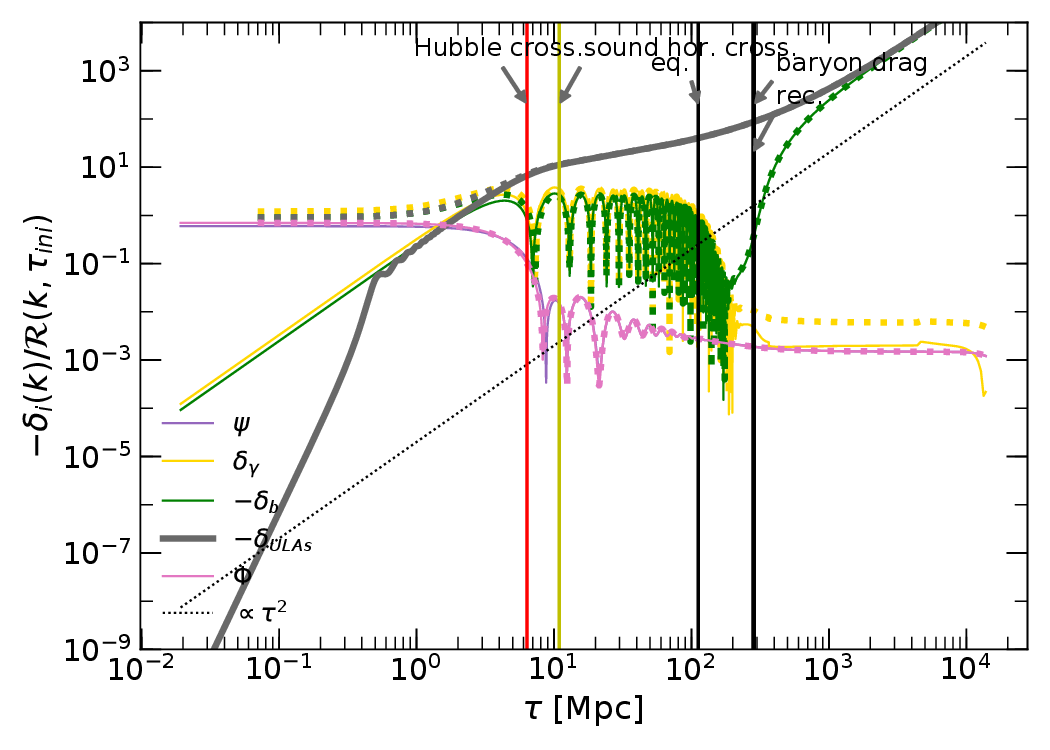}}			
	\caption[Evolution of complex FDM compared to ULAs]
	{\textbf{Evolution of complex FDM compared to real ULAs.} The evolution as function of $\tau$ of three characteristic DM perturbation modes, $k=10^2;10^1;10^0$ (from top to bottom) is shown for each, complex FDM (m$_{22}=30$) in the left-hand panels (solid orange) and ULAs (m$_{22}=0.8$) in the right-hand panels (solid gray). These two models have comparable transfer functions, see Fig.~\ref{fig:FDM-mtk}. The time of horizon crossing (red vertical lines) increases from top to bottom. The legends indicate further perturbation quantities (baryons in green, photons in yellow, and the metric potentials in light and dark pink). In addition, the corresponding perturbation quantities of all components of the CDM model are shown in dotted lines in all panels. While CDM modes never oscillate -- orange (left) and gray (right) dotted lines --, oscillations can be clearly seen in the top and middle panels for FDM and ULA modes (orange and gray solid lines, respectively). 
	}
	\label{fig:FDM-Tk-tau}
\end{figure*}

\begin{figure*} [!htbp]		
	{\includegraphics[width=0.49\textwidth]{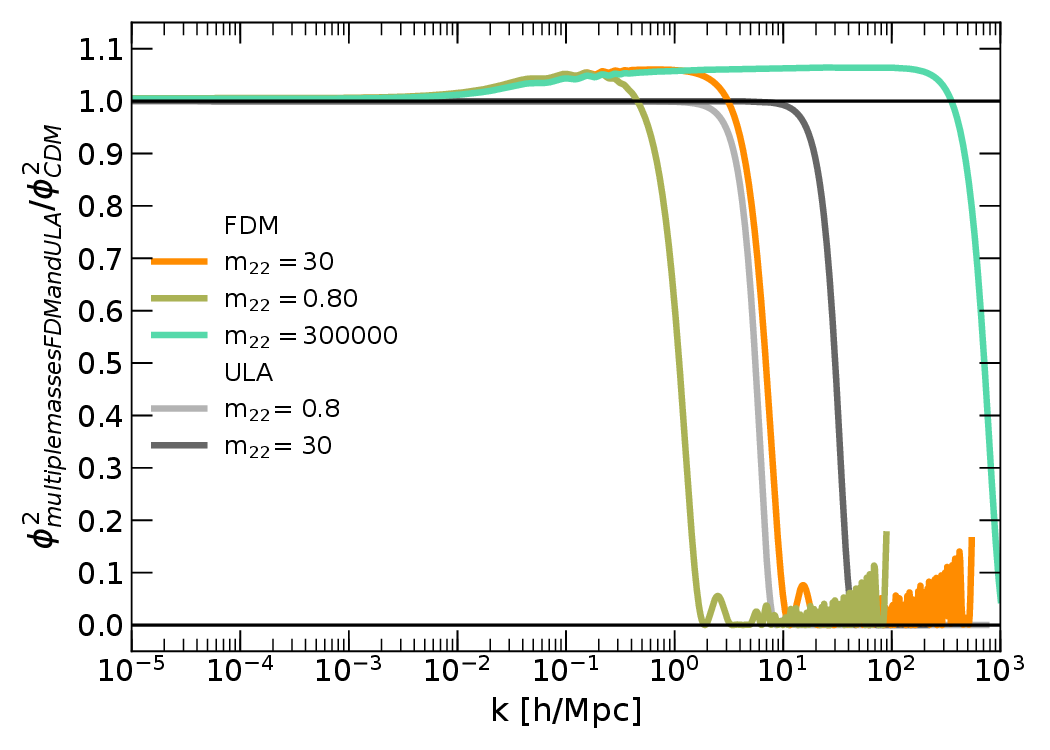}}
	\includegraphics[width=0.49\textwidth]{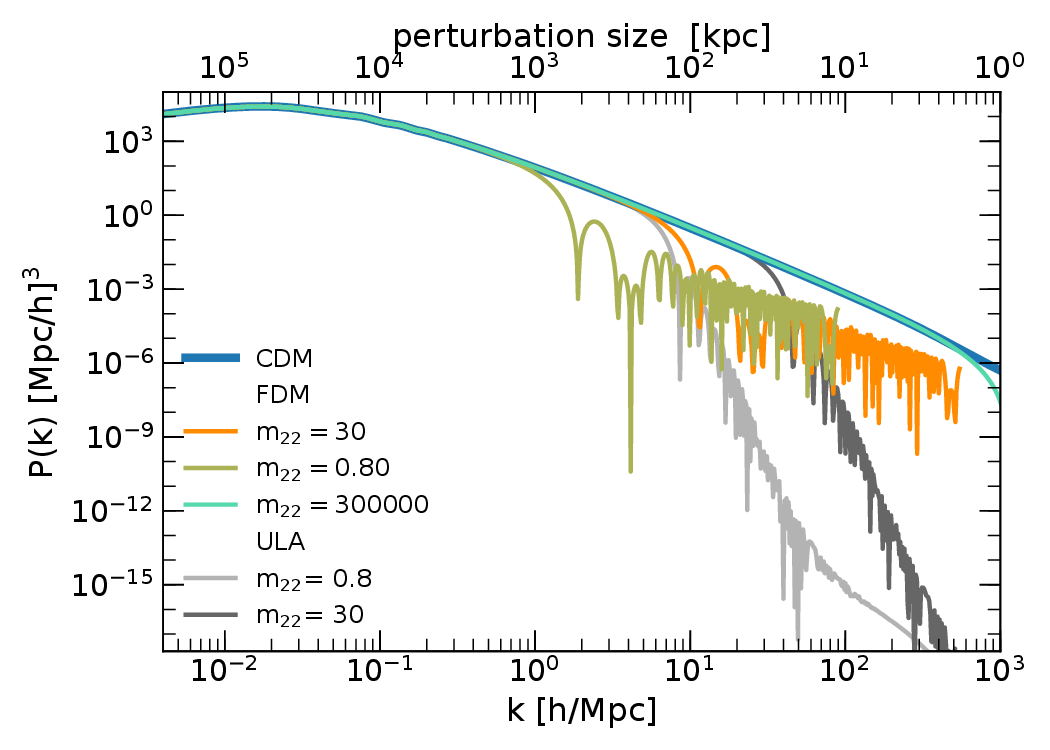}	
	\caption[Transfer functions and matter power spectra of complex FDM and ULA models with different particle mass]
	{\textbf{Transfer functions and matter power spectra of complex FDM and real ULA models with different particle mass}. In the left-hand panel, the colored solid lines display the transfer functions of complex FDM models with masses, $m_{22}=30; 0.8; 3\cdot 10^5$, each relative to CDM. The stiff phase of the complex FDM models is suppressed, i.e. cut off at high $k$. The gray lines display two ULA models with $m_{22}=0.8; 30$.
	In the right-hand panel, the colored solid lines display the corresponding matter power spectra of the same complex FDM models as in the left-hand panel; again cut off at high $k$. The gray lines display the corresponding power spectra of the ULA models of the left-hand panel. The CDM power spectrum is shown as the thick blue line (which is overlapped to a great extent by the high-mass FDM model with $m_{22}=3\cdot 10^5$). Note again the expanded $y$-axis range down to very small power.
	}
	\label{fig:FDM-mtk}
\end{figure*}

 Now, Fig.~\ref{fig:FDM-Tk-tau} shows the evolution, as a function of conformal time $\tau$, of three characteristic perturbation modes for complex FDM (m$_{22}=30$) and real ULAs (m$_{22}=0.8$). Although these models have a different mass, they have a comparable evolution of the transfer function (also seen in the left-hand panel of Fig.~\ref{fig:FDM-mtk}). The top left-hand panel displays the evolution of $k=10^2$~1/Mpc perturbations, which cross the horizon at $\tau \sim 10^{-1}$. The evolution of the FDM perturbation (orange solid line) shows acoustic oscillations, as the mode is sub-Jeans until $\tau \sim 1$, after which it becomes super-Jeans and the evolution approaches that of CDM. The ULA model (solid gray line in the top right panel) enters its fast oscillation regime at about the time of horizon crossing; the mode is also sub-Jeans. As the oscillation frequency increases very rapidly, the oscillation period gets much smaller than the Hubble time. In contrast to the evolution of complex FDM, the amplitude of the acoustic oscillation of the ULA model remains constant for a very long time, as it remains in the Jeans-filtering regime until $\tau \simeq 10^2$, when the mode becomes super-Jeans, as the Jeans mass shrinks less rapidly than that of the corresponding complex FDM model. This suppression of very small structure has also implications for the matter power spectrum of this model, displayed by the shallow slope of the matter power spectrum at high $k$, shown in the right-hand panel of Fig.~\ref{fig:FDM-mtk} and already discussed above.

 Furthermore, we see again the effect of the stiff phase onto the evolution of perturbations outside the horizon that we already pointed out in the previous section, namely that the initial amplitudes are not equal between (S)FDM and the other components, in contrast to CDM. In this model here, the stiff phase ends at $\tau \sim 10^{-2}$, and there is no radiation-domination prior to this time, a usual assumption in relativistic perturbation theory. Nevertheless, the FDM modes converge to those of CDM upon horizon entry. This deviance to CDM with respect to the super-horizon behavior could be a source of problems for models with even smaller particle masses, but these models are ruled out anyway by the constraints set by \hyperlinkcite{Li2014}{\PaperLi}. 
 
 In addition, we see a similar deviance for ULAs, for which our above explanation would not apply, since the Universe is radiation-dominated then. Although the initial $(w=-1)$-phase might play a role, we see no way to clarify this question here, also because we do not use our own modification of CLASS for the ULA runs, but rather the version provided by \citet{UrenaLopez2016}.
 
 Now, let us look at the middle panels, which display the evolution of a $k=10^1$~1/Mpc perturbation. For complex FDM, the transition to super-Jeans and horizon crossing coincide, hence the evolution of complex FDM is CDM-like. The same mode in the ULA model enters the horizon approximately at the transition to the CDM-like phase of the \acidx{EOS}, and it is still sub-Jeans. Again, the acoustic oscillations show a constant amplitude, but the evolution gets CDM-like as soon as the mode becomes super-Jeans. Similar results hold true for the evolution of the $k=10^0$~1/Mpc perturbations, shown in the bottom panels. Both modes enter the horizon in the CDM-like phase of the respective \acidx{EOS} for FDM and ULAs, hence they display no oscillations (except for tiny wiggles in the ULA mode, bottom right). Both modes exhibt a CDM-like evolution in their growth, as they are both already super-Jeans at the time of horizon entry.\par

\begin{figure} [!htbp]		
	{\includegraphics[width=0.49\textwidth]{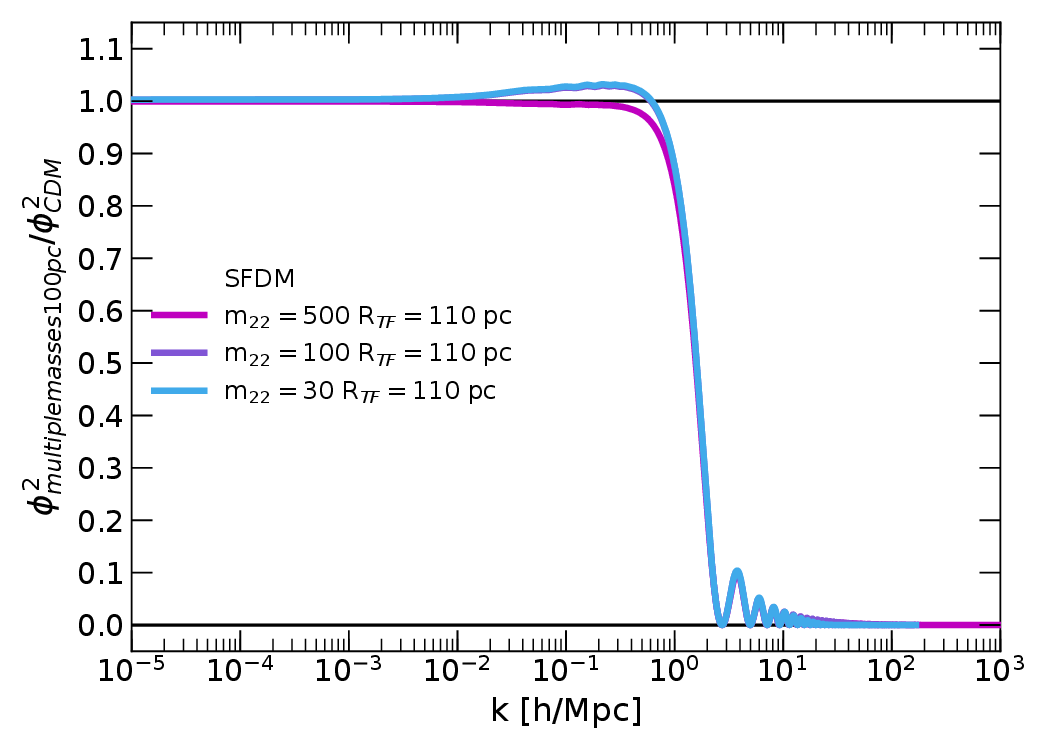}}
	{\includegraphics[width=0.49\textwidth]{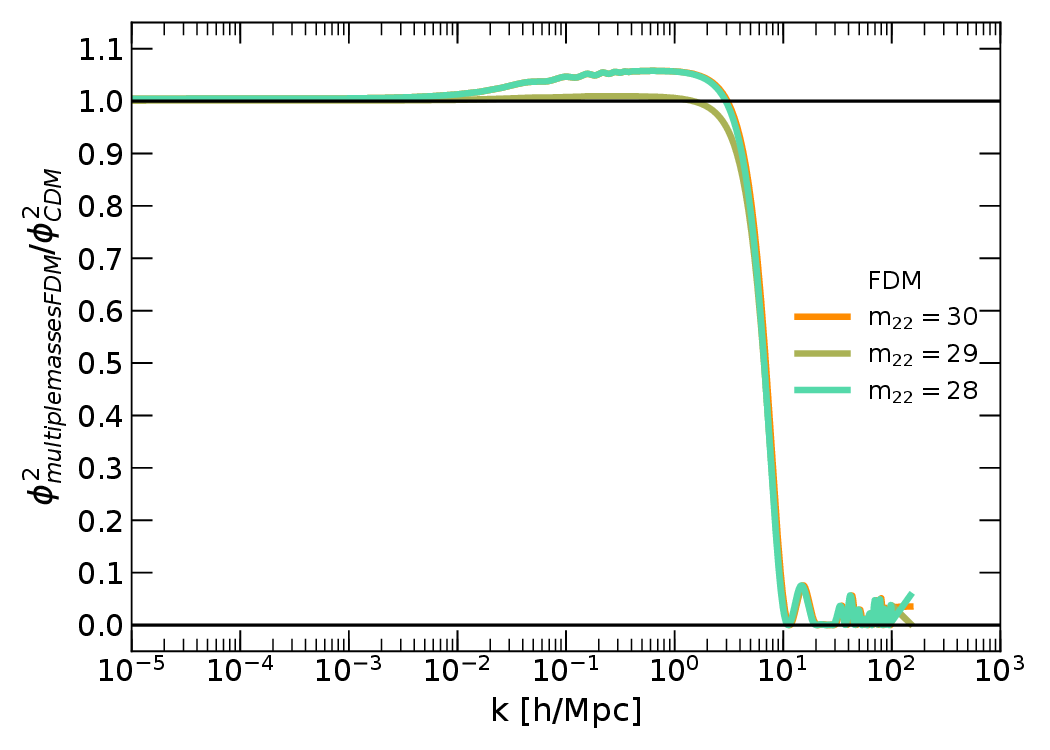}}	
	\caption[Transfer functions of the \boldmath{$R_{TF}=110$}~pc SFDM models and complex FDM models.]
	{\textbf{Transfer functions of the $R_{TF}=110$~pc SFDM models and FDM models.} The top panel displays SFDM models with particle masses $m_{22}=30;\:100;\:500$, but same $R_{TF}=110$~pc. The bottom panel displays FDM models with particle masses $m_{22}=30;\:29;\:28$. All plots are at $z=0$. The top panel may suggest, that there is a trend from high $m$ to low $m$, that suppresses the ``overshooting'' of the transfer functions. But using a finer sampling of masses in the bottom panel, we can see that there is a ``periodic'' behavior in the presence of the overshooting: particle masses $m_{22}= 30$ and $m_{22}=28$ display overshooting, whereas $m_{22}= 29$ does not.
	}
	\label{fig:mTk-multi-masses}
\end{figure}

The left-hand panel of Fig.~\ref{fig:FDM-mtk} compares our results to the ULA reference models of \hyperlinkcite{Shapiro2021}{\PaperSDR} (note that in their paper, they are labeled as ``FDM''), by computing the transfer functions and normalizing them to CDM (see also our earlier Fig.~\ref{fig:mTk-multi-SFDM}). Again, some complex FDM models show an ``overshooting'' in relation to the CDM transfer function. We analyzed the problem and found that it depends upon the particle mass and is somewhat ``periodic".
Figure~\ref{fig:mTk-multi-masses} demonstrates the ``periodic nature'' of the overshooting of the transfer functions for some SFDM and FDM models, since we found no trend from small to large masses or vice versa of this feature.
It can be seen that for perturbations with $k \lesssim 10^{-2}$ there is no overshooting.  Figure~\ref{fig:mTk-multi-SFDM} displayed the normalized transfer functions for SFDM, showing a fast decline from $1$ to $0$, followed by an oscillation, whose amplitude converges to zero. In contrast to that, for complex FDM in Fig.~\ref{fig:FDM-mtk}, we see increasing amplitudes of the oscillations following the decline. This is due to the fact, that the reference models (apart from the one with $m_{22}=3\cdot 10^5$), do not meet the cosmological constraints set by \hyperlinkcite{Li2014}{\PaperLi} and the stiff phase ends much too late, where the validity of our numerical implementation is not any longer guaranteed\footnote{In order to simulate models with such late-ending stiff phase up to high $k$ would require to modify critical core parts of CLASS, but since such models are excluded by BBN constraints, we do not pursue such a modification. Our main point here is to show the dependence on model parameters of the overall spectra.}. So, we have to truncate the transfer function at the point where the stiff phase (``slow oscillation'') goes over to the ``fast oscillation regime''. The right-hand panel of Fig.~\ref{fig:FDM-mtk} displays the corresponding matter power spectra, which are also truncated at this point. Similar to the results for SFDM, we see an early cutoff in the power spectra followed by a shallower slope for complex FDM, compared to that of the ULAs. The complex FDM model with particle mass $m_{22}=3\cdot 10^5$ meets the cosmological constraints (stiff phase ends before $a_{nuc}$, Fig.~\ref{fig:FDM-EOS}), thereby displaying a matter power spectrum that shows no significant differences compared to CDM up to $k \lesssim 10^3$ h/Mpc.\par
\begin{figure*} [!htbp]	
	{\includegraphics[width=0.32\textwidth]{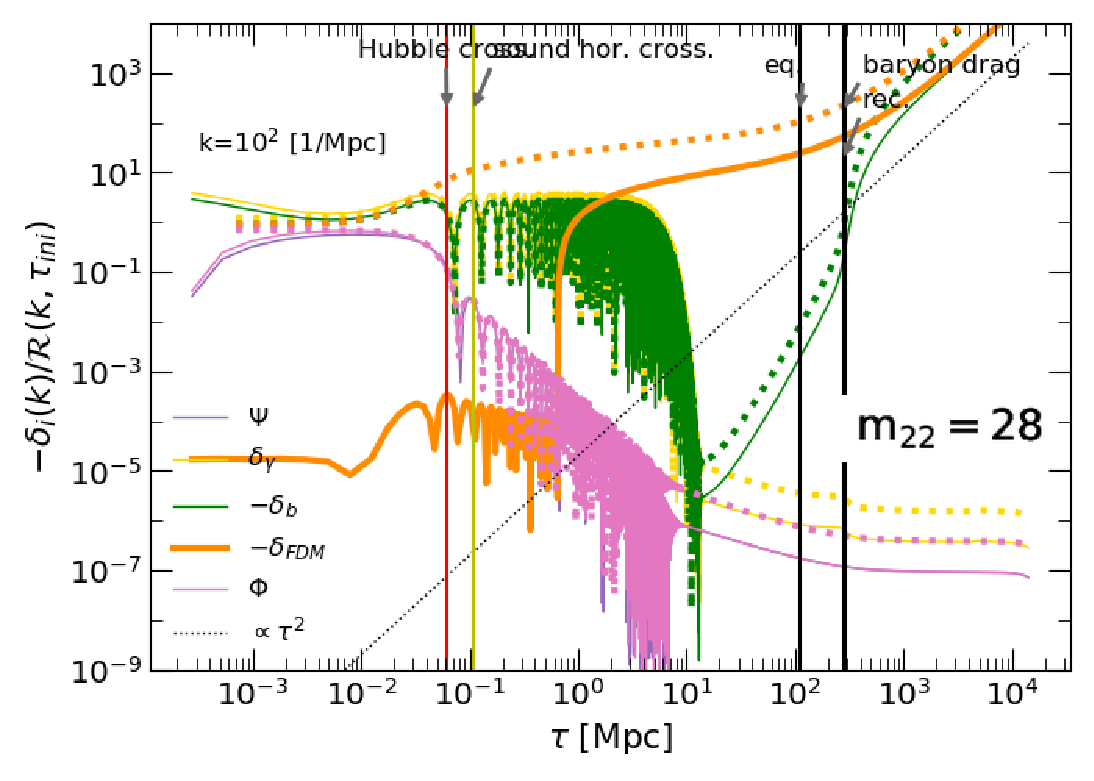}}
	{\includegraphics[width=0.32\textwidth]{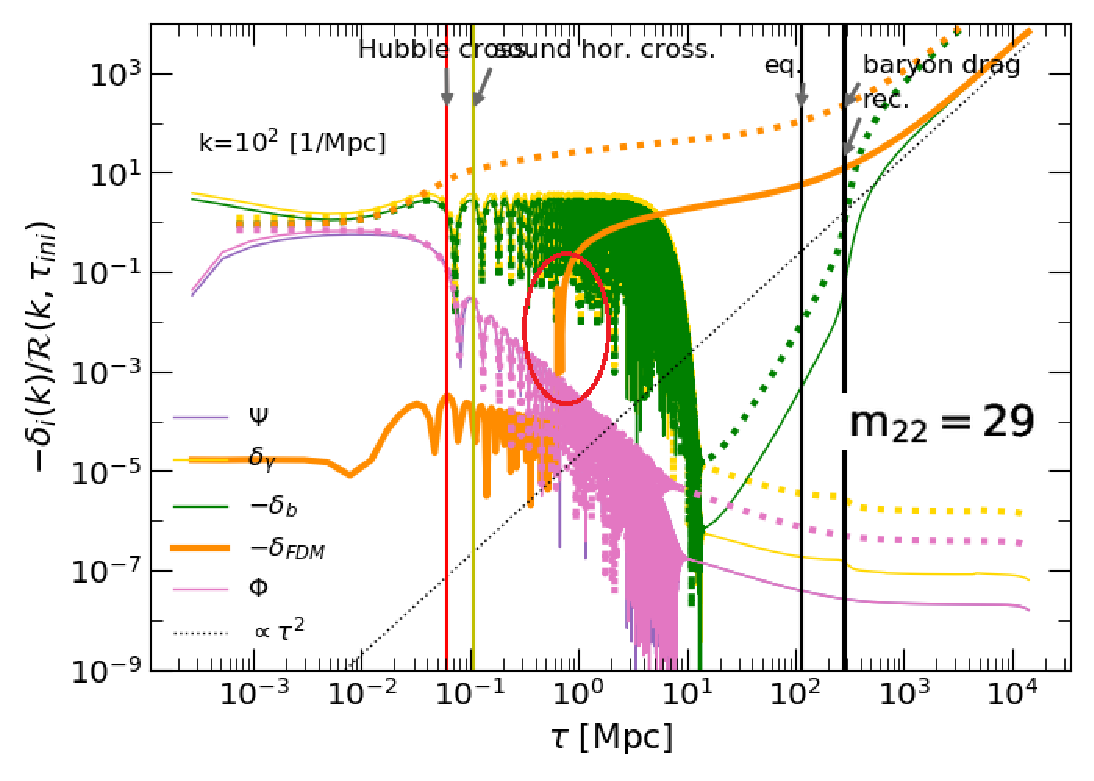}}
	{\includegraphics[width=0.32\textwidth]{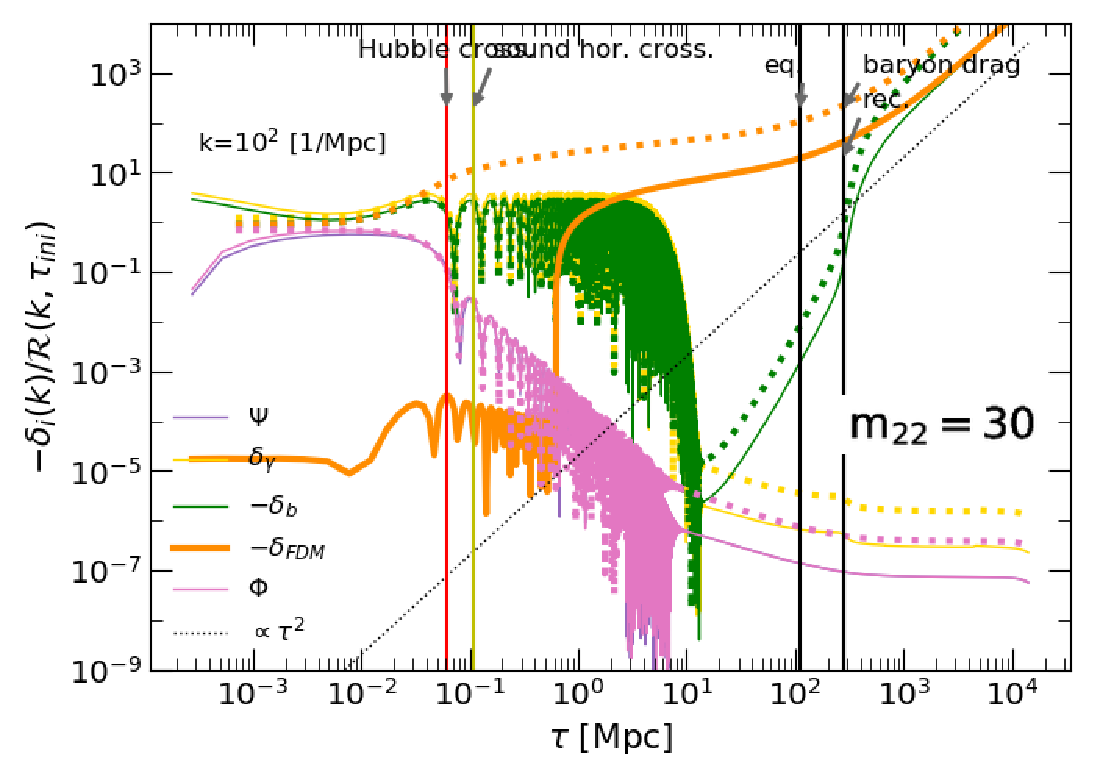}}		
	\caption[Overshooting from superpositioning of accoustic oscillations]
	{\textbf{Overshooting from superpositioning of accoustic oscillations}. The three panels show the temporal evolution of the same complex FDM perturbation mode (thick orange line), but for three different particle masses, $m_{22}=28;29;30$, from left to right. The first and the last panel exhibit cases where overshooting in the transfer function happens; the middle panel is a case without overshooting: the red circle marks the relevant event, when the transition to the super-Jeans regime occurs during a downward oscillation of the scalar field; further explanations are found in the main text. Other perturbation quantities are labeled in the legends, see also caption to Fig.\ref{fig:FDM-Tk-tau}. Perturbation quantities in the CDM model are all displayed with dotted lines (CDM in orange).
	}
	\label{fig:FDM-overshooting-tau}
\end{figure*}

As mentioned before, we see a strong dependence of the ``overshooting'' of the transfer functions of complex FDM on particle mass, see Fig.~\ref{fig:mTk-multi-masses}. Given that the feature is somewhat ``periodic'' and is not due to a trend from small to high mass, we think it might stem from some ``superpositioning'' of oscillations, which is illustrated in Fig.~\ref{fig:FDM-overshooting-tau}. The orange solid line indicates the evolution of complex FDM modes. The initial oscillation following horizon crossing (the red vertical line) of the sub-Jeans mode stops, once the mode transitions to super-Jeans. The point in time, when this happens depends upon the particle mass, because it determines the oscillation frequency of the SF. Depending on whether this transition takes place during an ascending edge of the oscillation ($m_{22}=28;30$ in the left-hand and right-hand panels), or during a descending edge (middle panel) of the oscillation will determine if overshooting occurs or not. Overshooting never happens for modes $k \lesssim 10^{-2}$, because they enter the horizon in the CDM-like phase, where no oscillations occur.\par

\section{Implications for Small-Scale Structure in \boldmath$\Lambda$SFDM: discussion}\label{sec:implications}

In \hyperlinkcite{Li2014}{\PaperLi}, the constraints for the SFDM model with repulsive \acidx{SI} were determined by requiring that the evolution of the background Universe meets cosmological constraints, notably that the characteristic stiff phase of complex SFDM ends before BBN is over, as well as the requirement for the \acidx{EOS} of SFDM to become CDM-like before $a_{eq}$, calibrated to the observational value obeyed by the best-fit $\Lambda$CDM model. The fiducial SFDM model, focused on in \hyperlinkcite{Li2014}{\PaperLi}, with parameters $m_{22}=30$ and $R_{TF}=1.1$~kpc fulfilled these constraints. Also, we checked this fiducial model in our analysis here, and we could see that, different from CDM, it exemplifies a cutoff in the matter power spectrum. On the other hand, the CMB temperature anisotropy spectrum up to high mode numbers $l$ is almost indistinguishable from CDM. In the meantime, \citet{Dawoodbhoy2021} analyzed the formation and inner structure of halos formed in 1D spherical collapse simulations in the \acidx{SFDM} model with strongly repulsive \acidx{SI} (SFDM-TF). Upon comparison with galaxy data, they found that models with $R_{TF} \gtrsim 1$~kpc are required in order to pass the cusp-core and too-big-to-fail tests.
This finding for the cusp-core problem was confirmed recently by \citet{Pils2022}, who studied the dynamical impact of baryons in halos by calculating adiabatic contraction. Again, kpc-size core radii were able to reproduce galaxy data well.
So, from this perspective, SFDM with strongly repulsive SI is a viable alternative to standard \acidx{CDM} by resolving its small-scale problems. However, an analysis of the linear regime of structure formation of SFDM-TF was subsequently published by \hyperlinkcite{Shapiro2021}{\PaperSDR}, who found that halo formation in models with high SI (equivalent to large $R_{TF}$) is actually much stronger suppressed than previously realized, questioning the ability of SFDM-TF to resolve the small-scale problems. Therefore, we complemented that work by running CLASS simulations here, not only to solve now the exact linear perturbation equations of SFDM, but also to include the impact of the other constituents of the Universe in a much more accurate manner. Hence, in addition to the fiducial model of \hyperlinkcite{Li2014}{\PaperLi}, we also investigated here other \acidx{SFDM} model parameters, notably such ones with reduced \acidx{SI} strength (i.e. smaller  $R_{TF})$, as were studied by \hyperlinkcite{Shapiro2021}{\PaperSDR}. However, in that paper the particle mass $m$ by itself was not a free parameter, because the linear structure formation analysis there was restricted to the TF regime applied after the early stiff phase (to which that analysis was blind). In other words, the only parameter that mattered was the combination $\lambda/m^2$ in the expression for $R_{TF}$.  On the other hand, our CLASS models here are not blind to the stiff phase, and the models we probe have to conform to the cosmological constraints, derived by \hyperlinkcite{Li2014}{\PaperLi}. Therefore, we adapted a particle mass for each model of SFDM, such that the constraints are fulfilled. Foremost, the stiff phase of all models ends at the same time, early enough before BBN is over. In order to investigate the influence of the repulsive \acidx{SI} in a clean way, we also included the model for complex FDM (without SI) with the same particle mass ($m_{22}=30$) as the original fiducial \acidx{SFDM} model. As stressed earlier, in this model the stiff phase ends too late (i.e. after BBN), but for the sake of a better understanding of the results we included it nevertheless. 
Also, for the sake of comparison between complex vs. real scalar fields, we included ULAs. The parameters for complex SFDM, complex FDM and real ULA models are again summarized in Table \ref{tab:implications}.\par
\begin{table} [!htbp] 
	\caption{Model parameters for complex SFDM, complex FDM and real ULAs \label{tab:implications}}
	\begin{ruledtabular}
		\begin{tabular}{rrlr}
			m [eV/c$^2$]          &  m$_{22}$  & $\syidx{SYMlambdasi}/(m c^2)^2$ [eV$^{-1}$ cm$^3$] & $R_{TF}$\\
			\hline
			~                     & ~     & SFDM with repulsive SI    &~\\	
			$3.0 \times 10^{-21}$ & 30    & $2 \times 10^{-18}$       & 1.1 kpc\\
			$5.0 \times 10^{-20}$ & 500   & $2 \times 10^{-20}$       & 110 pc\\
			$5.0 \times 10^{-19}$ & 5000  & $2 \times 10^{-22}$       & 11 pc\\		
			$5.0 \times 10^{-18}$ & 50000 & $2 \times 10^{-24}$       & 1 pc\\
			\hline	
			~                     & ~     & complex FDM               &~\\						
			$3.0 \times 10^{-21}$ & 30    & --                        & --\\
			$8.0 \times 10^{-23}$ & 0.8   & --                        & --\\
			$3.0 \times 10^{-17}$ & $3\cdot 10^5$ & --                        & --\\
			\hline
			~                     & ~     & ULA                       &~\\		
			$3.0 \times 10^{-21}$ & 30    & --                        & --\\	
			$8.0 \times 10^{-23}$ & 0.8   & --                        & --\\
		\end{tabular}
	\end{ruledtabular}
\end{table}
\begin{figure} [!htbp]
	{\includegraphics[width=\PW\columnwidth]{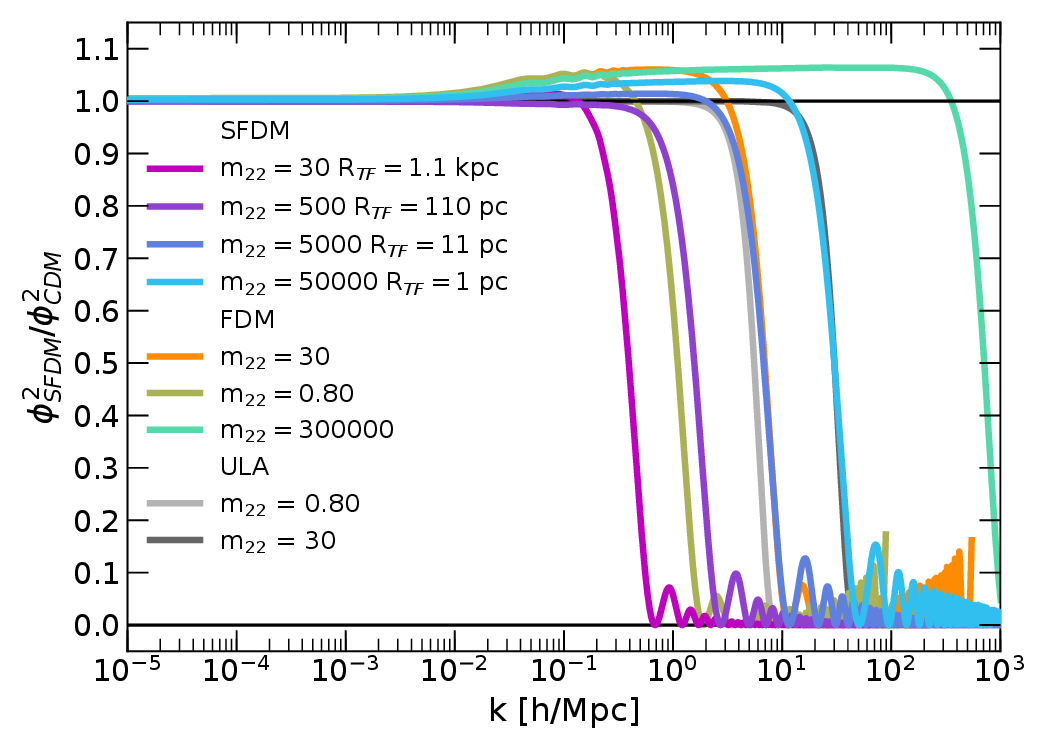}}	
	\caption[Transfer functions for multiple SFDM, FDM and ULA models]
	{\textbf{Transfer functions for multiple SFDM, FDM and ULA models}. 
	The colored solid lines display the transfer functions of SFDM with strength of \acidx{SI} corresponding to $R_{TF}=1.1$~kpc; $110;\;11;\:1$~pc, and complex FDM models with masses $m_{22}=30;\:0.8;\:3\cdot 10^5$, each relative to CDM. The stiff phase of the complex FDM models is suppressed in the plot, i.e. cut off at high $k$. The gray lines display two ULA models with $m_{22}=0.8; 30$.
	}
	\label{fig:SFDM-mTk-multi}
\end{figure}

\begin{figure} [!htbp]
	{\includegraphics[width=\PW\columnwidth]{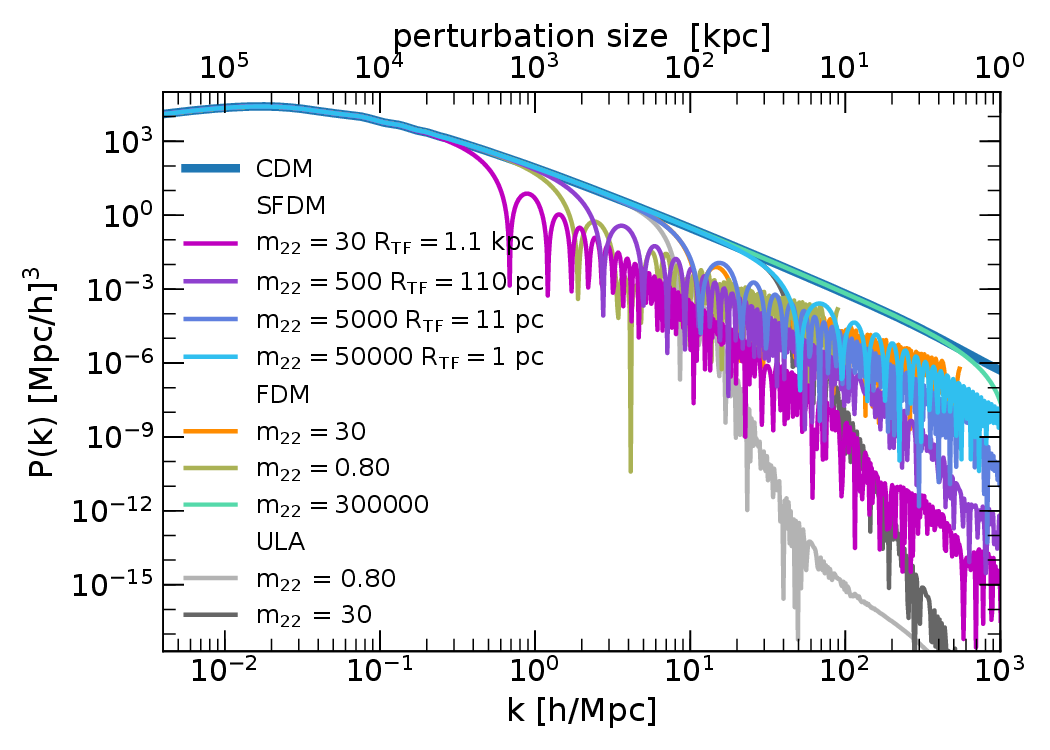}}			
	\caption[Matter power spectra for multiple SFDM, FDM and ULA models]
	{\textbf{Matter power spectra for multiple SFDM, FDM and ULA models}. The power spectra are displayed for SFDM models with different strength of \acidx{SI} corresponding to $R_{TF}=1.1$~kpc (magenta); $110$~pc (purple); $11$~pc (dark blue); $1$~pc (light blue) and for complex FDM models with different masses $m_{22}=0.8$~(light green); $30$~(orange).
	The CDM power spectrum is shown with a thick blue line, which overlaps to a great extent with the high-mass FDM model $m_{22}=3\cdot 10^5$. The gray lines display two ULA models with $m_{22}=0.8; 30$. Note the expanded $y$-axis range down to very small power.
	}
	\label{fig:SFDM-mPk-multi}
\end{figure}

Figure \ref{fig:SFDM-mTk-multi} compares all the transfer functions relative to that of CDM presented in one plot. In performing the comparison, we can ignore the ``overshooting'' of the transfer functions exhibited by some models, because we have seen above, that it has no effect on the power spectra. Indeed, the corresponding power spectra can be seen in Fig. \ref{fig:SFDM-mPk-multi}. For SFDM models, the cutoff occurs at smaller $k$, the larger $R_{TF}$, i.e. structure at increasingly larger and larger spatial scales is suppressed, the higher $R_{TF}$. Conversely, for smaller $R_{TF}$, the cutoff occurs at higher $k$. This way, the favored range of SI or $R_{TF}$ is subject to observational constraints.  Our results confirm the previous findings of \hyperlinkcite{Shapiro2021}{\PaperSDR}.

Now, if we look at the complex FDM model (orange solid line) with mass $m_{22}=30$, we find that its cutoff occurs at the same scale as the corresponding SFDM model with $R_{TF}=11$ pc, $m_{22}=5000$ (blue solid line), i.e. FDM corresponds to an SFDM model which has a significantly smaller structure-suppression scale, a result that we pointed out already upon comparison of the respective matter power spectra.
Furthermore, the same cutoff occurs for the ULA model with a particle mass $m_{22} = 0.8$ (light gray line). This last correspondence of similar cutoff scales between low-$m$ ULA models on the one hand, and small-$R_{TF}$ SFDM models on the other hand, has been found in \hyperlinkcite{Shapiro2021}{\PaperSDR}. In fact,
our results are in agreement with their relationship (71) that describes this correspondence, and it can be also seen upon comparison of our transfer functions to their Figure 6 (top panel). 


Let us elaborate now on the difference in the cutoff scale between complex FDM models and real ULA models of the \textit{same mass}.
For complex FDM, the cutoff occurs at smaller $k$ than for real ULAs, and the reason is the larger Jeans mass of complex FDM models, as follows. The Jeans scale for the ULA models is determined by that pressure which stems from the large-scale effects of quantum pressure, corresponding to $\lambda_{deB}$. This same pressure is also at play for complex FDM models. However, here the particles gain additional kinetic energy from the phase of the complex SF, which also accounts for the early stiff phase in the background evolution of complex FDM (or SFDM, for that matter). Now, this additional contribution to the kinetic energy, as a function of the frequency $\omega$ of the phase of the SF, evolves initially $\propto a^{-3}$, followed by a constant value depending on the particle mass (see Appendix B in \hyperlinkcite{Li2014}{\PaperLi}). This additional contribution to the kinetic energy of the particles is the source for the overall higher pressure, yielding a larger Jeans scale for complex FDM models, which results in a cutoff at smaller $k$, compared to real-field ULA models of same mass. This can be seen clearly by comparing complex FDM with $m_{22} = 30$ (solid orange line) to a real ULA model with identical mass (dark gray line). The latter displays the same cutoff as the SFDM model with $R_{TF} = 1$~pc, $m_{22}=5\cdot 10^4$. The same effect can be seen for the corresponding models with mass $m_{22}=0.8$, which shifts the cutoff for complex FDM (solid green line) to smaller $k$, close to that of the SFDM model with $R_{TF}=110$~pc, $m_{22}=500$ (solid purple line). The fact that the stiff phase ends too late for this particular FDM model has no impact on this discussion, because the stiff phase affects much smaller spatial scales, i.e. scales much below the respective cutoff. Also, we emphasize again that our results confirm previous findings of \hyperlinkcite{Shapiro2021}{\PaperSDR} concerning the differences between SFDM and ULA models, as seen in their Figure 6 (top panel).\par

Figure \ref{fig:SFDM-mPk-multi} displays the matter power spectra for the same models, as in Fig.~\ref{fig:SFDM-mTk-multi}, where the range of the $y$-axis is again expanded down to very small power. As expected, the cutoff in the power spectra reflects the behavior of the corresponding transfer functions. We stress that the cutoffs occur at physical scales quite larger than the corresponding values for $R_{TF}$ suggest. For example, for the fiducial model with $R_{TF}=1.1$~kpc, the cutoff in the power spectrum occurs at $\sim 3 \times 10^4$~kpc. Furthermore, the slope of the matter power spectrum toward high $k$ is almost identical between different SFDM models. On the other hand, those complex FDM models, which exhibit a too long lasting stiff phase, display a shallower falloff of power. This can be also seen in Fig.~\ref{fig:SFDM-mTk-multi}, as the amplitudes for the relative transfer functions of the complex FDM models increase toward high $k$. In contrast, the complex FDM model of high mass, $m_{22}=3\cdot 10^5$, that fulfills the cosmological constraints, evolves almost exactly like CDM, as expected. In addition, we see wiggles in the envelope of the matter power spectra of SFDM and FDM toward high $k$, which correspond to those seen in the transfer functions, a feature well known for scalar field DM models, in general. As in the previous section, we can see that the initial very sharp cutoff for the ULA models also flattens toward high $k$ (around $k \gtrsim 10^2$).  
We believe this happens, because of the long-lasting period of growth suppression of sub-Jeans perturbations from the time of horizon entry before becoming super-Jeans, as already discussed above.
\begin{figure} [!htbp]
	{\includegraphics[width=\PW\columnwidth]{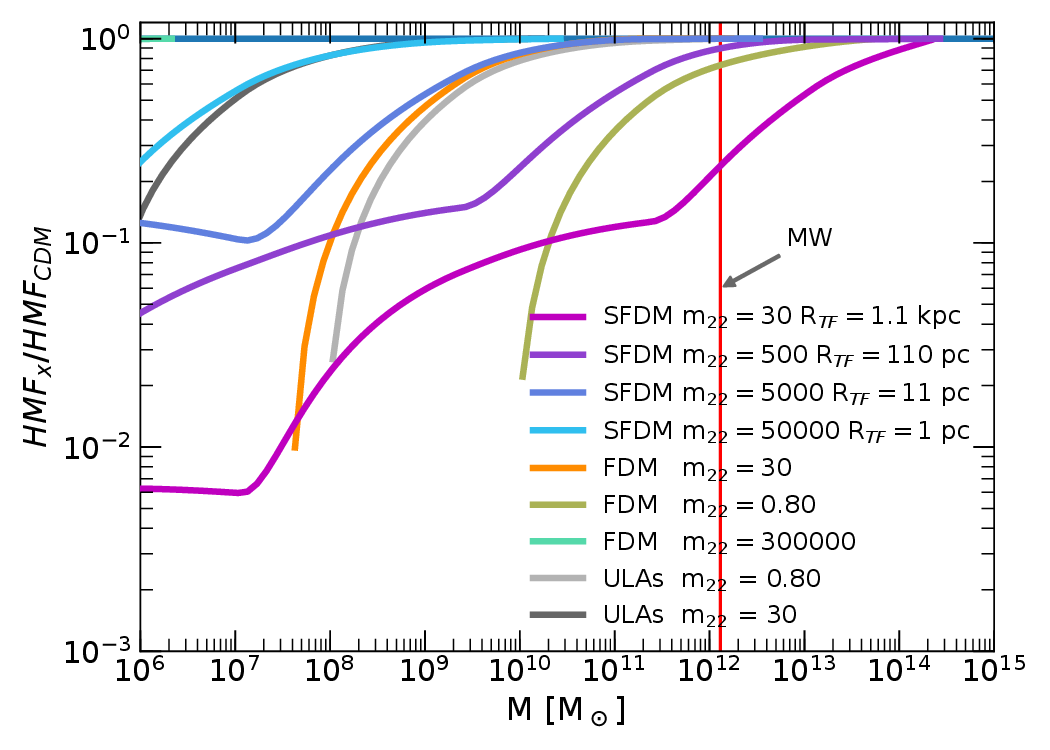}}			
	\caption[Suppression of structure relative to CDM]
	{\textbf{Suppression of structure relative to CDM}. The halo mass function (HMF) shows the suppression of structure relative to CDM for various models of SFDM, FDM and ULAs, as indicated by the legend. The vertical red line indicates the halo mass of the Milky Way galaxy, according to \citet{Posti2019}. This plot may be compared to Figure 8 in \hyperlinkcite{Shapiro2021}{\PaperSDR}.
	}
	\label{fig:SFDM-suppression}
\end{figure}

In any case, the feature identified for SFDM in \hyperlinkcite{Shapiro2021}{\PaperSDR} and confirmed in our work here, whereby an initial cutoff in power is followed by a shallower falloff thereafter, is also happening with ULA models, but at very much smaller power than for SFDM (this is the main reason that we show such an extended plot range in the $y$-axis of the power spectra). This feature is probably of no significance to observational constraints for ULAs, given the small powers where it occurs.  

However, that feature has an impact on complex-field FDM and SFDM models, in terms of the amount of structure suppression that can be observationally probed. As in \hyperlinkcite{Shapiro2021}{\PaperSDR}, we compare in Fig.~\ref{fig:SFDM-suppression} our results to observable quantities by computing the Press-Schechter (\citet{Press1974}) halo mass function (HMF). We use the LSS-Toolkit by \citet{Diemer2018}, based on the matter power spectra gained from our simulation runs of the various models. Figure \ref{fig:SFDM-suppression} displays them normalized to the HMF of CDM. We see the same correlation between the individual models, like in our matter power spectra. Comparing our results to those seen in Figure~8 of \hyperlinkcite{Shapiro2021}{\PaperSDR}, we recognize a ``wavelike'' feature in the HMF for the SFDM models. Detailed analysis has shown that these stem from the wiggles in the envelope of the corresponding matter power spectra at high $k$. In contrast to \hyperlinkcite{Shapiro2021}{\PaperSDR}, we used a standard sharp-$k$ filter in the computation, instead of the calibrated filter that they used, but since our aim is an overall comparison of the HMF, we do not further pursue this point. The red vertical line in the figure indicates the mass of the Milky Way's DM halo (M$_{200}=1.3 \times 10^{12}$~M$_\odot$ from observations of Gaia and the Hubble Space Telescope, see \citet{Posti2019}). It can be clearly seen, that the fiducial SFDM model (R$_{TF}=1.1$~kpc, $m_{22}=30$) of \hyperlinkcite{Li_2014}{\PaperLi} shows a too strong suppression of structure, as even scales of the size of the Milky Way are suppressed, compared to CDM. Again, this confirms the earlier findings of \hyperlinkcite{Shapiro2021}{\PaperSDR}. The upper bound on the SI strength has to be around $\lambda/(mc^2)^2 = 2 \times 10^{-22}$~eV$^{-1}$ cm$^3$, corresponding to small core radii of $R_{TF} = 11$~pc and $m_{22} \approx 5 \times 10^{3}$, in order to allow structures of size the Milky Way to form in an abundance expected from CDM. Therefore, as in \hyperlinkcite{Shapiro2021}{\PaperSDR} we may conclude that SFDM cannot resolve the small-scale problems of CDM. However, as discussed in \hyperlinkcite{Shapiro2021}{\PaperSDR}, the unconditional HMF has its own severe limitations, for it does not take into account the detailed merger histories of halos, nor tidal effects and other dynamical impacts over time. And the fact that the suppression of structure in SFDM below the cutoff ensues with a milder slope, compared to FDM or ULAs, allows for more nuances in the detailed abundances of structure at sub-kpc scales, a point that is also discussed in detail in \hyperlinkcite{Shapiro2021}{\PaperSDR}, and which requires more future investigations. 

Finally, the overall conclusion, however, seems to be further strengthened by the even more recent work of \citet{Hartman2022}, who investigated the impact of several SFDM models on large-scale observables, also modifying CLASS. In contrast to us, they derived a different fluid approximation of SFDM (considering only SI via $R_{TF}$ and neglecting the stiff phase), based on the Madelung equations (\citet{Madelung1927}) and applied to cosmological perturbation theory. They worked out constraints on the model parameters by applying Markov Chain Monte Carlo methods to fit the models to observational data obtained from the CMB, BAOs, growth factor measurements, and type Ia supernovae distances. As a result, they found an upper bound for SI at $R_{TF} \approx 500$~pc within $1\sigma$, and that SI with $R_{TF} \approx 1$~kpc is disfavored by $2.4\sigma$.
So, taken together, all these studies, along with ours, point to the conclusion that additional baryonic feedback physics would be required in SFDM-TF to explain observations at small galactic scales, after all. 

Now, what about complex FDM, or real ULA models? As found in \hyperlinkcite{Shapiro2021}{\PaperSDR} and confirmed here (see discussion above), a given FDM/ULA model with mass $m$ corresponds to a SFDM-TF model with certain $R_{TF}$, which is much smaller than the de Broglie length, $\lb \sim 1/m$, belonging to this FDM/ULA model. That means that a similar cutoff in the matter power spectra (or HMFs) between SFDM-TF and FDM/ULA will imply a different conclusion with regard to the expected core size of equilibrium halos in the respective models. For the same cutoff, the implied core size is much smaller for SFDM-TF. Therefore, observables from linear theory have an immediate impact onto the ability of SFDM-TF to resolve the small-scale problems. Of course, the same mass $m$ of FDM/ULA models constrained by power spectra, affects their $\lb$, with consequences for core radii, too. But the constraints are less potent in that case, because kpc-size cores are not quite excluded from linear observables.  On the other hand, recent literature has posed more severe upper-bound estimates for the $\lb$ of FDM/ULAs, using the size of Local Group dwarf galaxies in \citet{Nadler2021}, or certain ultrafaint galaxies such as Eridanus II in \citet{Marsh2019}. These works point to values of $m \sim 10^{-21}$ eV/c$^2$, i.e. close to those found earlier using Lyman-$\alpha$ forest measurements by \citet{Irsic2017}, \citet{Armengaud2017}. It is these latest constraints that seem to suggest that FDM/ULA models may also be subject to smaller halo core sizes than originally hoped, questioning their ability to resolve the small-scale problems.  
Before we leave this section, we also point out that some studies conclude that the small-scale issues of CDM may well be resolved using baryon feedback, see e.g. \cite{smallscale1, Fattahi2016, Sawala2017}. Therefore, similar feedback scenarios may have to be invoked for SFDM, in case the conclusions put forth here hold up in the future.


\section{Summary}\label{sec:conclusion}

In this paper, we studied the evolution of the background and linear structure formation of $\Lambda$SFDM universes by expanding upon previous literature and performing a comparative study of model parameters of interest, where the SFDM particle mass is typically ultralight, $m \gtrsim 10^{-22}$ eV/c$^2$. In order to carry out our calculations, we modified the open-source Boltzmann code CLASS. Our emphasis was on complex-field scalar field dark matter (SFDM) with a strongly repulsive, quartic self-interaction (SI), also called SFDM-TF, and complex-field fuzzy dark matter (FDM) without SI. We compared our results pertaining to each of these models using our modified CLASS code with real-field ultralight axion models (ULAs) without SI; the ULA runs were performed using the modified CLASS code by \citet{UrenaLopez2016}. 

In general, SFDM models have become popular as alternatives to the standard, collisionless cold dark matter (CDM) paradigm, because the characteristic scale in SFDM, below which structure formation is suppressed, can be of kpc-size, depending upon particle parameters. This way, SFDM could resolve the so-called small-scale problems that have plagued CDM for nearly three decades (cusp-core, missing-satellite, too-big-to-fail problems). However, recent studies have put ever smaller upper-bound limits on the suppression scale, and it remains to be seen whether SFDM can resolve the small-scale challenges, without the resort to the astrophysics of baryonic galactic feedback.

Our paper complements the work by \hyperlinkcite{Li2014}{\PaperLi}) who studied the background evolution of complex $\Lambda$SFDM models, which differ from $\Lambda$CDM in that CDM is replaced by a SFDM component of given mass $m$ and SI. It was established there that complex SFDM undergoes three distinctive phases in the Universe, an early stiff phase where SFDM is the dominating cosmic component, followed by a radiation-like phase during the radiation-dominated epoch of the Universe, and finally SFDM dominates again in the matter-dominated epoch, which is followed by $\Lambda$-domination thereafter. Thereby, new cosmological constraints from BBN and the CMB were derived for the background Universe. The stiff phase is a general characteristics of complex SFDM models. The implementation into CLASS of this early stiff phase prior to radiation-domination turned out to be a major challenge of our investigation here, but once established we were able not only to reproduce previous results, such as those of \hyperlinkcite{Li2014}{\PaperLi}, but also to expand them to calculate CMB and matter power spectra for various $\Lambda$SFDM models.
This way, our work also complements the recent results by \hyperlinkcite{Shapiro2021}{\PaperSDR}, who applied a semi-analytical approach to the linear structure formation by including SFDM-TF and radiation perturbations, deriving novel constraints on the particle parameters. In particular, they showed that
the characteristic suppression scale of SFDM-TF, that is related to the central core radius $R_{TF}$ of halos, is more severely constrained than previously thought. More precisely, in order to come up with a present-day halo mass function close to CDM above a reasonable mass scale, the SFDM-TF particle parameters have to be chosen such that sub-kpc halo core radii, $R_{TF} \lesssim 1-100$ pc, result. Hence, the ability of SFDM-TF to resolve the small-scale problems has been under question. A similar conclusion has been drawn in  
\citet{Hartman2022}, who also find that sub-kpc core radii, $R_{TF} \approx 500$~pc, are preferred, using linear observables and other indicators. The results of our paper here confirm the previous findings of \hyperlinkcite{Shapiro2021}{\PaperSDR}.

Our most important findings can be highlighted, as follows:

\begin{itemize}
    \item We modified CLASS to incorporate the early stiff phase with equation-of-state $w = p/\rho = 1$, prior to radiation-domination, that is a characteristics of complex-field SFDM models. We confirmed earlier results with regard to the background evolution of $\Lambda$SFDM models, and compared them to real-field ULA models that lack a stiff phase. 
    \item The evolution of the background pressure in the oscillatory phase of the scalar field differs between complex and real fields: in the former case, pressure falls off more mildly, while for the latter the drop in pressure occurs almost immediately with the onset of the oscillation, see Fig.~\ref{fig:SFDM-EOS-rho} and Fig.~\ref{fig:FDM-UL-rho}.  
    \item Complex FDM and SFDM models share as a common pressure source the kinetic energy due to the phase of the complex field. This leads to a cutoff at lower $k$ in the matter power spectra, compared to real-field models of comparable particle mass. Also, if SI is included as a second source of pressure, the cutoff shifts to even larger spatial scales, i.e. to even lower $k$.
    Apart from these features in the matter power spectra, we found that the impact of the stiff phase is limited to perturbations of very small scale (below $\sim$ pc) that enter the Universe very early, and are hence irrelevant with respect to galaxy formation. 
    \item We confirmed the earlier findings by \hyperlinkcite{Shapiro2021}{\PaperSDR}, not only with respect to the cutoff just mentioned, but also with regard to the ensuing falloff (i.e. the slope) of the matter power spectra. The slopes are similar to CDM, because the evolution history of perturbations is similar. Since the post-stiff phase background evolution is almost identical, the horizon entry of modes, is similar to CDM. Moreover, after the suppression of growth, the amplitude of perturbations start growing at similar times, although based on different physical effects. In the context of SFDM this means that modes, that are suppressed early on, can recover later in time, implying that there remains more small-scale structure than the low-$k$ cutoff would suggest.
    \item We confirmed previous results of \hyperlinkcite{Li2014}{\PaperLi}, pertaining to the background evolution of complex-field SFDM models. In particular, we could show that complex FDM models are subject to tighter cosmological and linear constraints than real ULA models; FDM models require a much higher particle mass than ULAs. 
    \item We confirmed the sharp cutoff in the matter power spectra of ULAs, as known from previous literature, but we find in addition that this initial drop is also followed subsequently by a milder decline toward high $k$. However, this transitional feature occurs only at very low power, beyond the sensitivity of current observational probes.
    \item We found that SFDM models exemplify acoustic pressure oscillations (``SAOs''), akin to BAOs, but as a result of scalar field oscillations. We speculate these might produce similar ``ring-like imprints'' onto the dark matter large-scale structure, or onto a cosmic gravitational wave background, but we must leave this idea to future studies.  
\end{itemize}

\begin{acknowledgments}
T.R.-D. is supported by the Austrian Science Fund FWF through an Elise Richter fellowship, grant no. V 656-N28. 
\end{acknowledgments}

\end{document}